\documentclass[journal,12pt,onecolumn,draftclsnofoot,]{IEEEtran}
\usepackage{cite}
\usepackage{amsmath,amssymb,amsfonts}
\usepackage{algorithm}
\usepackage{algpseudocode}
\usepackage{graphicx}
\usepackage{textcomp}
\usepackage{xcolor}
\usepackage{hyperref}
\usepackage{multirow}
\usepackage{url}
\usepackage{subcaption}
\makeatletter
\g@addto@macro{\UrlBreaks}{\UrlOrds}
\makeatother
\def\BibTeX{{\rm B\kern-.05em{\sc i\kern-.025em b}\kern-.08em
		T\kern-.1667em\lower.7ex\hbox{E}\kern-.125emX}}

\newcommand{\real}{\,\mathbb{R}}
\newcommand*{\rom}[1]{\expandafter\@slowromancap\romannumeral #1@}

\begin{document}
	
	\title{Optical Coherence Tomography Image Enhancement via Block Hankelization and Low Rank Tensor Network Approximation
		}
	
			\author{\IEEEauthorblockN{Farnaz Sedighin\IEEEauthorrefmark{1}, Andrzej Cichocki\IEEEauthorrefmark{2} and Hossein Rabbani\IEEEauthorrefmark{1}} \\					
						\IEEEauthorblockA{\IEEEauthorrefmark{1}
							Medical Image and
							Signal Processing Research Center, School of advanced Technologies in Medicine, Isfahan University of Medical Sciences, Isfahan 8174673461, Iran\\}
						Email:  f.sedighin@amt.mui.ac.ir, h\_rabbani@med.mui.ac.ir\\
						\IEEEauthorblockA{\IEEEauthorrefmark{2}
						Systems Research Institute PAS, 00-847 Warsaw\\}
						Email: cichockiand@gmail.com
					\thanks{Corresponding author: Farnaz Sedighin}}
				\maketitle
	\begin{abstract}
		In this paper, the problem of  image super-resolution  for Optical Coherence
		Tomography (OCT) has been addressed.
		Due to the motion artifacts, OCT imaging is usually done with a
		low sampling rate and the resulting images are often noisy and have low resolution. Therefore, reconstruction of high resolution
		OCT images from the low resolution versions is an essential step
		for better OCT based diagnosis. In this paper, we  propose a novel
		OCT super-resolution technique using Tensor Ring decomposition 
		in the embedded space. A new tensorization method
		based on a block  Hankelization approach with overlapped
		patches, called overlapped patch Hankelization, has been proposed  which allows us to employ  Tensor Ring decomposition. The Hankelization method enables us to better exploit the inter connection of pixels and consequently achieve better super-resolution of images. The low resolution image was first patch Hankelized and then its Tensor Ring decomposition with rank incremental has been computed. Simulation results  confirm that the proposed approach is effective in OCT super-resolution.
		\\	
	

\end{abstract}
	\section{Introduction}
	Optical Coherence Tomography (OCT) is an important non-invasive imaging technique  used for retina disease diagnosis \cite{huang1991optical,fercher1996optical,fercher2003optical,spaide2018optical,wang2022ee,karthik2023convolution, medhi2023improved, diao2023classification, abbasi2019three,pekala2019deep,dong2022spatially}. Many eye and brain diseases such as Age related Macular Degeneration (AMD), diabetic retinopathy or other non-diabetic diseases can be diagnosed using high quality OCT images \cite{virgili2015optical}.

	 For preventing motion artifacts resulting from blinking or patient movements, OCT images are usually captured in a low sampling rate \cite{daneshmand2020super},  which results in low resolution and noisy OCT images. However, for a more accurate OCT based diagnosis, the resolution of these images should be increased and high quality images should be prepared. Considering this fact, de-noising and super-resolution of OCT images have been investigated by many papers \cite{li2017statistical, amini2017optical, daneshmand2020super, daneshmand2020reconstruction}.
	
	 In order to increase the resolution of the low resolution OCT images, many approaches have been proposed \cite{daneshmand2020super,wang2018super, yuan2020axial,huang2019simultaneous,fang2013fast, qiu2021n2nsr,cao2020super,das2020unsupervised,achim2023space}. In \cite{fang2013fast}, an approach called  Sparsity Based Simultaneous Denoising and Interpolation (SBSDI) has been proposed which is based on the sparse representation of the OCT images. In SBDI, denoising and super-resolution of OCT images have been done simultaneously. An approach based on using Alternating Direction Method of Multipliers (ADMM) and Forward Backward Splitting (FBS) has been proposed in \cite{wang2018super}. Learning Dictionaries from high resolution OCT images and imposing sparsity for OCT super-resolution has been proposed in \cite{scrivanti2021non}.
	 
	  Deep learning methods have been also exploited for OCT super-resolution \cite{yuan2020axial,qiu2021n2nsr,cao2020super}. In \cite{qiu2021n2nsr}, semi-supervised U-net and DBPN have been used for OCT super-resolution and their performances have been verified. \cite{cao2020super} used a new deep learning based super-resolution approach by modifying Generative Adversarial Network (SR-GAN). Generative Adversarial Network has been also exploited in \cite{das2020unsupervised, yuan2022image} for OCT super-resolution.
	  
	    Ghaderi, et.al, have been proposed two approaches called LOw rank First Order Tensor based Total Variation (LOFOTTV) and  LOw rank Second Order Tensor based Total Variation (LOSOTTV) for OCT super-resolution. The approaches have been exploited the low rank assumption of a tensor resulting from combination of patches of the 3D OCT image and FOTTV (SOTTV).

	In many of biomedical signal and image processing research areas, captured data is of higher order dimension (3rd order or more). In addition, sometimes, several modalities from a physical phenomenon have been simultaneously recorded and analyzed together. An example is when  several B-scans from one subject have been recorded. In such scenarios, recording and representing the data in a matrix is inefficient or sometimes, inconvenient. Moreover, in some cases transferring a raw data into higher order spaces can increase the processing performance in comparison to the processing of low order data.  These facts have motivated researchers to move from matrices to tensors \cite{cichocki2014era,cichocki2015tensor, cichocki2016low,zhao2018spatial}. 
	
	Tensors are higher order arrays (3-rd or more) widely used for analyzing higher order datasets, especially images \cite{kolda2009tensor}. 	
	Tensor decomposition is an important approach for analyzing the data represented by a tensor \cite{kolda2009tensor, cichocki2015tensor}. In tensor decomposition, a higher order tensor is decomposed approximately into factor matrices or smaller and lower order core tensors \cite{kolda2008scalable}. The resulting latent variables, usually preserve the main information and structure of the original tensor and can be used for further analyzing and processing of the original tensor. 
	
	basic tensor decomposition approaches consist of  Canonical Polyadic (CP) and Tucker decompositions \cite{kolda2009tensor}. In CP decomposition, the original tensor decomposes into  sum of rank-one tensors. The number of rank-one tensors is known as the CP rank of the original tensor \cite{phan2013candecomp,sorber2013optimization}. In Tucker decomposition, an $N$-th order original tensor is decomposed into an $N$-th order core tensor and $N$ factor matrices \cite{kim2007nonnegative, zhou2015efficient}. In Tucker decomposition  the number of elements resulting from the decomposition of the original tensor, increases exponentially with the order of the original tensor, called curse of dimensionality \cite{cichocki2016low}. This limits using of Tucker decomposition for higher order tensors \cite{cichocki2016low}.
	
	For overcoming curse of dimensionality, a family of tensor decomposition algorithms, known as tensor networks, has been proposed \cite{cichocki2016low, pan2022tednet}. In tensor networks, the number of elements resulting from decomposition, increases linearly with the tensor order, and hence they are more proper for analyzing higher order tensors.

	Two important members of tensor networks are Tensor Train (TT) and Tensor Ring (TR) decompositions  \cite{oseledets2011tensor,zhao2016tensor,zhao2019learning}. These two tensor decomposition approaches have been widely used in different signal and image reconstruction approaches \cite{sedighin2021image,yu2019tensor,yokota2018missing,yuan2018tensor}. TT and TR decompositions are  basically applicable for higher order tensors. So in many of  algorithms the raw data is reformatted into a higher order tensor before applying TT or TR decomposition\cite{yokota2018missing}. Hankelization is a common approach for transferring a low order dataset  into a higher order one, which has been used as an initial step for many of TT or TR based  algorithm \cite{yokota2018missing,sedighin2020matrix}. The resulting higher order dataset is said to be in the embedded space. 
	
	Another important issue for efficient use of TT or TR decomposition is Selecting proper ranks (details will be discussed in the following subsections). Simulations show that the TR ranks cannot be selected to small or too large where both cases reduce the algorithm performance \cite{yuan2018higher}.  For overcoming this problem, several rank incremental approaches have been proposed \cite{yang2018tensor,sedighin2020matrix}. In rank incremental methods, instead of selecting fixed initial ranks, the TT or TR ranks are increased gradually during several iterations. This has been shown to be more effective than selecting fixed ranks \cite{yokota2018missing}.

	In this paper, we mainly focus on OCT super-resolution in the embedded space using patch Hankelization and  TR decomposition. This is the first time for applying Hankelization and TR decomposition for analyzing OCT images. 
	A new Hankelization approach, called overlapped patch Hankelization, which is based on a modification of patch Hankelization approach of \cite{sedighin2020matrix}, has been proposed. In this new Hankelization approach, in contrast to \cite{sedighin2020matrix}, the patches have overlap with each other. The amount of overlap can be set as an initial parameter for the algorithm. This can prevent (or improve) the mosaic view of the final result. After Hankelization the raw data into a high order tensor, TR decomposition with rank incremental is applied for the super-resolution and finally the resulting dataset in the embedded space is transferred back into the original image space. 
	
	Contributions of this paper can be summarized as
	\begin{itemize}
		\item{Introducing a modified Hankelization approach with overlapped patches which prevents the mosaic view of the output.}
		\item {The first time (to the best of our knowledge) develop a method for  OCT super-resolution by using TR decomposition with rank incremental.} 
		\item {Simultaneous denoising and super-resolution of OCT images by gradually increasing the TR ranks and  improving the quality of the resulting high resolution OCT in comparison to the other existing approaches.}
		
	\end{itemize}
	
	The paper has been organized as follows: Notations and preliminaries have been presented in Section.~\ref{not}. TT and TR decompositions are reviewed in Section.~\ref{tt-tr}. Section.~\ref{recon} is devoted to the tensor based reconstruction algorithms.  
	The proposed approach will be introduced in Section.~\ref{alg}. Simulation results and conclusion will be presented in Sections ~\ref{sim} and \ref{conc}, respectively.

	\section{Notations and preliminaries}
	\label{not}
	Notations used in this paper are adapted from \cite{cichocki2016low}. Vectors, matrices and tensors are denoted by small bold letters, capital bold letters and underlined capital bold letters, respectively. The $(i_1,i_2,\ldots,i_N)$-th element of $\underline{\textbf{X}}$ denoted by $x_{i_1,i_2,\ldots,i_N}$. Hadamard (element wise) product of two tensors ot two matrices is denoted by $\circledast$.  Mode-{n} unfolding of an $N$-th order tensor of size $I_1\times I_2 \times \cdots \times I_N$, denoted by $\textbf{X}_{(n)}$ which is an $I_n\times I_{n+1}\ldots I_NI_1\ldots I_{N-1}$ matrix. Mode-{n} canonical unfolding of tensor $\underline{\textbf{X}}$ is also defined as $\textbf{X}_{[n]}$ which is an $I_1\ldots I_n\times I_{n+1}\ldots I_N$ matrix. Transpose operator is denoted by $(.)^T$, the Frobenius norm of a tensor or matrix is denoted by $\|.\|_F$, and the trace operator is denoted by ``$tr$".
	 	
	\section{Tensor Train and Tensor Ring decompositions}
	\label{tt-tr}
	TT decomposition, factorizes an $N$-th order tensor $\underline{\textbf{X}} \in \real^{I_1\times I_2\times \cdots \times I_N}$ into a series of $N$ third order core tensors inter-connected in a train \cite{oseledets2011tensor}. The $n$-th core tensor is $\underline{\textbf{G}}^{(n)} \in \real^{R_{n-1}\times I_n\times R_{n}}$, where $I_n$ is the size of the $n$-th mode of the original tensor and $[R_0, R_1,\ldots, R_{N}]$ is the TT rank vector. In TT decomposition, $R_0=R_N=1$ and consequently, the first and the last core tensors are two matrices \cite{oseledets2011tensor}. 
	
\begin{figure}[t!]
			\centering
			\begin{subfigure}[t!]{1.0\linewidth}
			\centerline{
				\includegraphics[width=8cm, height=3.5cm, trim={7cm 6cm 5cm 3.5cm}, clip]{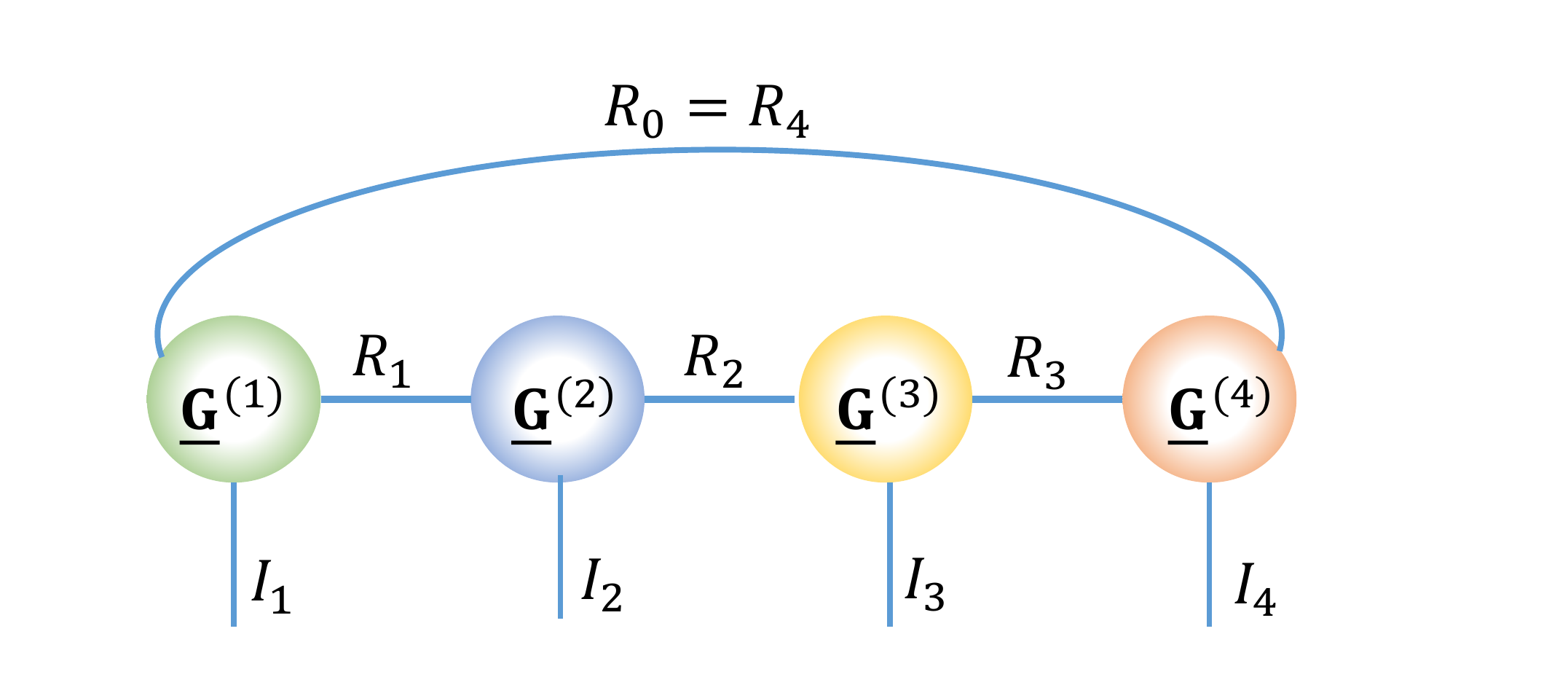}
						}
		\caption{}
		\label{tr}
		\end{subfigure}
	\hfill
		\begin{subfigure}[t!]{1.0\linewidth}
			\centerline{
				\includegraphics[width=8cm, height=3cm, trim={5cm 7cm 5cm 4.5cm}, clip]{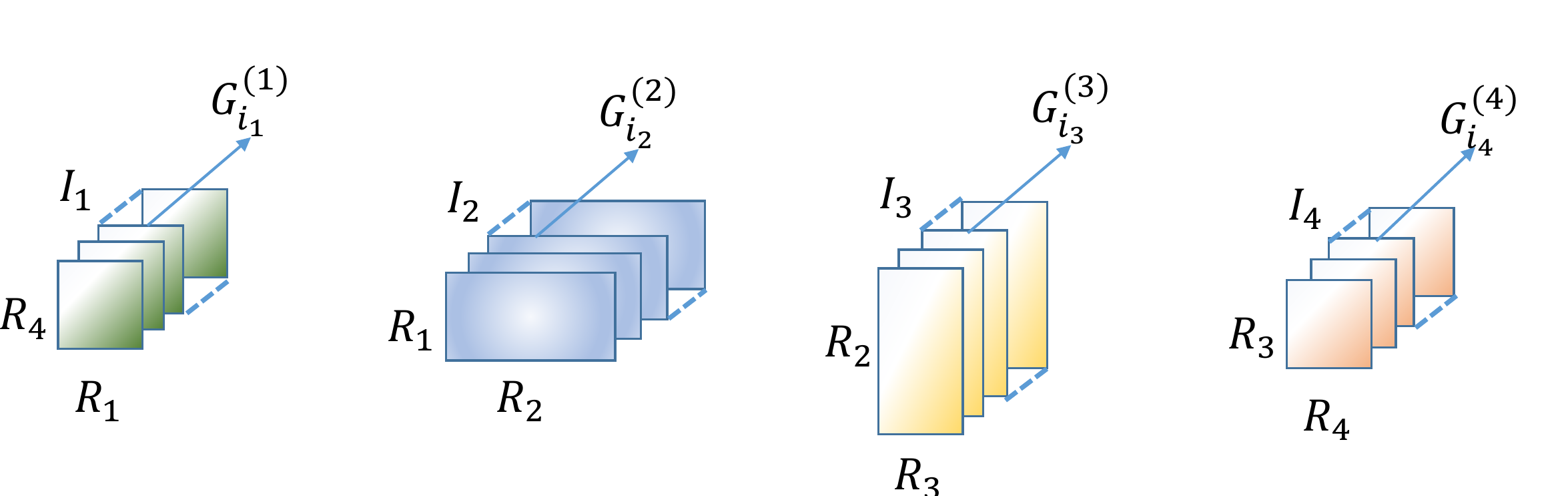}
			}
		\caption{}
		\label{tr-slice}
		\end{subfigure}
		\caption{(a): Tensor Ring (TR) decomposition of a 4-th order tensor. In special case  for $R_0=R_4=1$ the tensor ring simplifies to tensor train. (b): Representation of TR structure in form of slices. }
		\label{tr-general}
	\end{figure}

	In TR decomposition, similar to TT decomposition, an $N$-th order tensor is decomposed into a series of third order core tensors, where the core tensors are inter-connected in a chain \cite{zhao2016tensor}, as
	\begin{equation}
	\label{treq}
	\underline{\textbf{X}} \approx  \ll \underline{\textbf{G}}^{(1)}, \underline{\textbf{G}}^{(2)}, \ldots, \underline{\textbf{G}}^{(N)}\gg,
	\end{equation}	
	 where $\underline{\textbf{G}}^{(n)} \in \real^{R_{n-1}\times I_n\times R_{n}}$ is the $n$-th core tensor, $I_n$ is the size of the $n$-th mode and $[R_0, R_1,\ldots, R_{N}]$ is the TR rank vector. In contrast to TT decomposition, in TR decomposition, $R_0=R_N >1$. Using this definition, TR decomposition can be considered as a generalized version of TT decomposition \cite{zhao2016tensor}.  A schematic illustration of TR decomposition of a 4-th order tensor of size $I_1 \times I_2 \times I_3 \times I_4$, has been shown in Fig.~\ref{tr-general}.   In this figure, each colored circle shows a third order core tensor.
	Each element of a tensor with TR structure computed as:
	\begin{equation}
	\label{element}
	x_{i_1,i_2,\ldots,i_N}\approx  tr(\textbf{G}^{(1)}_{i_1}\textbf{G}^{(1)}_{i_2}\cdots \textbf{G}^{(1)}_{i_N}),
	\end{equation}
	where $\textbf{G}^{(n)}_{i_n}$ is the $i_n$-th lateral slice of the $n$-th core tensor.
	
	Determining the proper TR rank for the TR-based algorithms is an important and difficult issue which can highly affect the performance of the algorithm \cite{cheng2020novel, yuan2018higher}. Several algorithms like \cite{yokota2018missing, sedighin2020matrix}, use incrementally increasing ranks  approaches for finding the optimal ranks. Therefore, the ranks of the TR decomposition, or equivalently the size of the core tensors, are not fixed but are increased gradually in each iteration, until the desired approximation accuracy is achieved. The recent studies indicate  that algorithms with rank incremental usually outperform the fixed rank approaches \cite{yokota2018missing}. 
	
	Recent studies for natural image reconstruction and super-resolution have been demonstrated that Hankelization and tensorization of an original incomplete or noisy image, allow us to effectively exploit the internal correlations of pixels and patches of an image, and consequently increases the reconstruction performance \cite{yokota2018missing,shi2020block}. Hankelization is an approach for transferring a lower order dataset (in the signal space) into a higher order one (in the embedded space) based on the Hankel structure. Recall that in a matrix with Hankel structure, all of the elements in each skew-diagonal are the same. 
	Different Hankelization approaches have been used for signal and image reconstruction. In \cite{yokota2018missing}, a Hankelization approach based on multiplication of duplication matrix, by different modes of the original tensor, has been proposed. In \cite{sedighin2020matrix}, patch (block) Hankelization which is based on Hankelization of patches of an image (or blocks of a dataset) has been proposed. The patch Hankelization achieved by multiplying patch duplication matrix by different modes of the original dataset followed by a folding step. The resulting higher order tensors can then be effectively analyzed by TT or TR decompositions. 
	
	\section{Tensor based reconstruction approaches}
	\label{recon}
	As mentioned in the Introduction, tensors have been widely used for different signal and image reconstruction and completion.

		Tensor based image completion and reconstruction approaches are divided into two main categories.  The first group contains the rank minimization based approaches. These algorithms are based on minimizing the ranks of different unfoldings of the original (incomplete) tensor. These algorithms have been widely used in image reconstruction and super-resolution including OCT super-resolution \cite{liu2012tensor,bengua2017efficient, ding2019low, daneshmand2020super, long2019low, he2022low, zhou2017tensor,liu2018image,yu2022online,zhang2021new,huang2020provable,ding2022tensor}. 
		
		The algorithms of the second group are based on different decompositions of the original tensor \cite{wang2017efficient,yuan2018higher,yuan2018tensor,asante2021matrix}. 
		As mentioned earlier, the latent factors resulting from tensor decomposition, usually preserve the information of the original tensor and can reconstruct the incomplete or low resolution tensor. 
		
		Several reconstruction methods based on CP \cite{sorensen2019fiber, yokota2016smooth, wang2022bayesian} and Tucker decompositions \cite{yokota2018missing, yamamoto2022fast,yu2022low} have been proposed.  TT and TR decompositions have also been used for image and signal reconstruction and completion \cite{cai2022provable,liu2022efficient,sedighin2021image,bengua2017efficient}.
		Simulation results presented by several papers \cite{yokota2018missing, sedighin2021image}, show that usually the algorithms of the second group are more efficient for natural image reconstruction with slices missing comparing to the algorithms of the first group.  Moreover, to  best of our knowledge,  tensor decomposition, and specially TR decomposition, has not been used for OCT super-resolution.

\section{The proposed approach}
\label{alg}
\subsection{Overlapped patch Hankelization}
\label{han}
\begin{figure}[t!]
	\begin{minipage}[b]{1.0\linewidth}
		\centering
		\centerline{
			\includegraphics[width=7.5cm, height=3cm, trim={5cm 5cm 5cm 3.5cm}, clip]{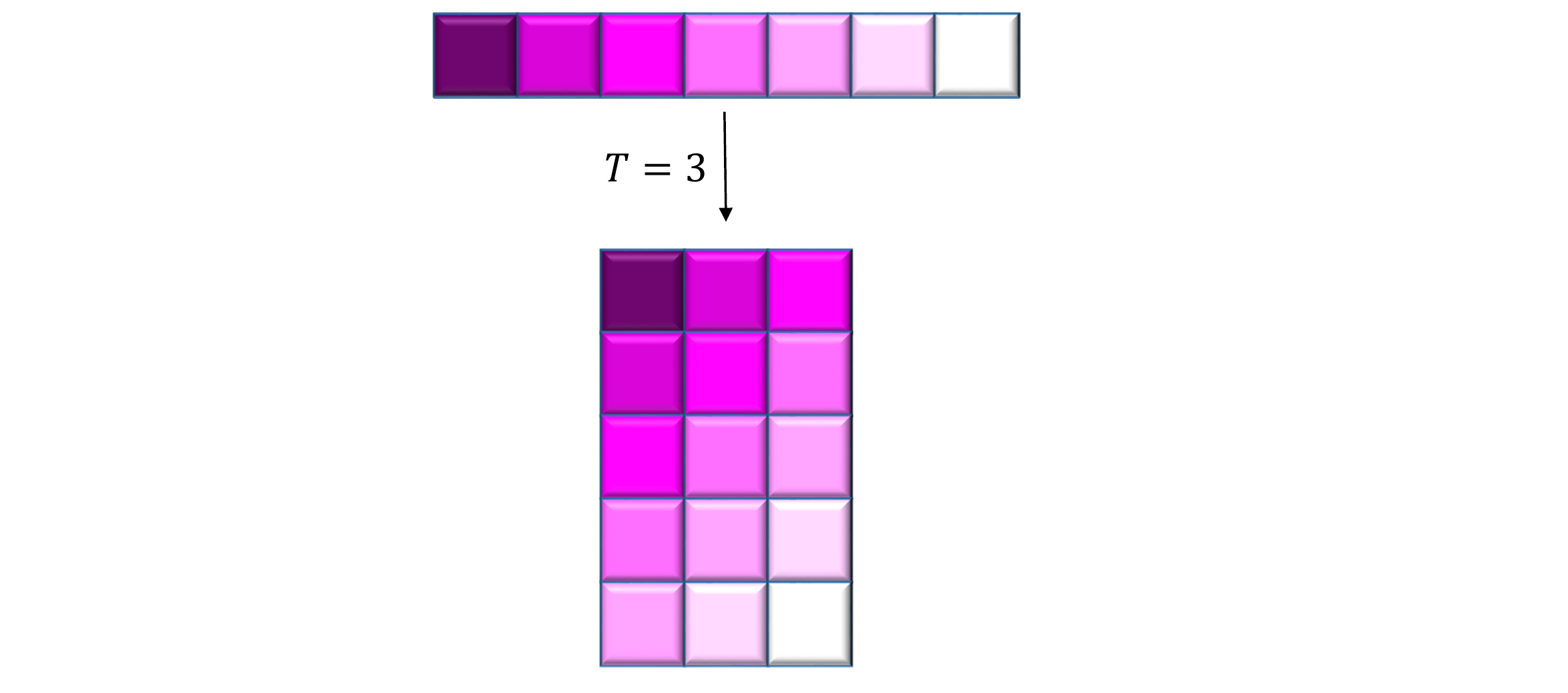}
		}
	\end{minipage}
	\caption{Hankelization of a one dimensional dataset by window size equal to 3.}
	\label{onehan}
\end{figure}

As mentioned earlier, Hankelization is an approach for transferring lower order datasets into higher order ones. In \cite{kalantari2018time,hassani2007singular,elsner1996singular}, Hankelization has been done for time series analysis. Hankelization of a one dimensional dataset is to reshape that dataset into a matrix with Hankel structure. This has been illustrated in Fig.~\ref{onehan}.

 In  recent studies \cite{yokota2018missing,yokota2020manifold}, Hankelization has been applied for  completion (reconstruction) and super-resolution of natural images. It has been demonstrated that by Hankelizing a dataset, correlations and connections between its elements and patches can be exploited more efficiently and this can improve the reconstruction power of the algorithms. In \cite{yokota2018missing} an approach  called MDT has been proposed for tensor reconstruction via  Hankelization of lower order datasets.  

\begin{figure}[t!]
		\centering
			\begin{subfigure}[t!]{1.0\linewidth}
				\centerline{
			\includegraphics[width=6cm, height=3cm, trim={6cm 3cm 5cm 5cm}, clip]{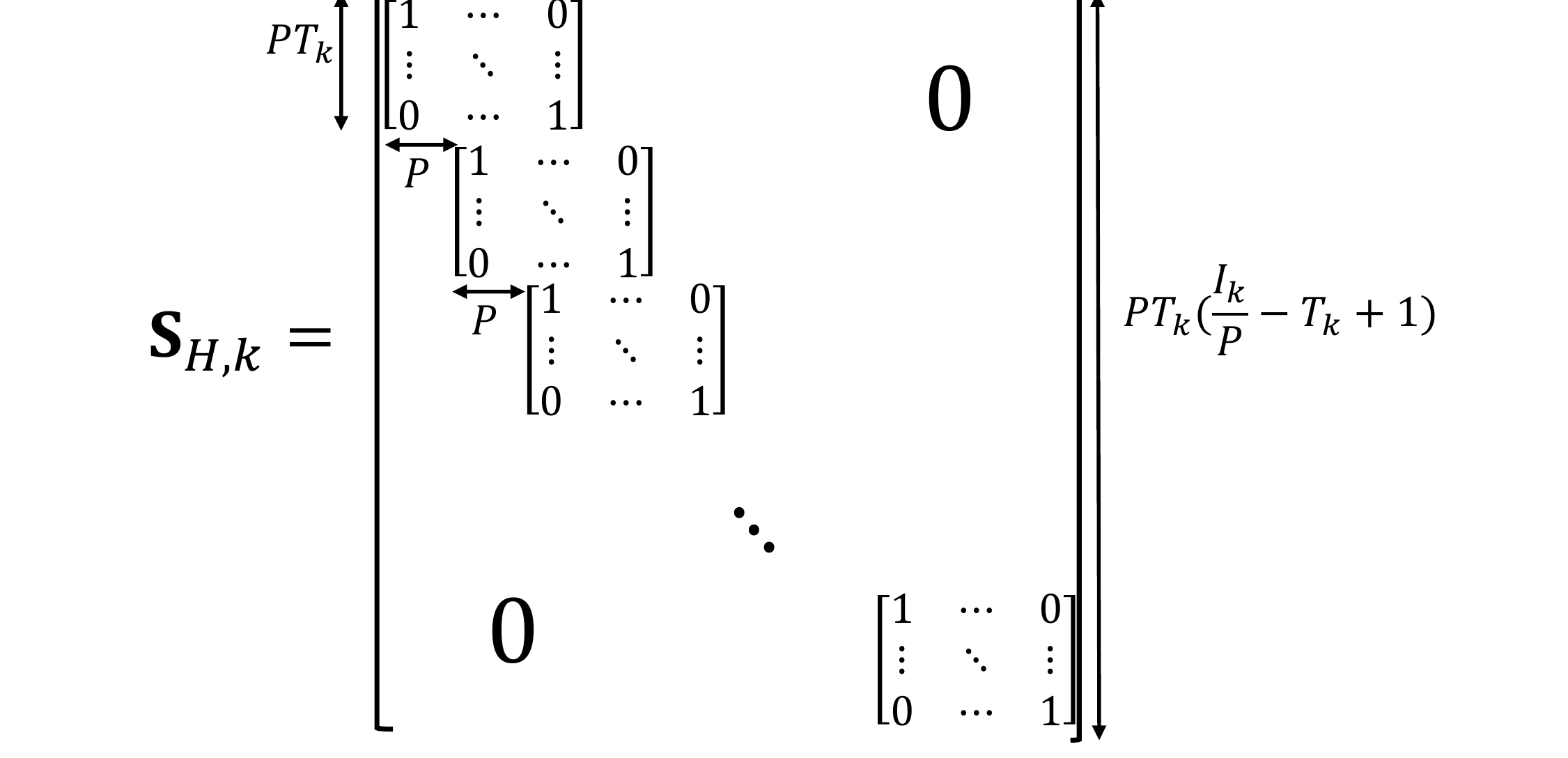}
					}
				\caption{}
			\end{subfigure}
		\hfill
		\begin{subfigure}[t!]{1.0\linewidth}
			\centerline{
			\includegraphics[width=9cm, height=6cm, trim={0cm 0cm 0cm 0cm}, clip]{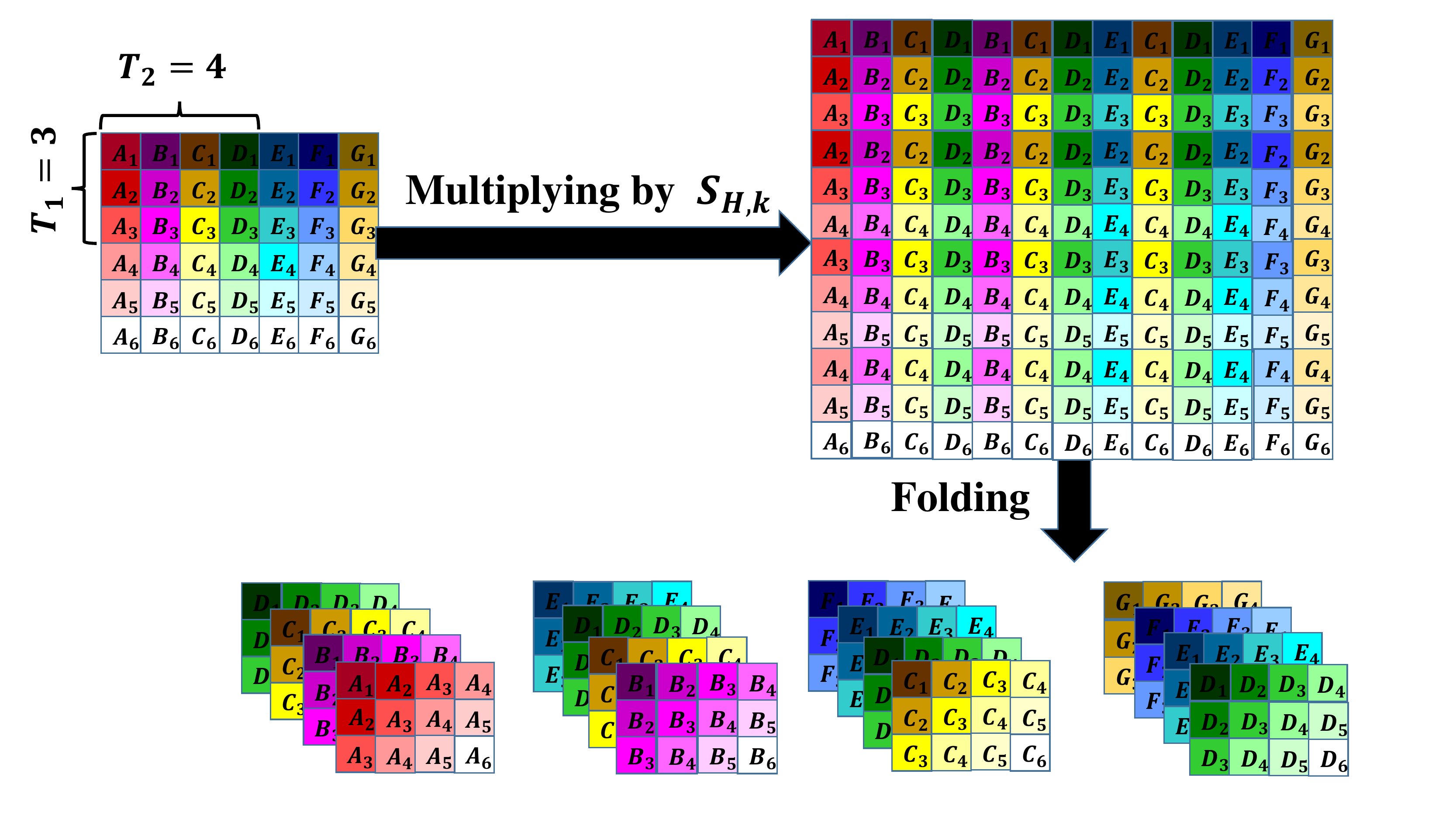}
					}
				\caption{}
		\end{subfigure}
	\caption{(a): A sample for a patch duplication matrix. (b): Block Hankelization of a matrix. Each colored square is a patch of the matrix.}
	\label{dup}
\end{figure}

Patch based Hankelization approach has been proposed previously by us in \cite{sedighin2020matrix}, which is based on the Hankelization of patches of the an image or a dataset instead of its individual elements. The Hankelization approach is based on the multiplication of a matrix, called patch duplication matrix, by an original matrix or tensor (representing an image), followed by folding, i.e tensorization of a huge block matrix. A sample for patch duplication matrix has been illustrated in Fig.~\ref{dup}.  The size of the patch duplication matrix ($\textbf{S}_{H,k}, k=1,2$) which is multiplied by the $k$-th mode of the matrix with size $I_1 \times I_2$ is $PT_k(I_k/P-T_k+1) \times I_k$, where $P$ is the patch size and $T_k$ is the window size for the $k$-th mode. 

\begin{figure}[t!]
	\begin{minipage}[b]{1.0\linewidth}
		\centering
		\centerline{
			\includegraphics[width=8cm, height=4cm, trim={6cm 6.5cm 6cm 5cm}, clip]{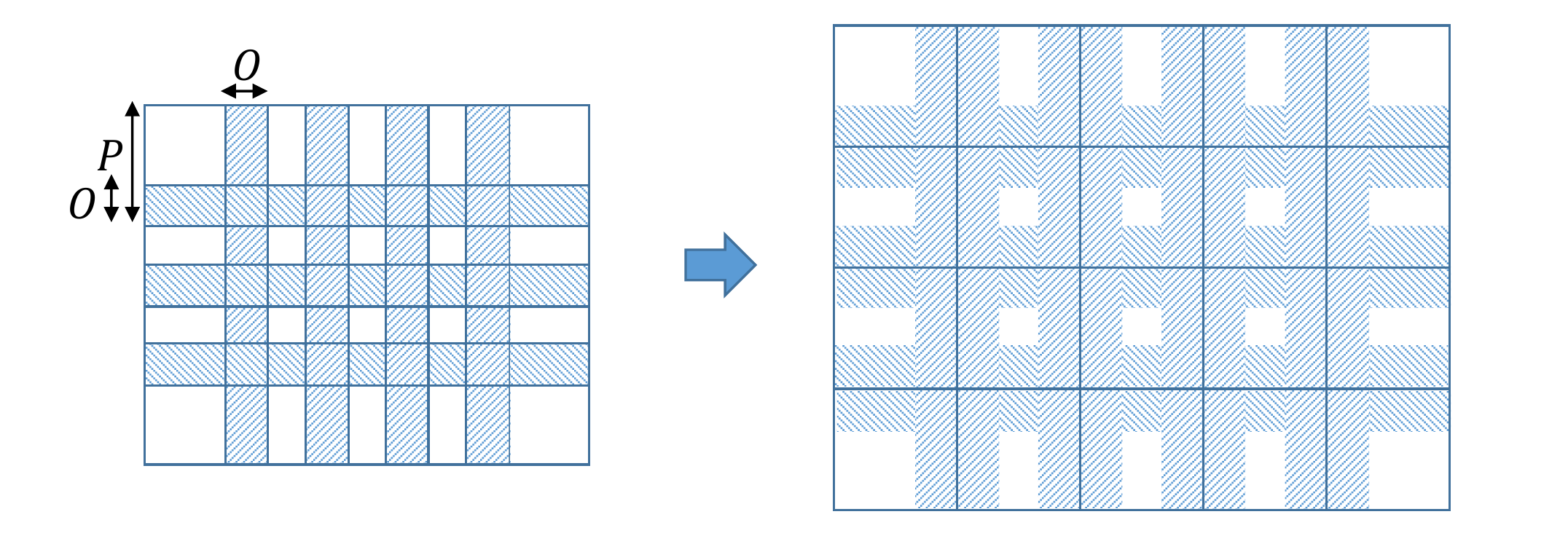}
		}
	\end{minipage}
	\caption{Patches (blocks) of a matrix dataset with overlap (left hand) and the larger matrix resulted from putting the patches together without overlap. Each patch is of size $P \times P$ and the size of the overlap between two neighbor patches is $O$ (rows or columns).}
	\label{opat}
\end{figure}

In patch Hankelization method of \cite{sedighin2020matrix}, the patches of image (or blocks of the dataset) have no overlap. The resulting 6-th order tensor from patch Hankelization of an $I_1\times I_2$ matrix with patch size $P$ and window size $\textbf{t}=[T_1, T_2]$ was of size $P \times P \times T_1 \times D_1 \times T_2 \times D_2$, where $D_1 = PT_1(I_1/P-T_1+1)$ and $D_2 = PT_2(I_2/P-T_2+1)$ \cite{sedighin2020matrix}. Non-overlapped patches can result in blurring of the final image. For overcoming this problem, in this paper, we have modified the previous version of patch Hankelization in a way that the patches have overlap with each other. A schematic view of a matrix with overlapped patches (before Hankelization) has been shown in  the left hand of Fig.~\ref{opat} where the shadowed parts show the overlapped elements of the consecutive patches.

The size of the patches is $P\times P$ and the overlap of neighbor patches is $O$. For overlapped patch Hankelization, the $I_1 \times I_2$ matrix of Fig.~\ref{opat} is first transformed into a $J_1\times J_2$ matrix where
$J_1 = P(\frac{I_1-P}{P-O}+1)$ and $J_2 = P(\frac{I_2-P}{P-O}+1)$. Note that, if $I_n-P$ is not dividable by $P-O$, the original matrix is zero padded (or padded by the last elements of the original tensor), in a way that $I_n-P$ will be dividable by $P-O$. This larger matrix (shown in the right hand of Fig.~\ref{opat}) resulted by arranging the patches together without overlap, where the last $O$ rows of each patch are the same as the first $O$ rows of the next patch and the last $O$ columns of each patch are the same as the first $O$ columns of the next patch. This new larger matrix is then simply patch (block) Hankelized with patches of size $P\times P$ and window size of $\textbf{t}=[T_1, T_2]$. The resulting six-th order tensor have the size of $P \times P \times T_1 \times D_1 \times T_2 \times D_2$, where $D_1 = PT_1(J_1/P-T_1+1)$ and $D_2 = PT_2(J_2/P-T_2+1)$. 

As mentioned before, after performing of low-rank TR decomposition in embedded space, we need to reconstruct the image in original space by so called de-Hankelization. De-Hankelization of the resulting 6-th order tensor is done in a similar way to \cite{sedighin2020matrix}, by averaging  the corresponding frontal slices and then averaging the corresponding skew-diagonals. De-Hankelization of six-th order tensor of size  $P \times P \times T_1 \times D_1 \times T_2 \times D_2$ with $D_1 = PT_1(J_1/P-T_1+1)$ and $D_2 = PT_2(J_2/P-T_2+1)$ results in a matrix $\textbf{X}_J$ with size $J_1 \times J_2$, which should again transformed into $\textbf{X}$ with size $I_1 \times I_2$. For this purpose, the parts of $\textbf{X}_J$ which does not have overlap with other parts (white parts in Fig.~\ref{opat}) will kept fixed and the parts which have overlapped with each other (shadowed parts in Fig.~\ref{opat}) are weighted averaged as
\begin{equation}
\label{avg1}
\begin{split}
\textbf{X}(:,(P-O)&(l-1)+1: (P-O)(l-1)+O) = \\
&\textbf{X}_J(:,(l-1)P-O+1:(l-1)P)\circledast\textbf{W} + \\
&\textbf{X}_J(:,(l-1)P+1:(l-1)P+O)\circledast(1-\textbf{W}),
\end{split}
\end{equation} 
and 
\begin{equation}
\label{avg2}
\begin{split}
\textbf{X}((P-O)&(k-1)+1: (P-O)(k-1)+O,:) = \\
&\textbf{X}_J((k-1)P-O+1:(k-1)P,:)\circledast\textbf{W}^{T} + \\
&\textbf{X}_J((k-1)P+1:(k-1)P+O,:)\circledast(1-\textbf{W})^{T},
\end{split}
\end{equation} 
where $l=1:J_2/P$, $k=1:J_1/P$ and $\textbf{W}$ is a weight matrix whose rows are the same and the $m$-th row defined as 
\begin{equation}
\begin{split}
\textbf{W}(m,u)= (O-u-1)/(O-1),   \quad \quad u=1,\ldots, O.
\end{split}
\end{equation} 

Note that, if the original $I_1\times I_2$ matrix has been initially zero padded, the size of the resulting $\textbf{X}$ will be more than $I_1 \times I_2$ and the extra elements which have been  added to the end of the rows and end of the columns should be removed.

The important point about selecting patch sizes is that, for larger images, we usually select larger patch sizes. In addition, the size of patches should be increased for noisy images, but can be selected smaller for clean images. However, selecting larger patches can result in a lower convergence ratio and consequently increase the simulation time.   
 
\subsection{New algorithm for OCT super-resolution}
In this section, we will introduce a new algorithm for OCT image super-resolution. The algorithm can be applied in both situations when only one target B-scan is available  or when  several B-scans close to the target B-scan are also available. When several B-scans close to the target B-scan are  available, the weighted average of the B-scans is computed, where usually the larger weight is dedicated to the target B-scan and smaller weights are dedicated to the other B-scans.  This results in a low sampling rate B-scan, $\textbf{X}_l$, with size $I_1\times I_2^l$ which is computed as
\begin{equation}
\label{avg}
\textbf{X}_l = \sum_{i=1}^M \alpha_i \textbf{X}_l^i,
\end{equation} 
where $\textbf{X}_l^i$ is the $i$-th B-scan, $\alpha_i$ is the weight of that B-scan and $M$ is the number of available B-scans, and
\begin{equation}
\label{weight}
	\sum_{i=1}^M \alpha_i \leq 1.
\end{equation} 

In the first stage of our algorithm, the missing A-scans are interpolated with rate $L$ using Spline method. $L$ is the super-resolution ratio and the resulting $\textbf{X}$ after interpolation will be of size $I_1\times I_2$ where $I_2=I_2^l\times L$.  The resulting matrix $\textbf{X}$ is used as an initialization for the algorithm.

Matrix $\textbf{X}$ is then patch Hankelized with overlapped patches of size $P$, overlap size of $O$ and window size of $\textbf{t}=[T_1,T_2]$, which results in a 6-th order tensor of size $P \times P \times T_1 \times D_1 \times T_2 \times D_2$. Now, similar to \cite{yokota2018missing,sedighin2021image},  the super-resolution can be performed by minimizing the following cost function:
\begin{equation}
\label{cost}
J(\boldsymbol{\theta})=\|\underline{\boldsymbol{\Omega}}_H\circledast(\underline{\textbf{X}}_H-\widehat{\underline{\textbf{X}}}_H(\boldsymbol{\theta}))\|^2_F,
\end{equation}
where $\underline{\boldsymbol{\Omega}}_H$ is the overlapped patch Hankelized version of ${\boldsymbol{\Omega}}$, which is a binary mask matrix with the size of $I_1\times I_2$, whose columns are 1 for the observed columns of $\textbf{X}$ and 0 for the interpolated columns of $\textbf{X}$. $\underline{\textbf{X}}_H$ is the overlapped patch Hankelized version of $\textbf{X}$ and $\widehat{\underline{\textbf{X}}}_H$ is an estimation of $\underline{\textbf{X}}_H$ which has a TR structure as:
\begin{equation}
\label{est}
\widehat{\underline{\textbf{X}}}_H \simeq \ll \underline{\textbf{G}}^{(1)}, \underline{\textbf{G}}^{(2)}, \ldots, \underline{\textbf{G}}^{(N)}\gg,
\end{equation} 
and $\boldsymbol{\theta}$ is the vector of unknown parameters.
Similar to \cite{yokota2018missing,sedighin2021image}, minimizing (\ref{cost}) can be replaced equivalently by minimizing the following cost function
\begin{equation}
\label{cost2}
\begin{split}
J(\boldsymbol{\theta}|\boldsymbol{\theta}^k)=\|\underline{\boldsymbol{\Omega}}_H\circledast(\underline{\textbf{X}}_H-\widehat{\underline{\textbf{X}}}_H(\boldsymbol{\theta}))\|^2_F + \|(\underline{\textbf{1}}-\underline{\boldsymbol{\Omega}}_H)\circledast(\widehat{\underline{\textbf{X}}}_H(\boldsymbol{\theta}^k)-\widehat{\underline{\textbf{X}}}_H(\boldsymbol{\theta}))\|^2_F,
\end{split}
\end{equation}
where $\underline{\textbf{1}}$ is a tensor of all 1, and $\widehat{\underline{\textbf{X}}}_H(\boldsymbol{\theta}^k)$ is the estimated tensor in the $k$-th iteration. The cost function (\ref{cost2}) can be written as
\begin{equation}
\label{cost3}
J(\boldsymbol{\theta}|\boldsymbol{\theta}^k)=\|\underline{\tilde{\textbf{X}}}_H-\widehat{\underline{\textbf{X}}}_H(\boldsymbol{\theta})\|^2_F,
\end{equation}
where $\underline{\tilde{\textbf{X}}}_H =\underline{\boldsymbol{\Omega}}_H\circledast  \underline{\textbf{X}}_H +(\underline{\textbf{1}}-\underline{\boldsymbol{\Omega}}_H)\circledast\widehat{\underline{\textbf{X}}}_H(\boldsymbol{\theta}^k)$.
 
Similar to many of TR based tensor reconstruction approaches, TR ranks for $\widehat{\underline{\textbf{X}}}_H$ or better to say the size of the core tensors are not determined in advance, but are increased gradually in each iteration. Since $\textbf{X}$ has been initially interpolated, the initial rank can be set larger than 1.  After estimation of $\widehat{\underline{\textbf{X}}}_H$ in each iteration, de-Hankelization has been applied for transforming the resulting tensor back into the original image space which results in $\widehat{\textbf{X}}$. Then smoothing by replacing each estimated element (corresponding to ${\boldsymbol{\Omega}}=0$) by the average of its eight neighbors is applied. Resulting  ${\textbf{X}}$ in each iteration is updated as
\begin{equation}
\label{update}
{\textbf{X}} = {\boldsymbol{\Omega}}\circledast {\textbf{X}} + ({\textbf{1}}-{\boldsymbol{\Omega}})\circledast \widehat{{\textbf{X}}},
\end{equation}
where ${\textbf{1}}$ is a matrix with the same size as ${\boldsymbol{\Omega}}$, whose all elements are 1. The updated $\textbf{X}$ is again patch Hankelized and the procedure is repeated for the new rank vector. The procedure is repeated until  the desired approximation accuracy is achieved or the maximum rank reaches its limit and $\widehat{\textbf{X}}$ will be  the output.

Moreover, in order to  further improve  the quality of the final results, the output of the algorithm for several (two or three) different patch sizes are averaged and the result accounts as the final output of the algorithm. So the final output of the algorithm computed as:
\begin{equation}
\label{fin}
\widehat{\textbf{X}}_\textrm{final}=\frac{\sum_{i=1}^{N_p}\widehat{\textbf{X}}_i}{N_p},
\end{equation}
where 	$\widehat{\textbf{X}}_i$ is the estimated output with patch size $P_i \times P_i$ and overlap $O_i$, $\widehat{\textbf{X}}_\textrm{final}$ is the final estimated B-scan with increased resolution and $N_p$ is the number of patch sizes. Using larger patch sizes will increase noise suppression, while smaller patch sizes convey more details to the final image. So averaging the results for different patch sizes can simultaneously decrease the noise and increase the details in the final image.

In addition, for further suppressing the noise, for some of the patch sizes, the sum of weights i.e., (\ref{weight}) for parts of B-scans which  mostly contain noise instead of information, is set less than 1 (say $0.95$). 
The pseudo-code of the proposed algorithm has been shown in Algorithm.~\ref{alg1}.
	
\begin{algorithm}[!t]
	%
	\caption{Pseudo-code of the proposed algorithm}
	\label{algorithm1}	
	\begin{flushleft} 
		\textbf{INPUT:} Low resolution B-scan  (or volume of B-scans) $\textbf{X}_l$ of size $I_1\times I_2^l$, super-resolution ratio ($L$), binary mask matrix $\boldsymbol{\Omega}$, patch size vector $\textbf{p}=[P_1,\ldots,P_{N_p}]$, window size $\textbf{t}=[T_1,T_2]$, overlap vector $\textbf{o}=[O_1,O_2]$, initial rank vector $\textbf{r}$ and maximum rank $R_{\textrm{max}}$.\\
		\textbf{OUTPUT:} B-scan with increased resolution $\widehat{\textbf{X}}_\textrm{final}$ of size $I_1 \times I_2$ where $I_2 = L\times I_2^l$. \\
	\end{flushleft}
	\begin{algorithmic}[1]
		\State If several B-scans are available, $\textbf{X}_l$ will be equal to average of the B-scans, otherwise,   $\textbf{X}_l$ is the available B-scan. 
		\State Interpolate $\textbf{X}_l$ with ratio $L$ using Spline method which results in $\textbf{X}$ with size $I_1 \times I_2$ and $I_2=L\times I_1^l$. 
		\State Apply smoothing by replacing each of the estimated elements (for $\boldsymbol{\Omega}=0$) by the average of its eight neighbors and keeping the observed elements (for  $\boldsymbol{\Omega}=1$) fixed.
		\For {$i=1:N_p$}
		\While {$\max(\textbf{r}_l) \leq R_{max}$ }
		\State Overlapped Patch Hankelize $\textbf{X}$  and binary mask matrix $\boldsymbol{\Omega}$, by block size $P_i\times P_i$, with overlap $O_i$ and window size $\textbf{t}=[T_1, T_2]$ which results in $\underline{\textbf{X}}_H$ and $\underline{\boldsymbol{\Omega}}_H$.
		\State Compute the TR decomposition of ${\underline{\textbf{X}}_H}$, i.e., $\widehat{\underline{\textbf{X}}}_H$ with rank vector $\textbf{r}$. 
		\State Increase the elements of the rank vector $\textbf{r}$ by 1. 
		\State De-Hankelize $\widehat{\underline{\textbf{X}}}_H$ to provide $\widehat{{\textbf{X}}}_i$.
		\State Apply smoothing by replacing each of the estimated elements (for $\boldsymbol{\Omega}=0$) by the average of its eight neighbors and keeping the observed elements (for  $\boldsymbol{\Omega}=1$) fixed. 
		\State ${{\textbf{X}}} = {\boldsymbol{\Omega}} \circledast {\textbf{X}} + ({\textbf{1}}-{\boldsymbol{\Omega}}) \circledast \widehat{{\textbf{X}}}_i$.
		\State Overlapped Patch Hankelize $\textbf{X}$  and binary mask matrix $\boldsymbol{\Omega}$, by block size $P_i\times P_i$, with overlap $O_i$ and window size $\textbf{t}=[T_1, T_2]$ which results in $\underline{\textbf{X}}_H$
		\EndWhile
		\EndFor
		\State $\widehat{\textbf{X}}_\textrm{final}=\frac{\sum_{i=1}^{N_p}\widehat{\textbf{X}}_i}{N_p}$
		
	\end{algorithmic}
	\label{alg1}
\end{algorithm}

\section{Simulation results}
\label{sim}
In this section, the effectiveness of the proposed algorithm for the super-resolution of OCT images has been investigated. 

Three OCT datasets has been used for evaluating the performance of the proposed algorithm. The first dataset, dataset1\footnote{\url{http://people.duke.edu/~sf59/Fang_TMI_2013.htm.}},  contains 18 volumes of $450 \times 900$ B-scans, where each volume contains 5 noisy and one reference B-scans \cite{fang2013fast} has been exploited. One of the noisy B-scans assumed as the target B-scan and the other 4 ones are the B-scans close to the target one. This dataset also contains a reference denoised B-scan for each subject, so PSNR (Peak signal to noise Ratio) and SSIM (Structural SIMilarity) of the results can also be computed.  Dataset2\footnote{\url{https://misp.mui.ac.ir/en/topcon-3d-oct-diabetic-data -denoising-0}} \cite{kafieh2014three}, contains three dimensional OCT images of size $612\times 500\times 128$ of several diabetic patients. Dataset3\footnote{\url{https://misp.mui.ac.ir/en/oct-basel-data-0}}  \cite{tajmirriahi2021modeling}, contains noisy B-scans of healthy, Diabetic and non-diabetic patients. Each category contains several subjects where 1 to 300 B-scans of size $300 \times 300$ have been taken for each subject.
Note that Dataset2 and Dataset3 do not contain any reference images for the subjects, hence PSNR and SSIM of the outputs cannot be computed. 
  
  The proposed approach has been compared with the spline interpolation, LRFOTTV and LRSOTTV algorithm of \cite{daneshmand2020super}, 3D-SBSDI \cite{fang2013fast} and MDT \cite{yokota2018missing}. For MDT, the window size was set in a way that the best results achieved. For a fair comparison, the input to the MDT was the averaged B-scan  of (\ref{avg}).

  In the first simulation, performances of the algorithms have been compared for the reconstruction of artificially subsamples OCT images of dataset1. For the proposed algorithm, the patch sizes have been set equal to $15$ with overlap equal to $7$, $10$ for overlap equal to $6$ and $7$ with overlap equal to 4 and the window size was set equal to $[2,2]$. The initial rank vector varies from $[1,1,1,1,1,1]$ to $[5,5,5,5,5,5]$ and maximum rank varies from $[5,5,5,5,5,5]$ to $[9,9,9,9,9,9]$. For the MDT approach, the window size was set equal to [8,8].

\begin{figure*}[t!]
	\centering
	\centerline{
		\small
		\begin{tabular}{lccc}
			\quad \quad \quad  \quad Original noisy image&Incomplete image& Spline interpolation\\
			\includegraphics[scale=0.15, trim={5.5cm 5.5cm 7.7cm 1.7cm}, clip]{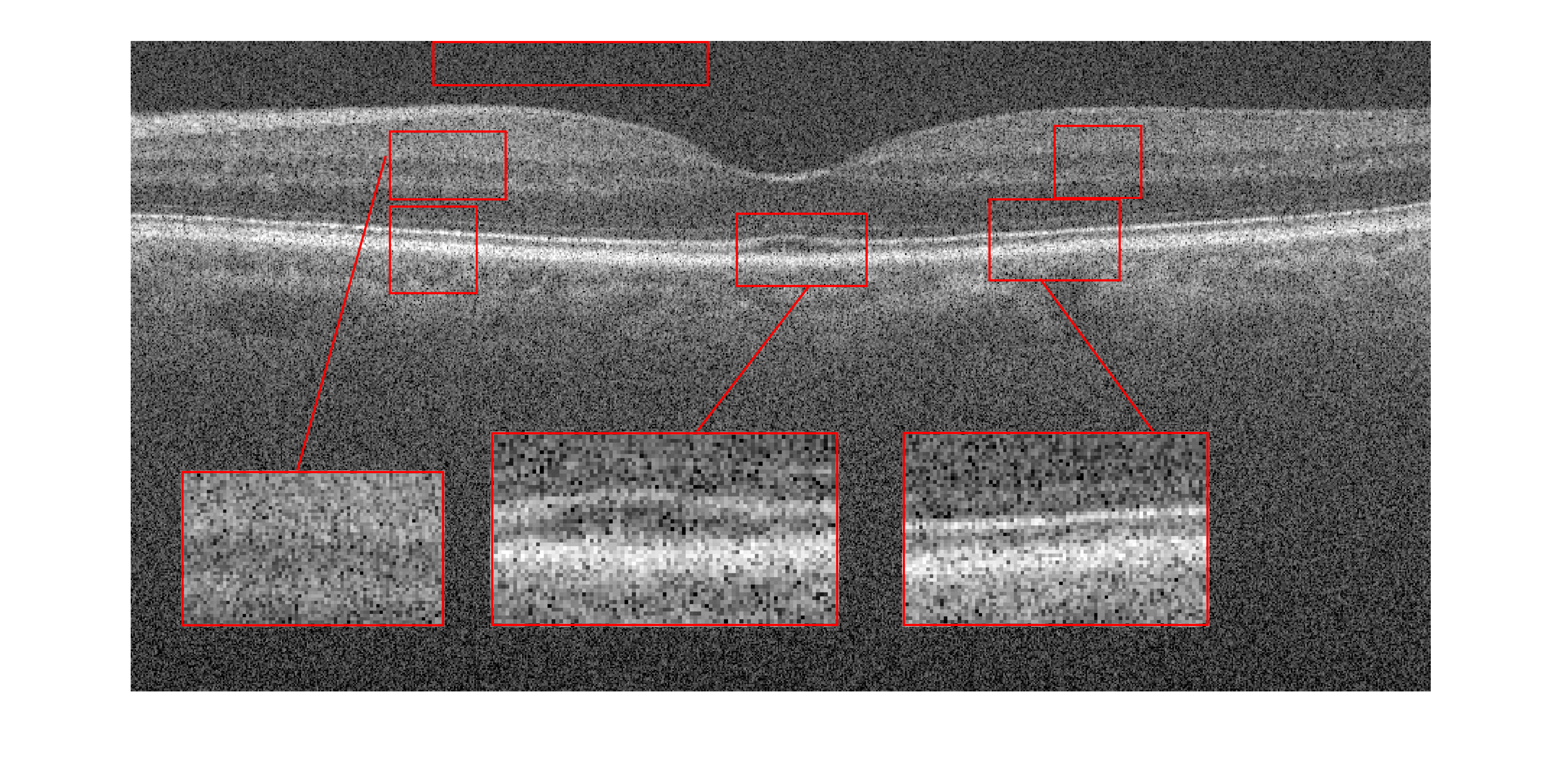}&
			\includegraphics[scale=0.15,trim={5.5cm 5.5cm 7.7cm 1.7cm}, clip]{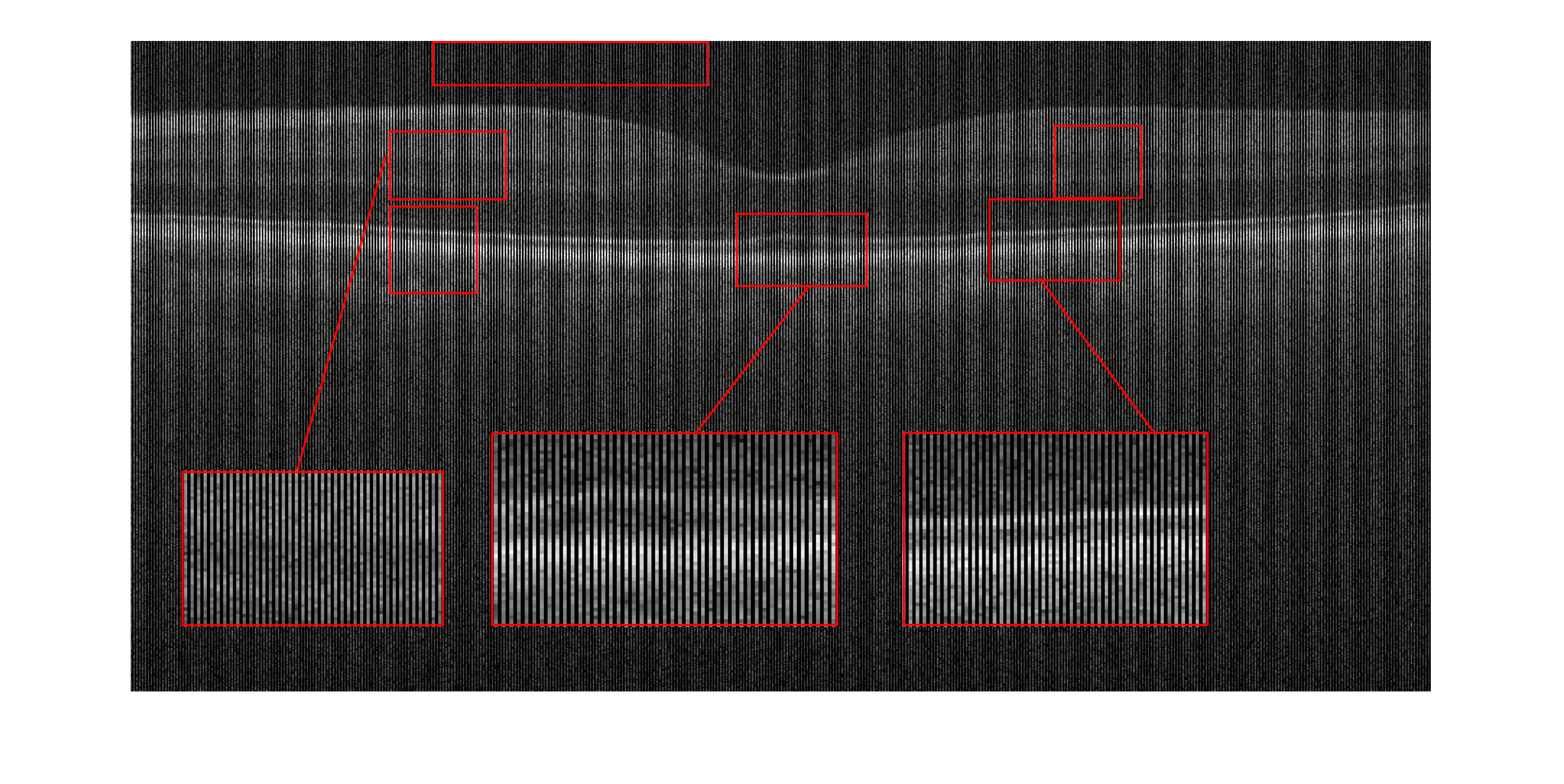}&
			\includegraphics[scale=0.15,trim={5.5cm 5.5cm 7.7cm 1.7cm}, clip]{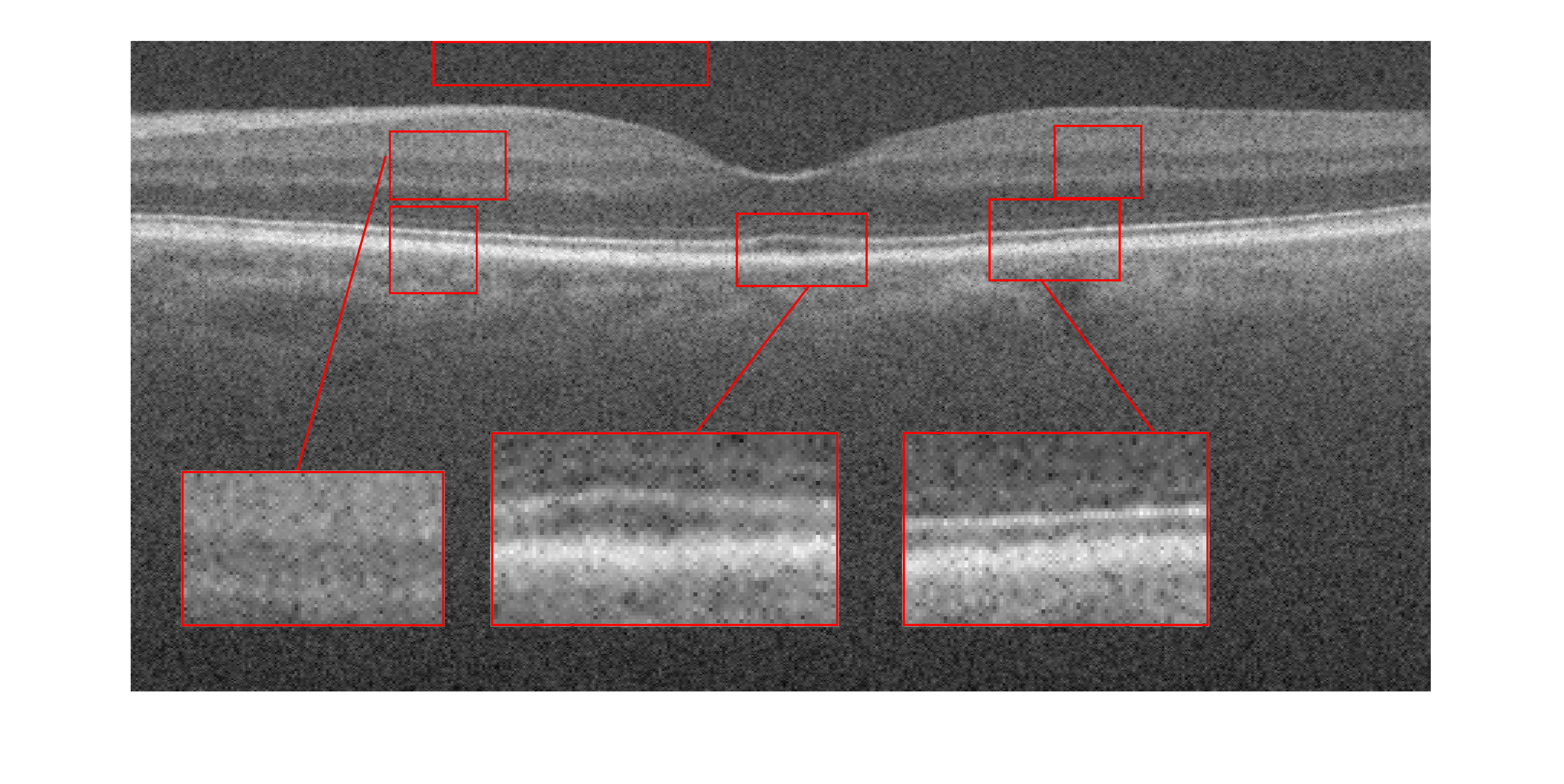}\\
			\quad \quad \quad \quad \quad$[17.7827,0.0810]$&$[11.4556,0.0251]$&$[23.6096,0.2750]$\\
			\quad \quad \quad \quad  \quad \quad \quad SBSDI \cite{fang2013fast}&LRFOTTV \cite{daneshmand2020reconstruction}& LRSOTTV \cite{daneshmand2020reconstruction}\\
			\includegraphics[scale=0.15, trim={5.5cm 5.5cm 7.7cm 1.7cm}, clip]{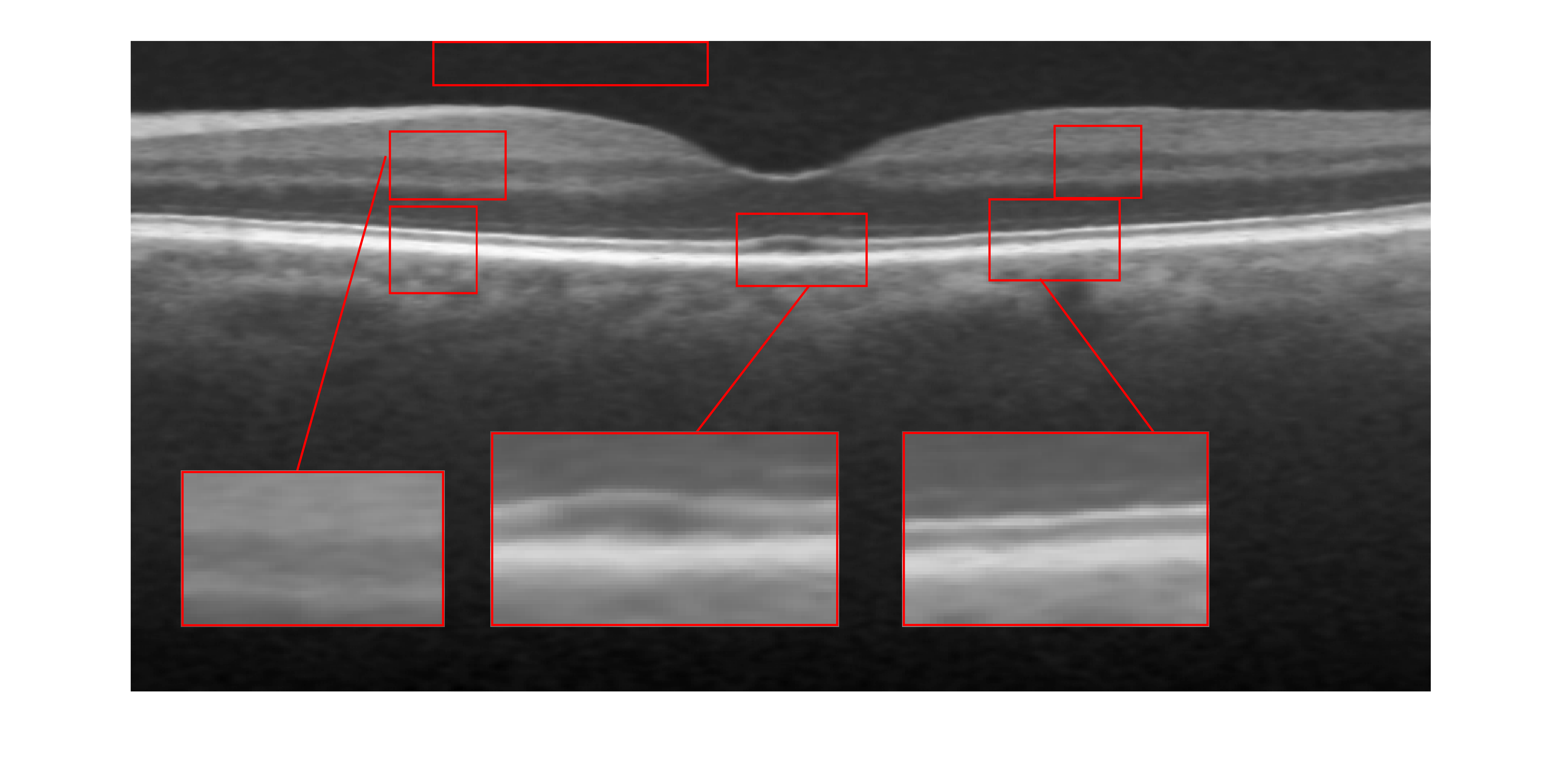}&
			\includegraphics[scale=0.15,trim={5.5cm 5.5cm 7.7cm 1.7cm}, clip]{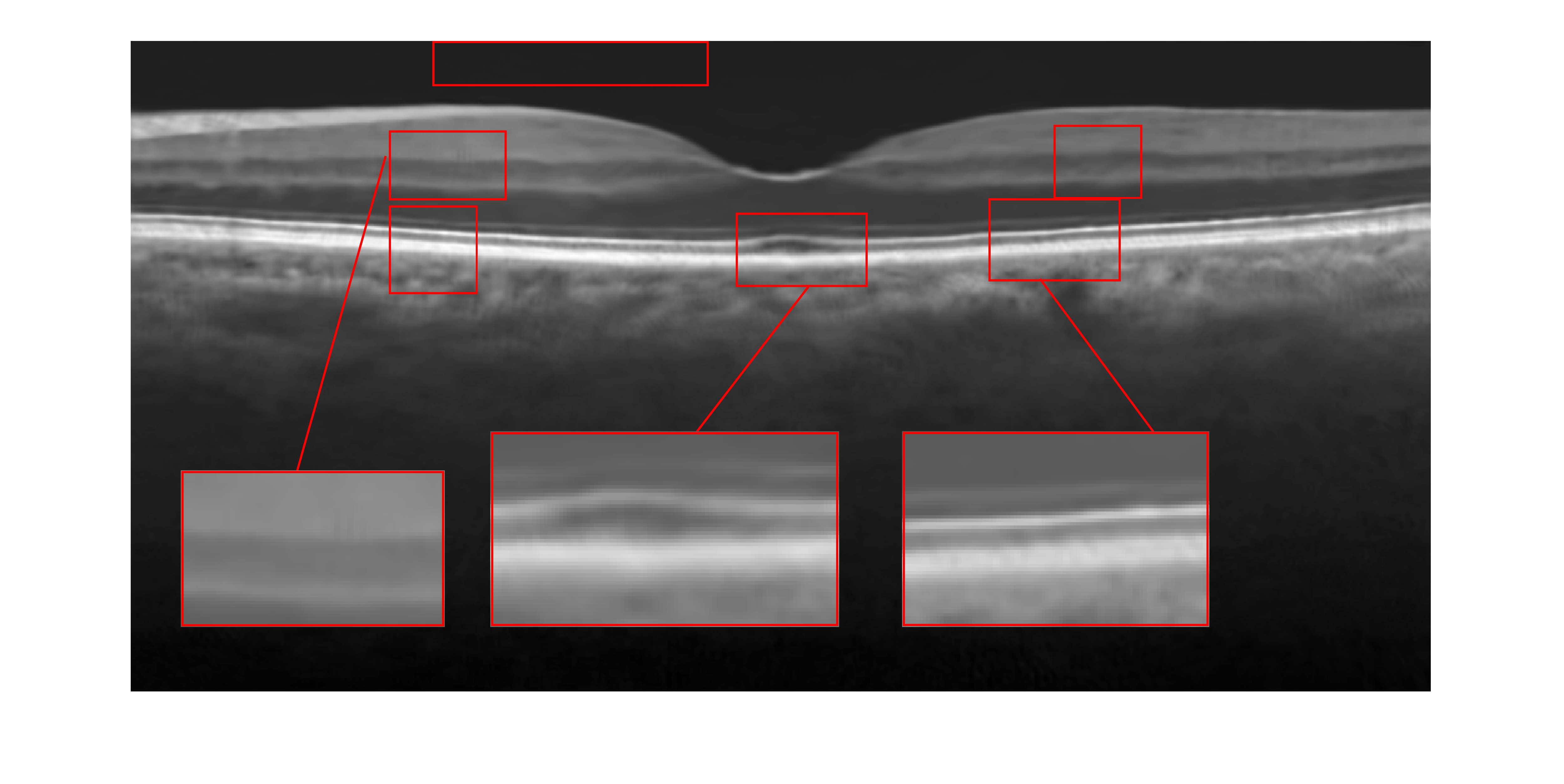}&
			\includegraphics[scale=0.15,trim={5.5cm 5.5cm 7.7cm 1.7cm}, clip]{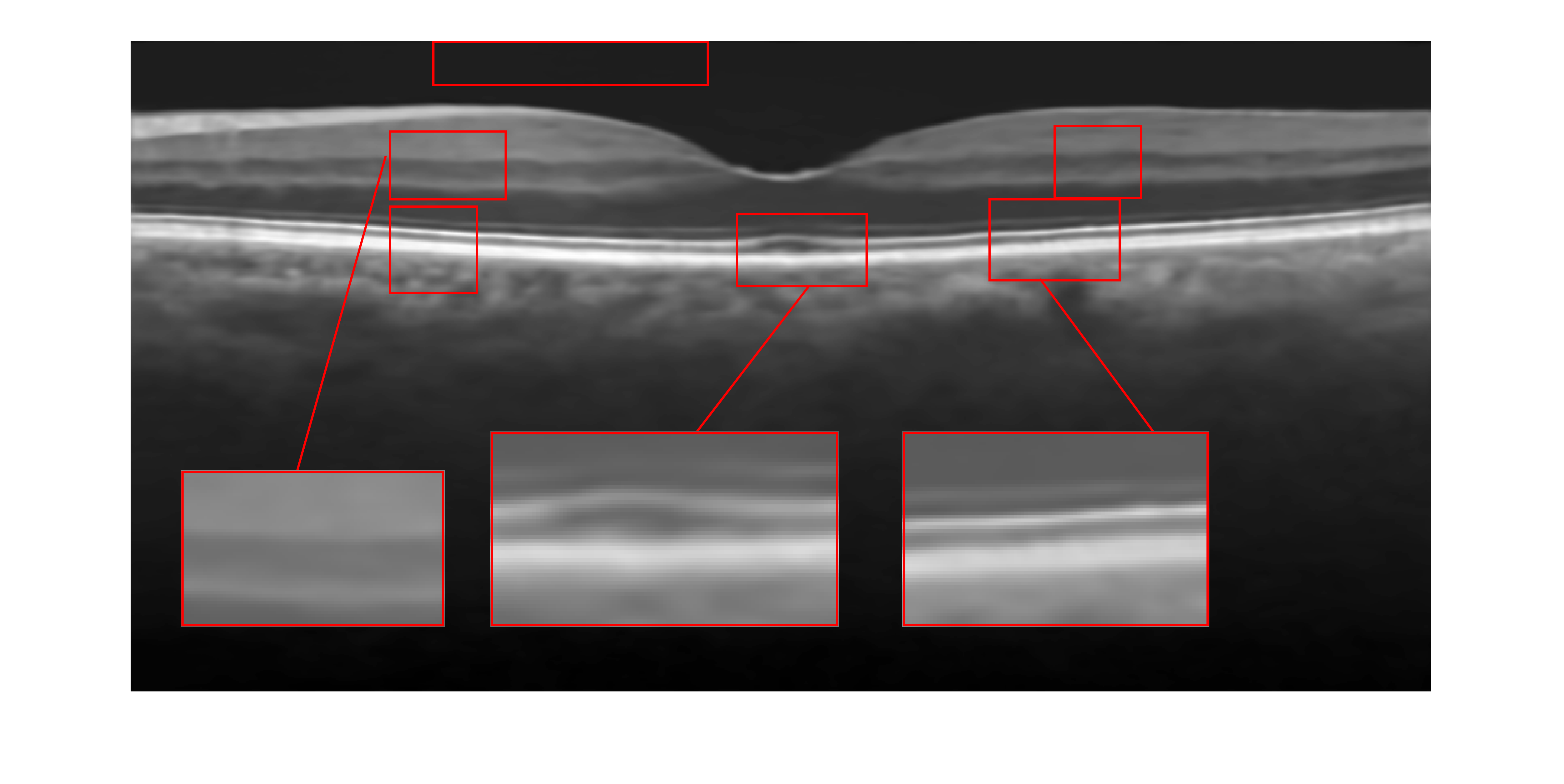}\\			
		\quad \quad \quad \quad \quad	$[28.7111,0.6941]$&$[28.5247,0.6980]$&$[28.7440,0.7037]$\\
		\quad \quad \quad  \quad \quad \quad \quad \quad	MDT \cite{yokota2018missing}& Proposed& Reference image\\
			\includegraphics[scale=0.15, trim={5.5cm 5.5cm 7.7cm 1.7cm}, clip]{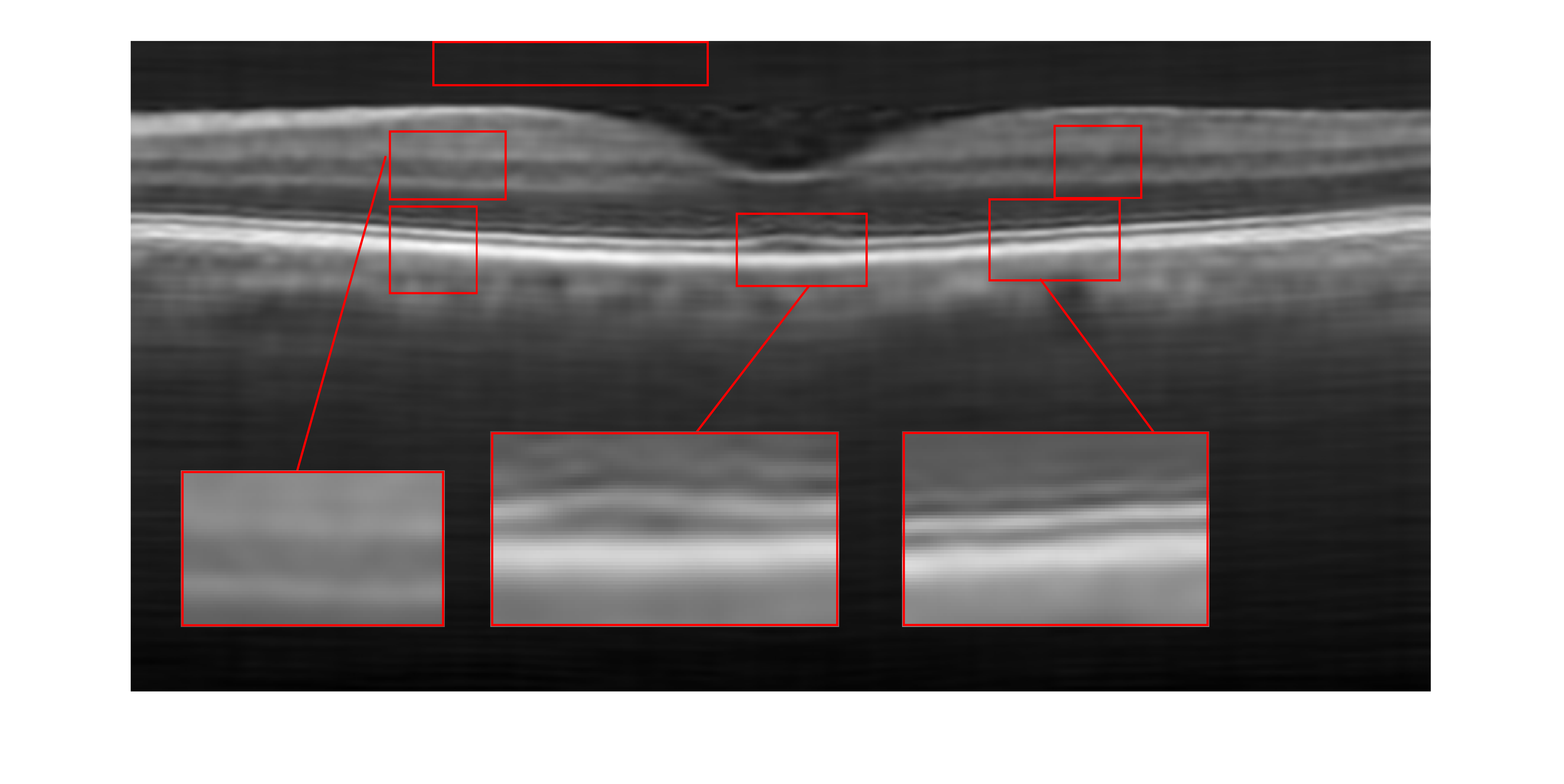}&
			\includegraphics[scale=0.15, trim={5.5cm 5.5cm 7.7cm 1.7cm}, clip]{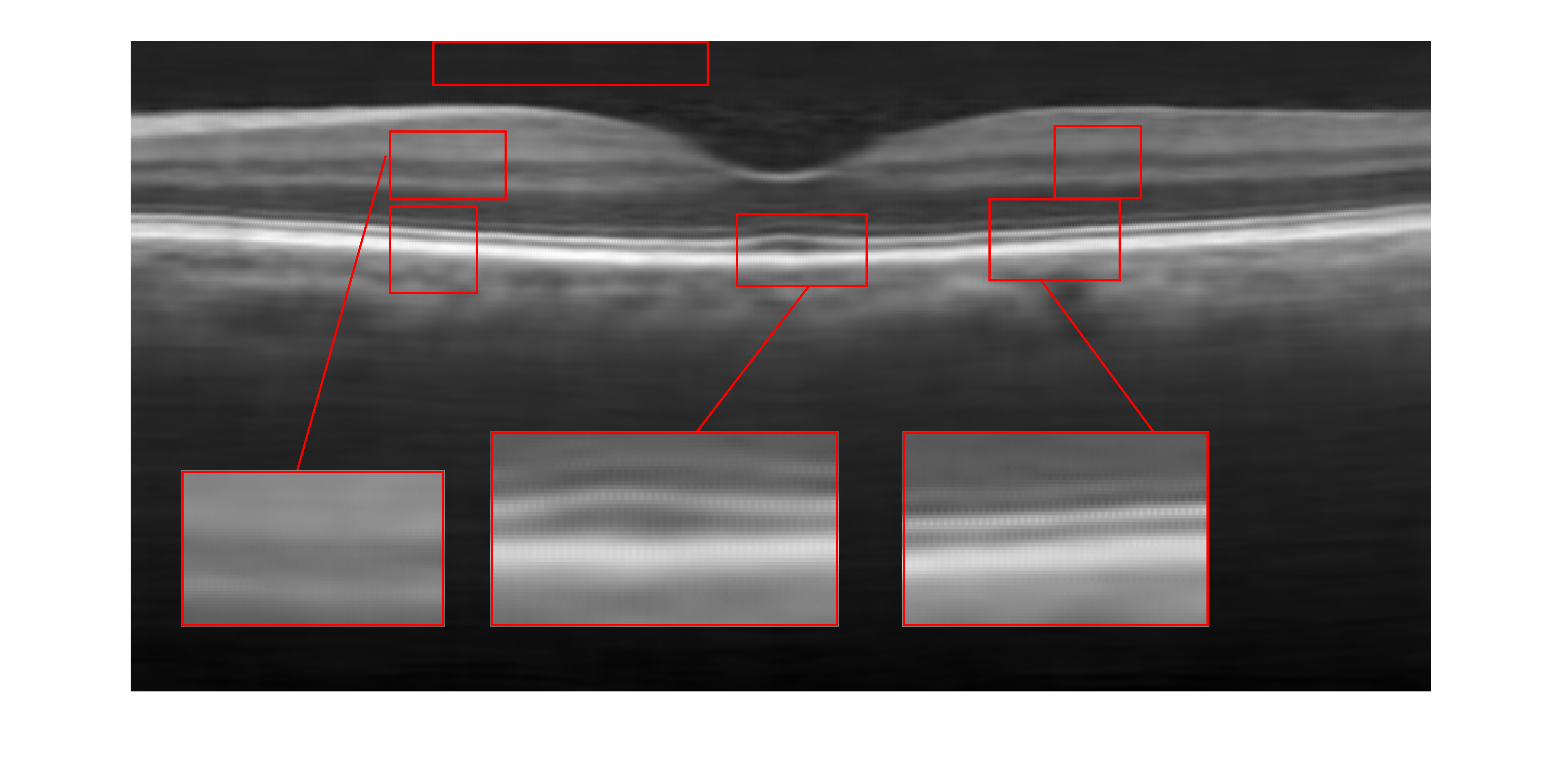}&
			\includegraphics[scale=0.15,trim={5.5cm 5.5cm 7.7cm 1.7cm}, clip]{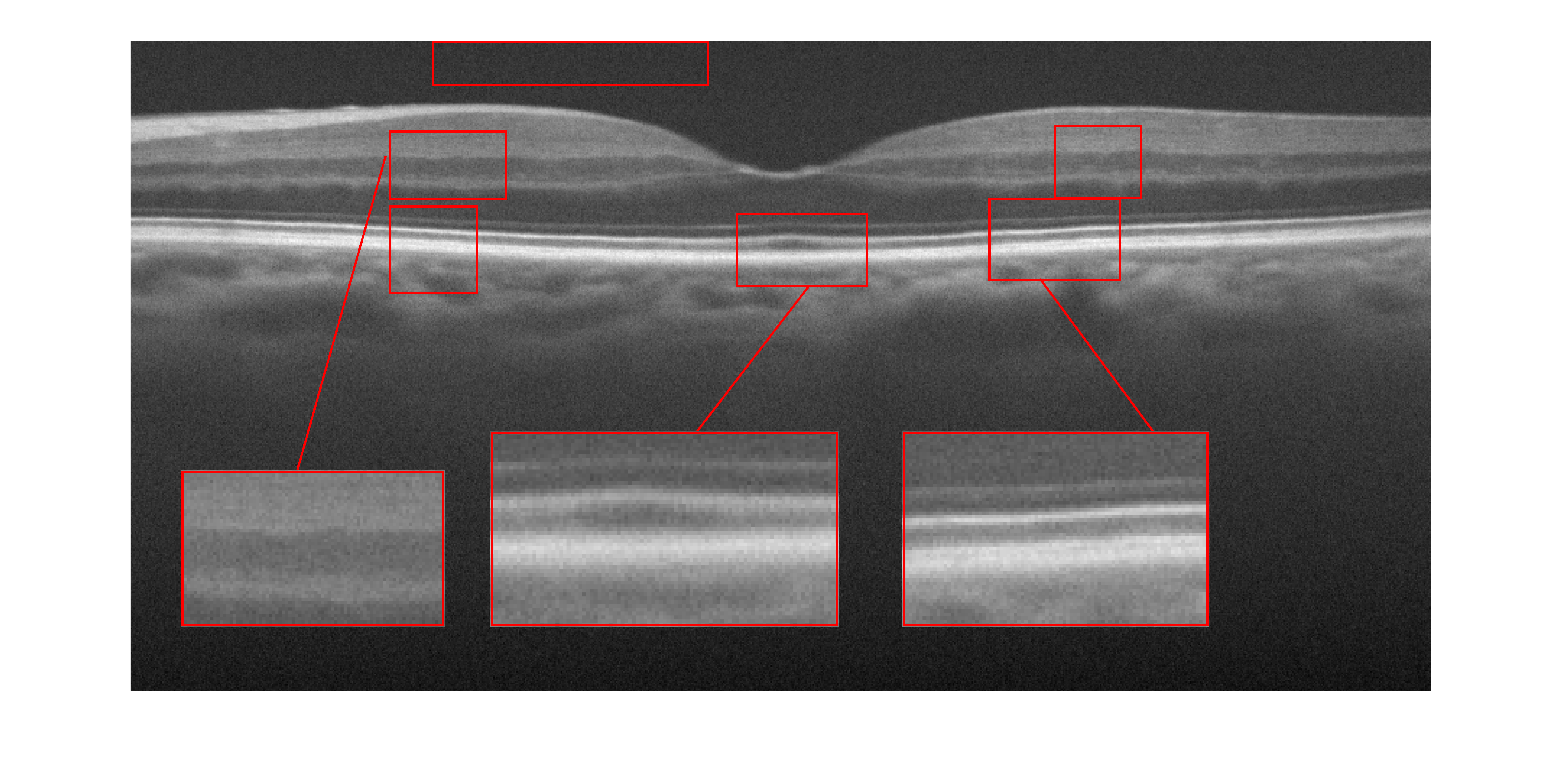}\\
		\quad \quad \quad \quad \quad \quad	$[28.8132,0.6986]$&$\mathbf{[29.2272,0.7047]}$&\\
			\end{tabular}
	}
\caption{Comparison of the performance of the algorithms for the super-resolution of an OCT image of dataset1 with $50\%$ missing ratio of A-scans. PSNR (in \upshape{dB}) and SSIM corresponding to each image have been written in brackets beneath each image.
}
	\label{sim1}
\end{figure*}

For $P=10$, the weights ($\alpha_i$) are the same as $P=15$ and $P=7$, but for lower parts of B-scans which mostly contain noise the summation of weights set equal to $0.95$. Sample output for each algorithm in addition to the PSNR  and SSIM  resulting from each algorithm (in comparison to the reference images) have been shown in Fig.~\ref{sim1}. The averaged  PSNR and SSIM for the supper-resolution of 18 volumes of B-scans of dataset1 exploiting the 6 algorithms have been also shown in Table.~\ref{res1}. The results show the higher performance of the proposed algorithm in comparison to the other approaches.

\begin{table*}[h!]
	\caption{Averaged PSNR and SSIM of the algorithms for the super-resolution of artificially missed volumes of B-scans of dataset1 with missing ratio=$50\%$.}
	\centering
	\centerline{
		\small
		\begin{tabular}{c|ccccccccc}
			& Spline interpolation& SBSDI& MDT& LRFOTTV& LRSOTTV& Proposed approach\\	
			\hline
			PSNR& $22.35712\pm 0.9122$ &$28.29\pm 2.5955$&$28.2139\pm 2.4$&$28.3513 \pm 2.7859$&$28.4809 \pm 2.7586$&$\mathbf{28.7129}\pm \mathbf{2.5904}$\\
			\hline
			SSIM& $0.2248\pm 0.0216$ &$0.69 \pm 0.0293$&$0.6894\pm 0.0253$&$0.6930 \pm 0.0345$&$0.6988 \pm 0.0314$&$\mathbf{0.6989}\pm \mathbf{0.0271}$\\	
		\end{tabular}
	}
	\label{res1}
\end{table*}

Super-resolution of artificially missed B-scans of dataset1 with $66\%$ missing ratio of A-scans has been investigated.  The  averaged PSNR's and SSIM's have been shown in Table.~\ref{psnr66}. As the results show, the proposed algorithm provides  higher  PSNR and SSIM for the reconstructed  B-scans comparing to the other algorithms.

\begin{table*}[h!]
	\caption{Averaged PSNR and SSIM of the algorithms for the super-resolution of artificially missed volumes of B-scans of dataset1 with missing ratio=$66\%$.}
	\centering
	\centerline{
		\small
		\begin{tabular}{c|ccccc}
			& Spline interpolation& SBSDI& MDT&  Proposed approach\\	
			\hline
			PSNR& $22.3245\pm 0.8887$ &$28.5018\pm 2.5285$&$27.0479\pm 2.3458$&$\mathbf{28.6634}\pm \mathbf{2.4793}$\\
			\hline
			SSIM& $0.2355\pm 0.0199$ &$0.6951 \pm 0.0235$&$0.6669\pm 0.0322$&$\mathbf{0.6994}\pm \mathbf{0.0196}$\\	
		\end{tabular}
	}
	\label{psnr66}
\end{table*}

\begin{figure*}[h]
	\centering
	\centerline{
		\begin{tabular}{lccccc}
		 Original noisy image& \quad Incomplete image & \quad \quad Spline interpolation & \quad MDT \cite{yokota2018missing} & \quad Proposed algorithm\\ 
			\includegraphics[scale=0.18, trim={14.5cm 6cm 18.2cm 1.7cm}, clip]{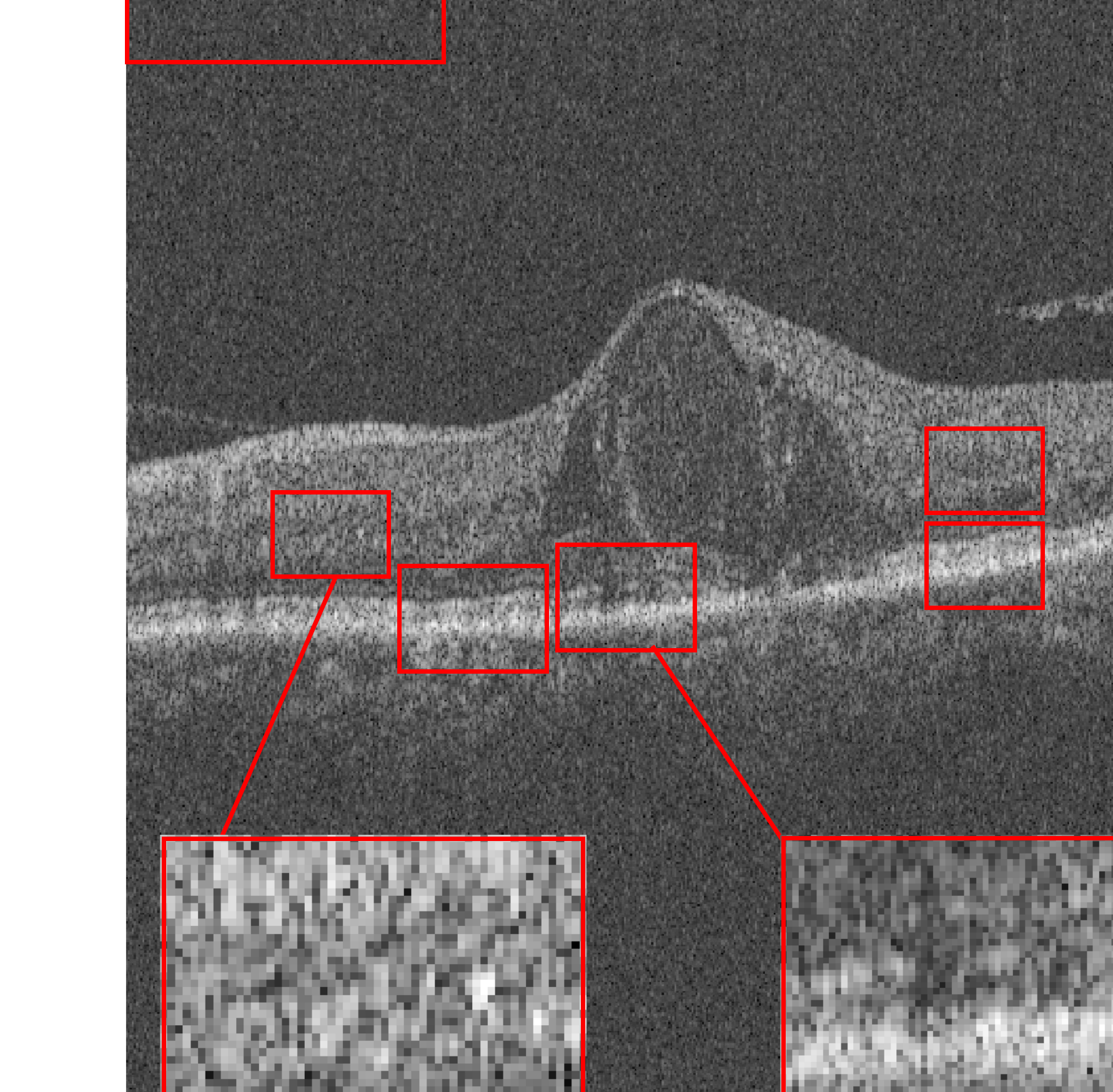}&\includegraphics[scale=0.18, trim={14.5cm 6cm 18.2cm 1.7cm}, clip]{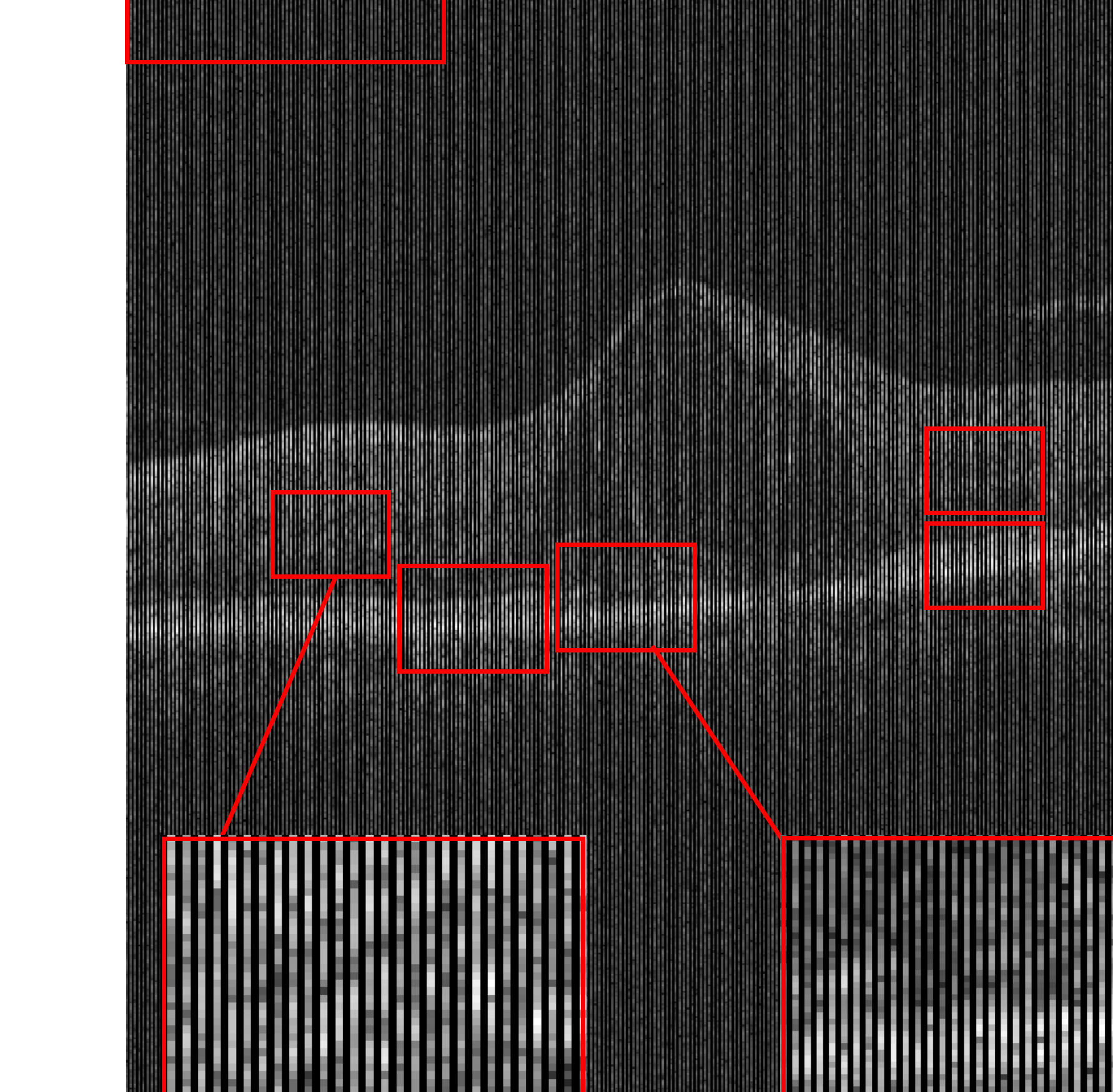}&\includegraphics[scale=0.18, trim={14.5cm 6cm 18.2cm 1.7cm}, clip]{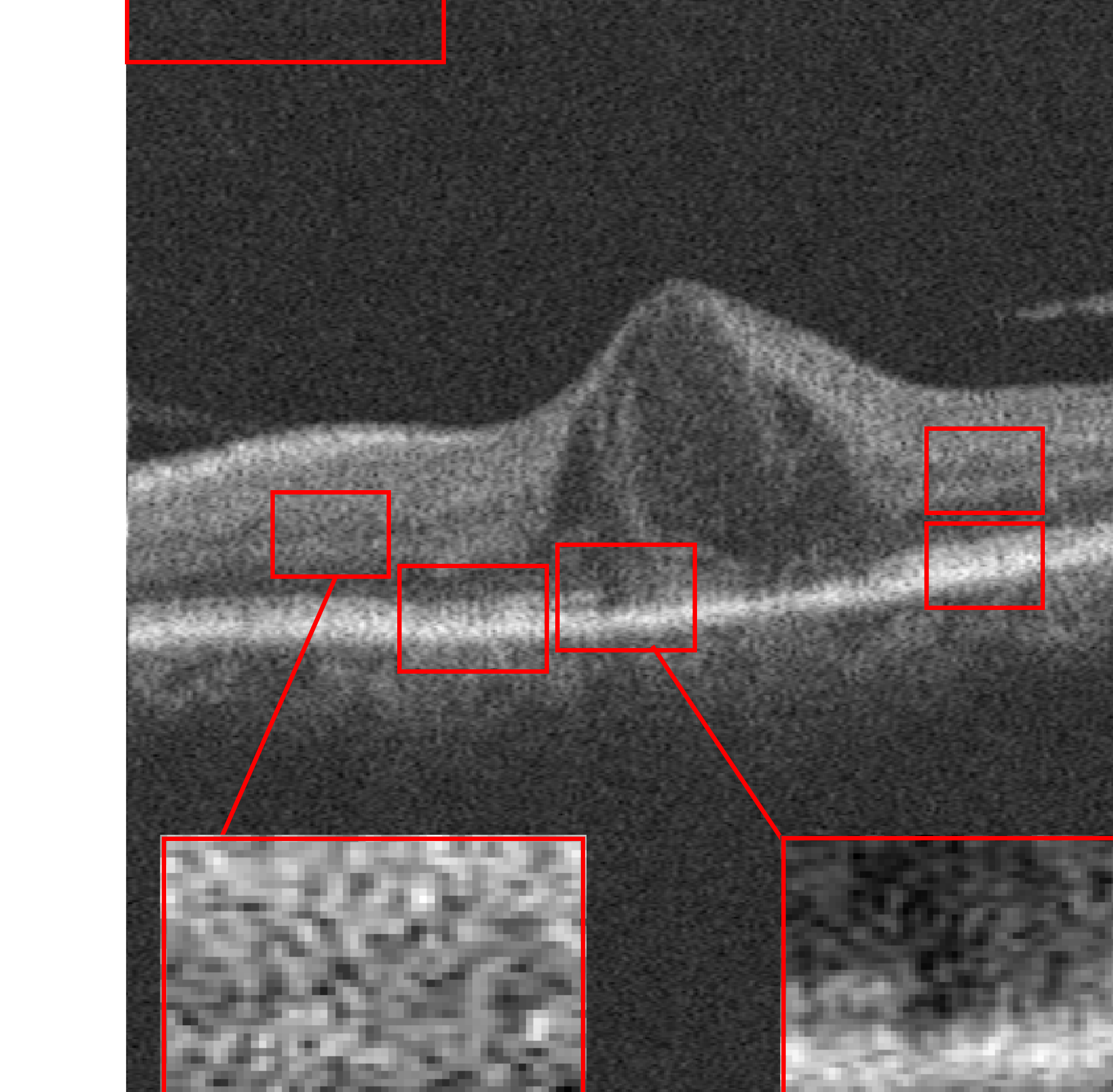}&
			\includegraphics[scale=0.18, trim={14.5cm 6cm 18.2cm 1.7cm}, clip]{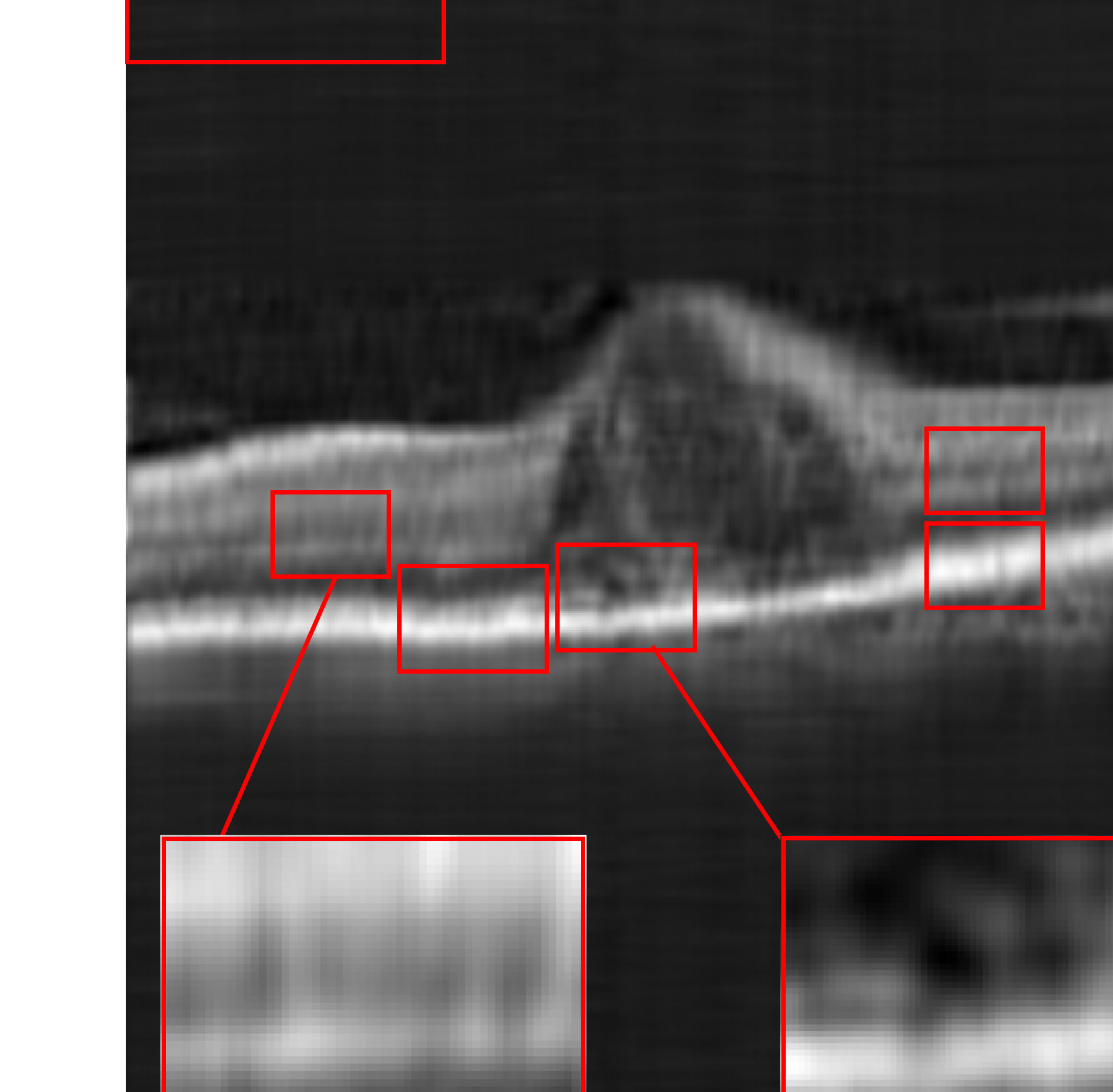}&\includegraphics[scale=0.18, trim={14.5cm 6cm 18.2cm 1.7cm}, clip]{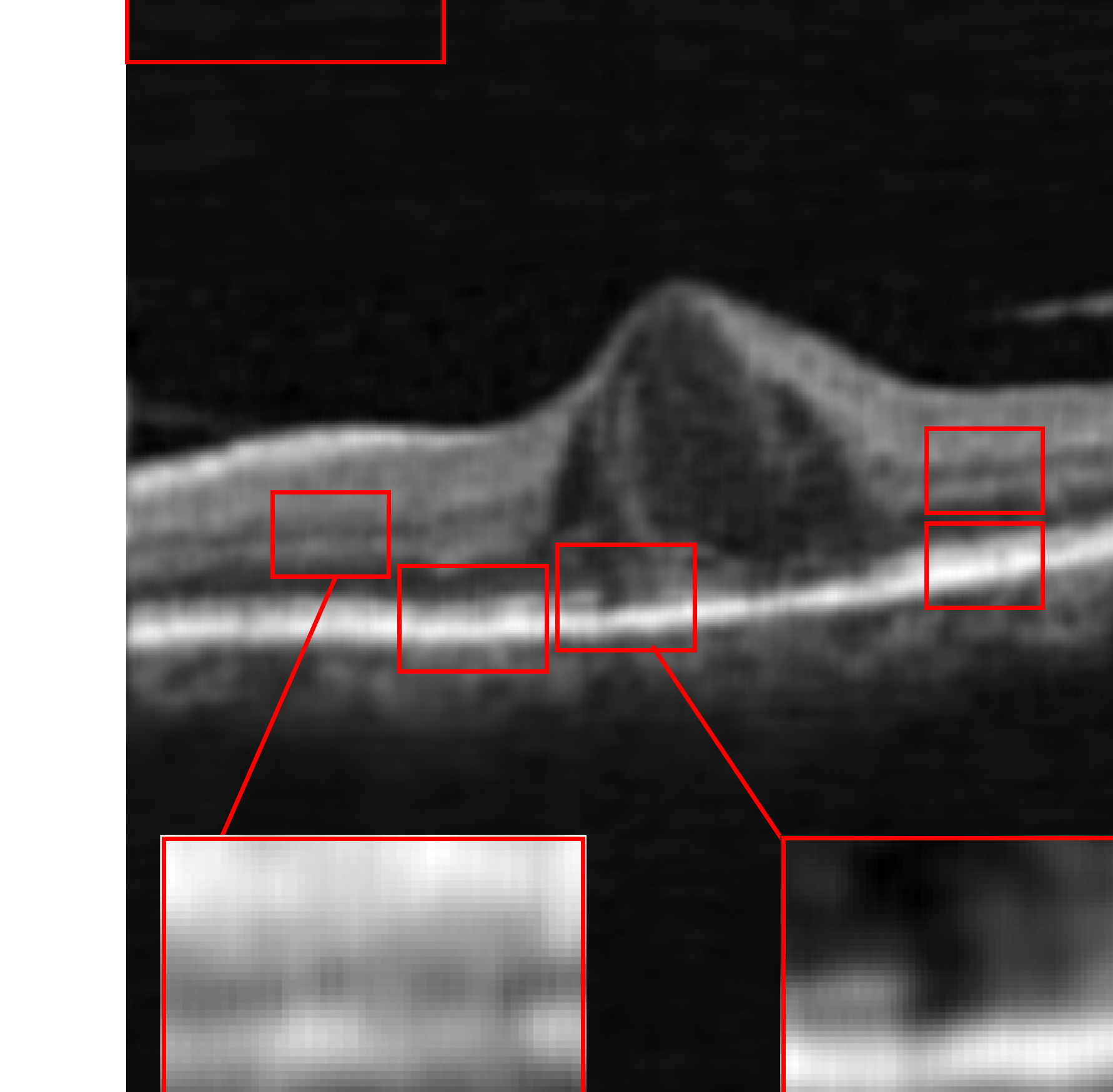}\\
			\quad \quad  CNR=1.9861&CNR=0.4667&CNR=3.0882&CNR=3.7875&\textbf{CNR=3.8238}\\
			\includegraphics[scale=0.18, trim={14.5cm 6cm 18.2cm 1.7cm}, clip]{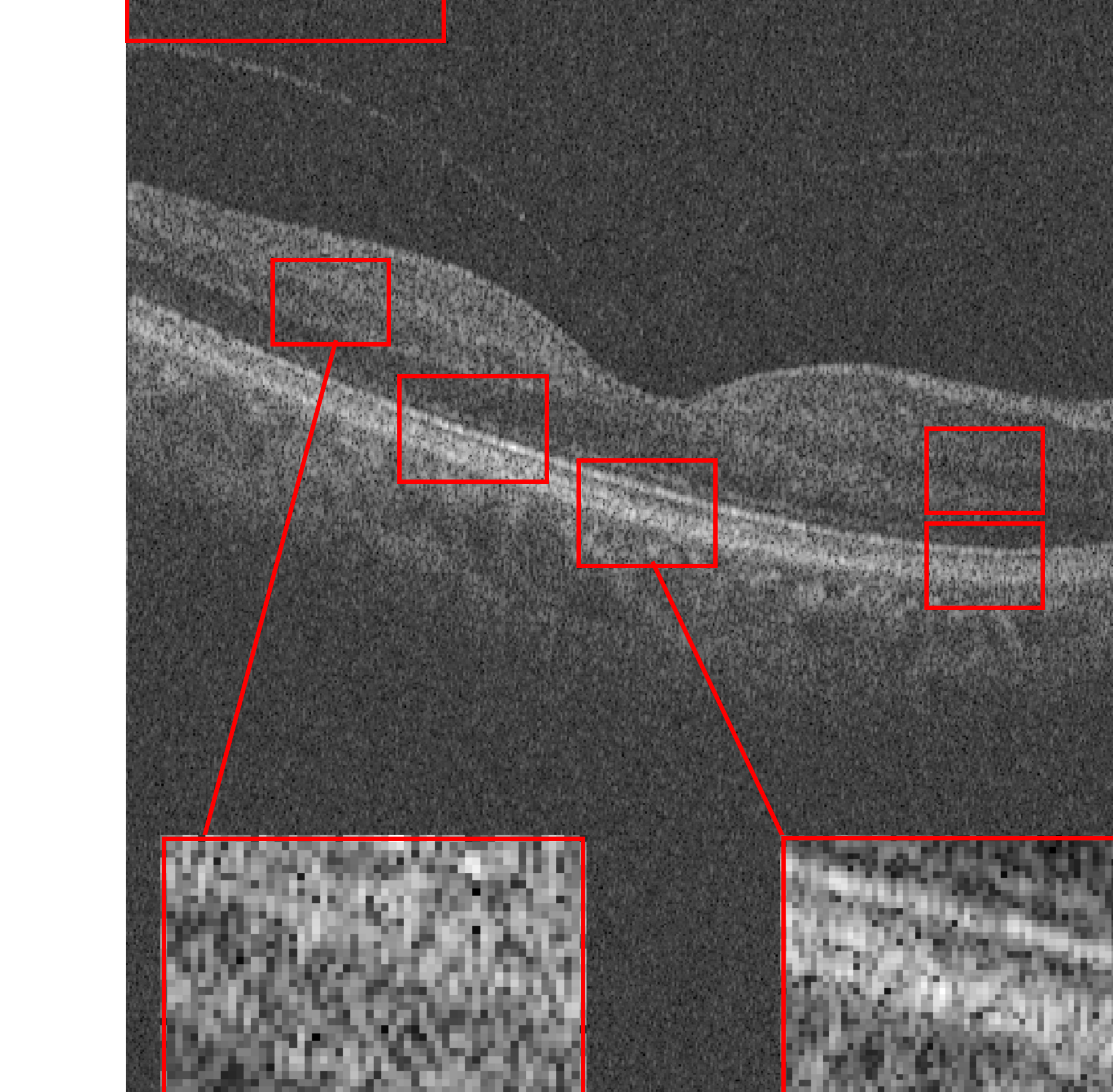}&\includegraphics[scale=0.18, trim={14.5cm 6cm 18.2cm 1.7cm}, clip]{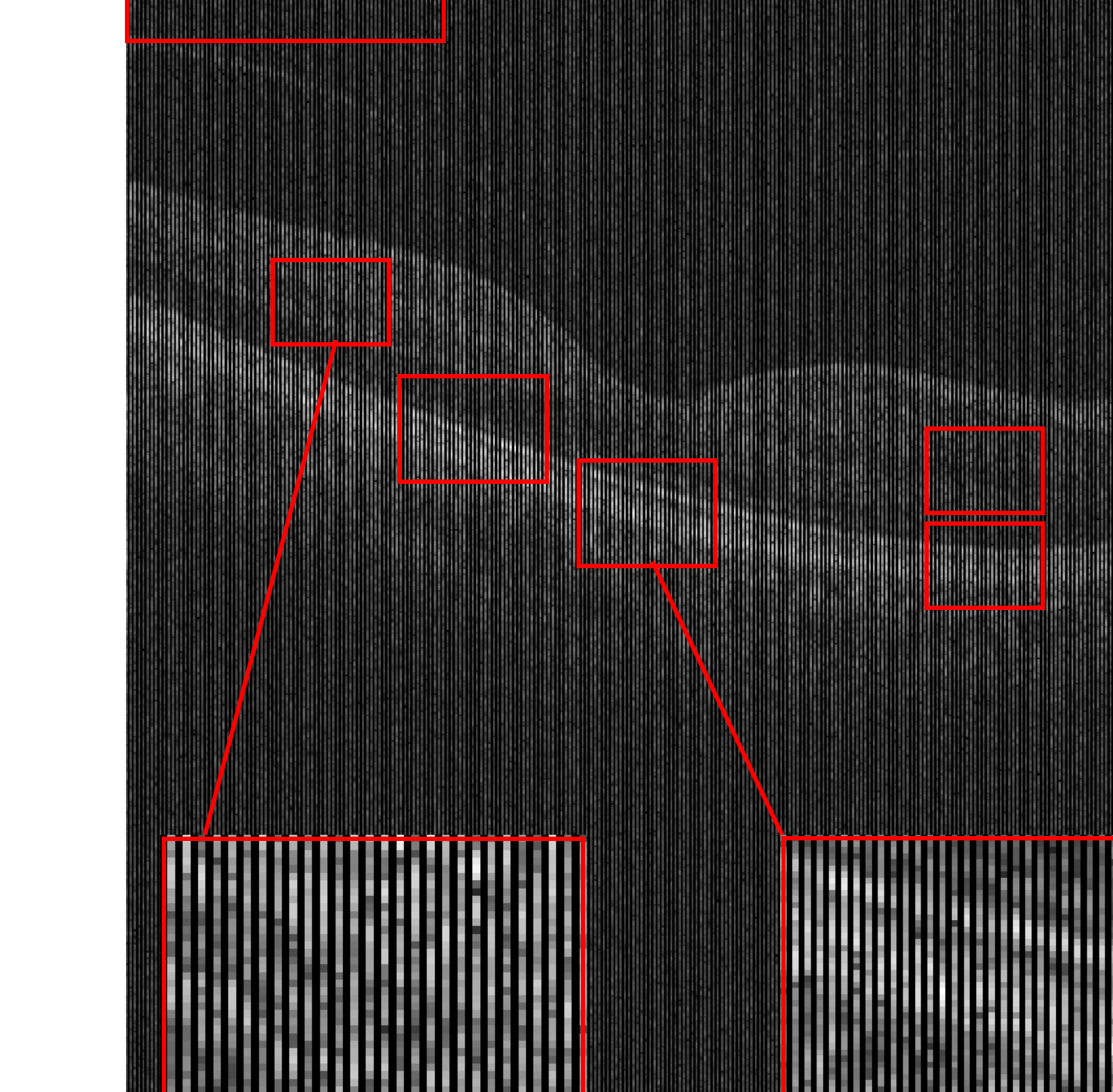}&\includegraphics[scale=0.18, trim={14.5cm 6cm 18.2cm 1.7cm}, clip]{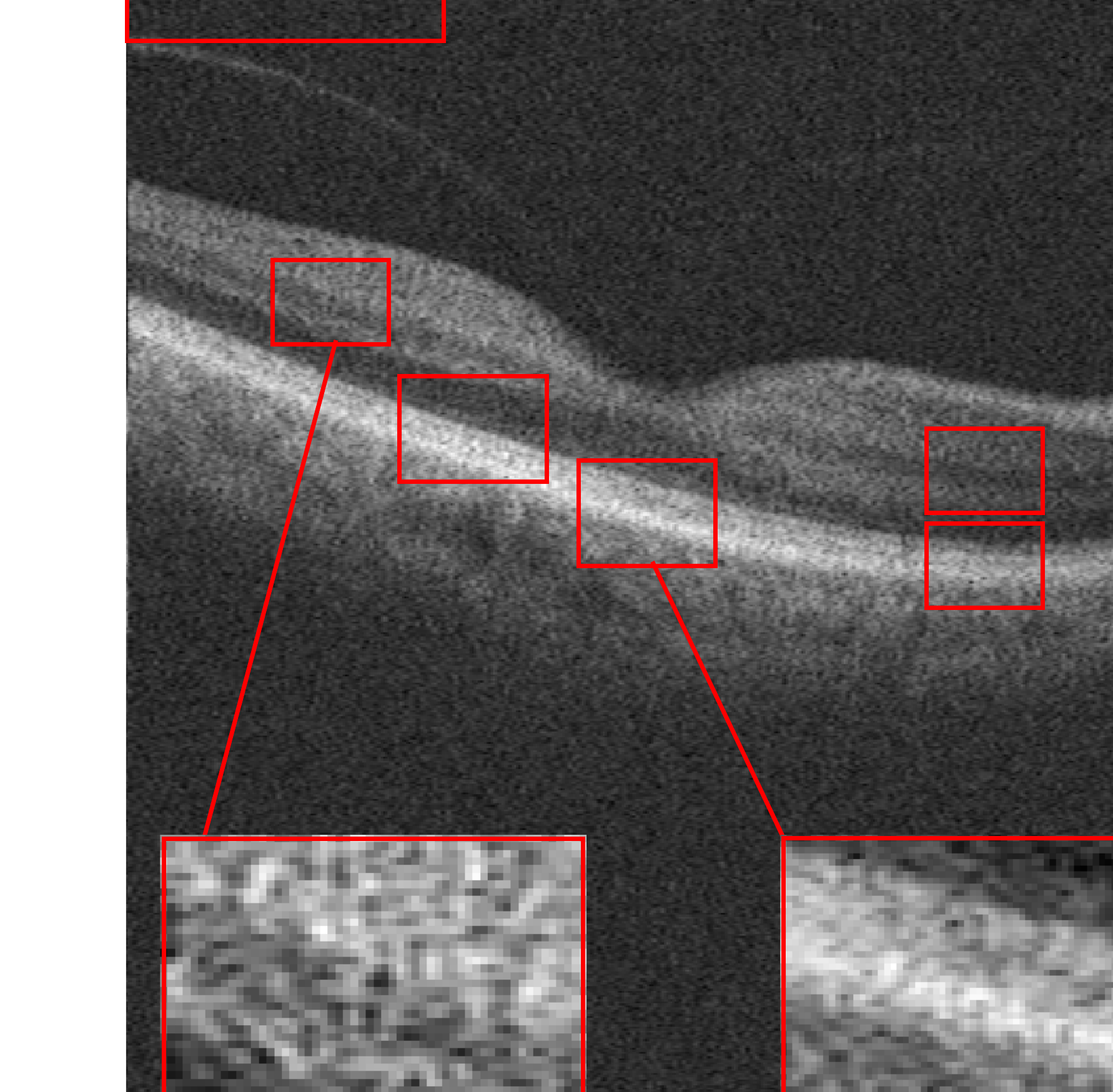}&
			\includegraphics[scale=0.18, trim={14.5cm 6cm 18.2cm 1.7cm}, clip]{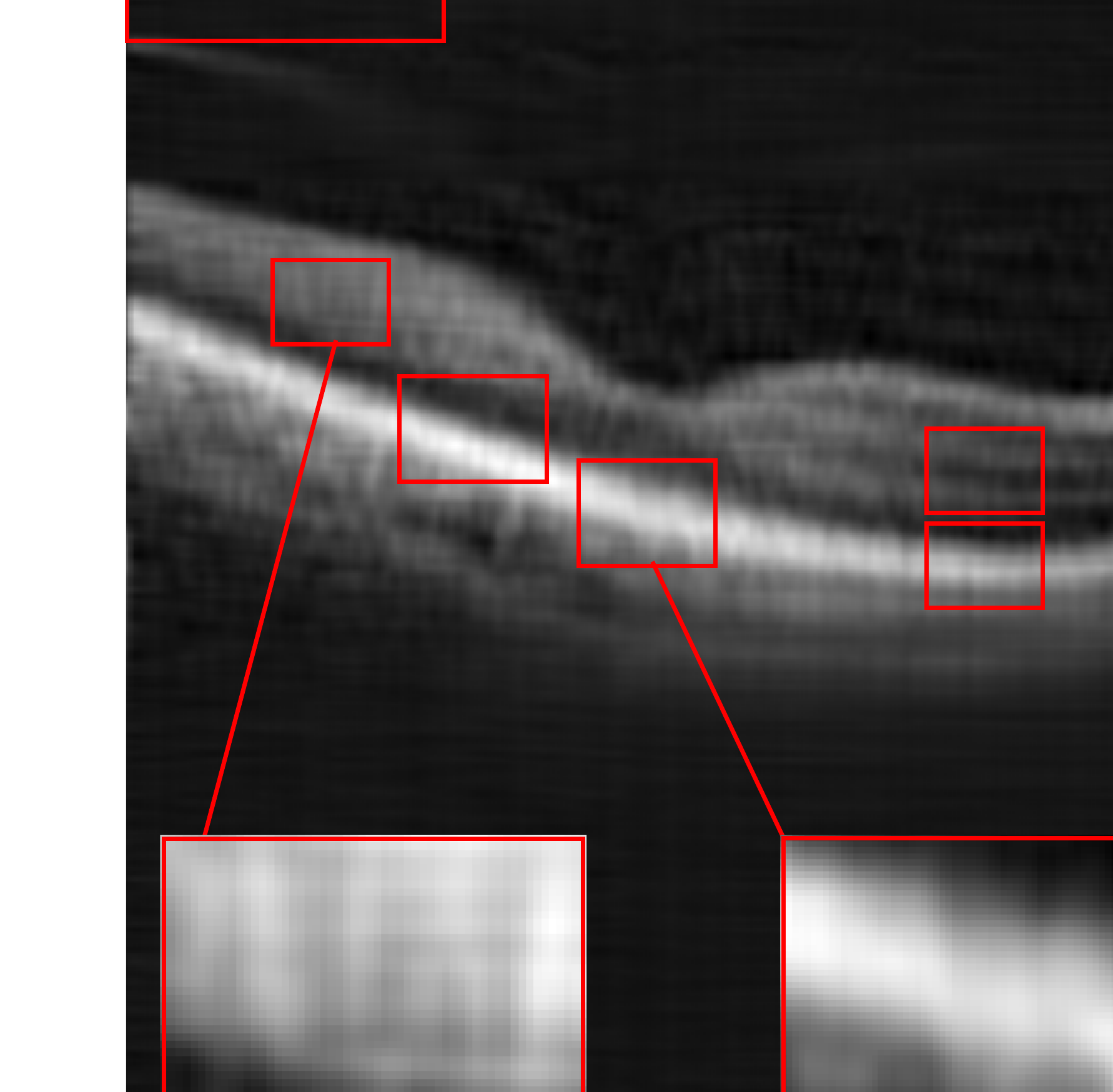}&\includegraphics[scale=0.18, trim={14.5cm 6cm 18.2cm 1.7cm}, clip]{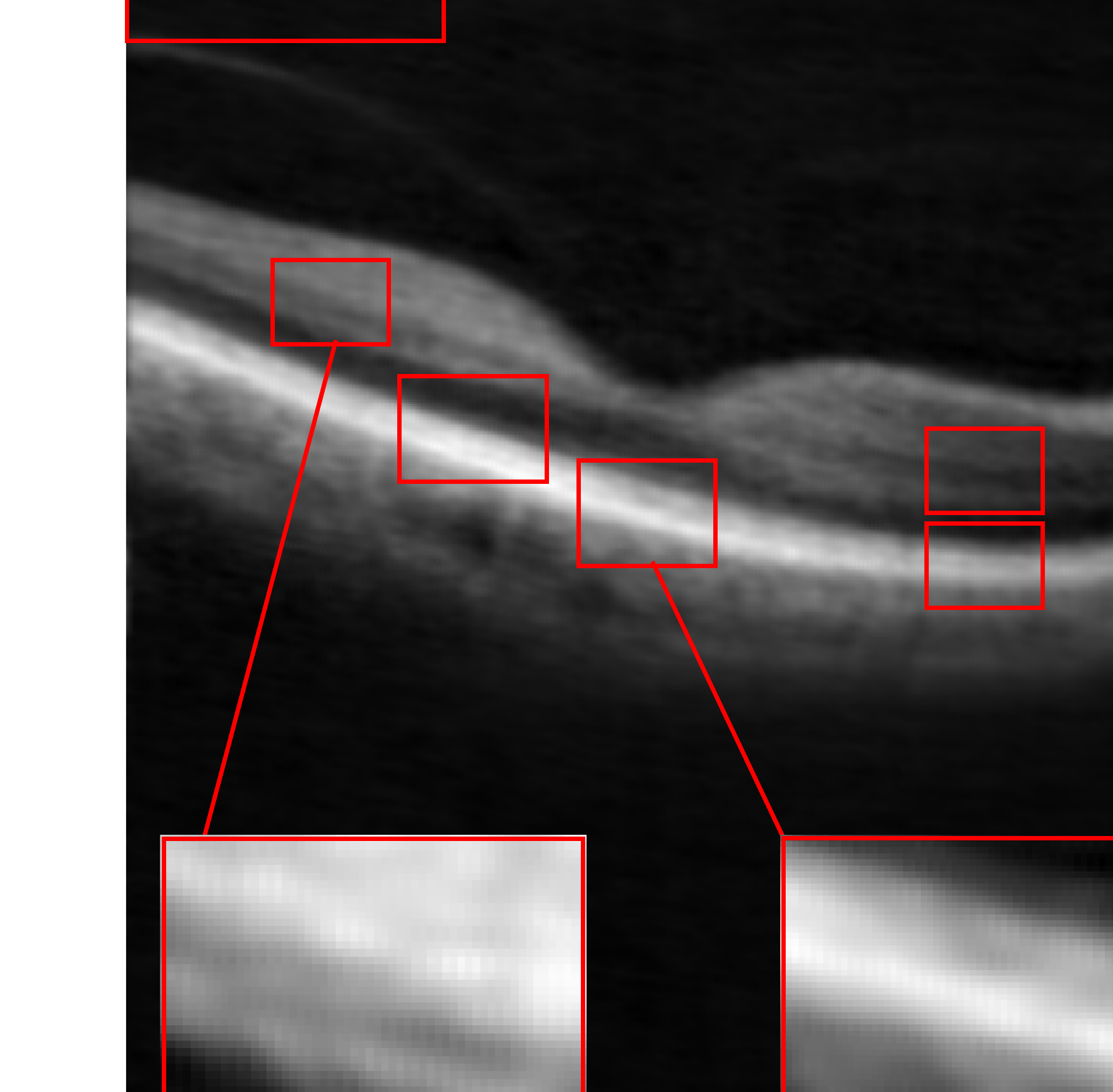}\\
			\quad \quad  CNR=1.7055&CNR=0.4120&CNR=2.6569&CNR=3.5883&\textbf{CNR=3.6502}\\
			\includegraphics[scale=0.18, trim={14.5cm 6cm 18.2cm 1.7cm}, clip]{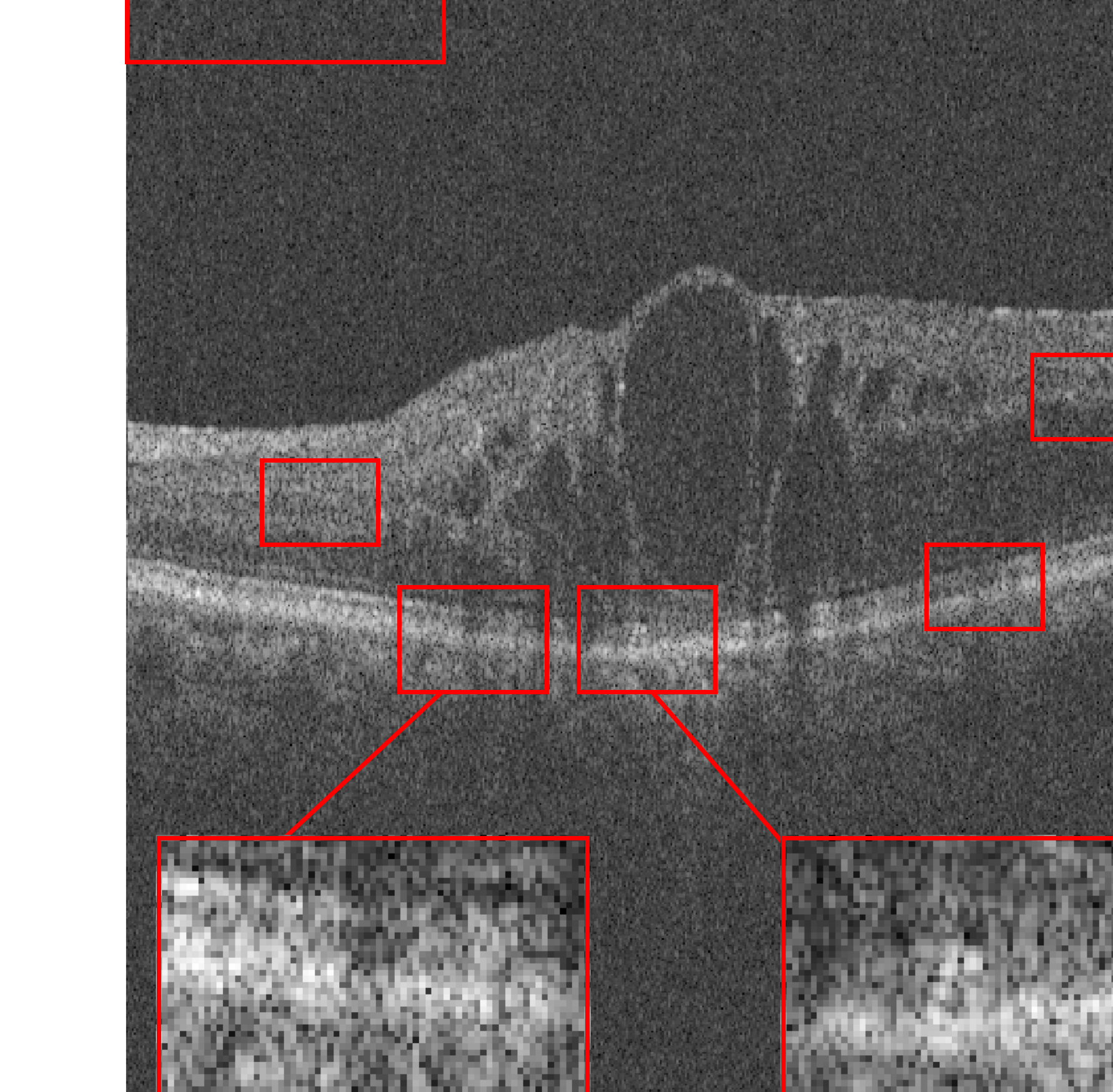}&\includegraphics[scale=0.18, trim={14.5cm 6cm 18.2cm 1.7cm}, clip]{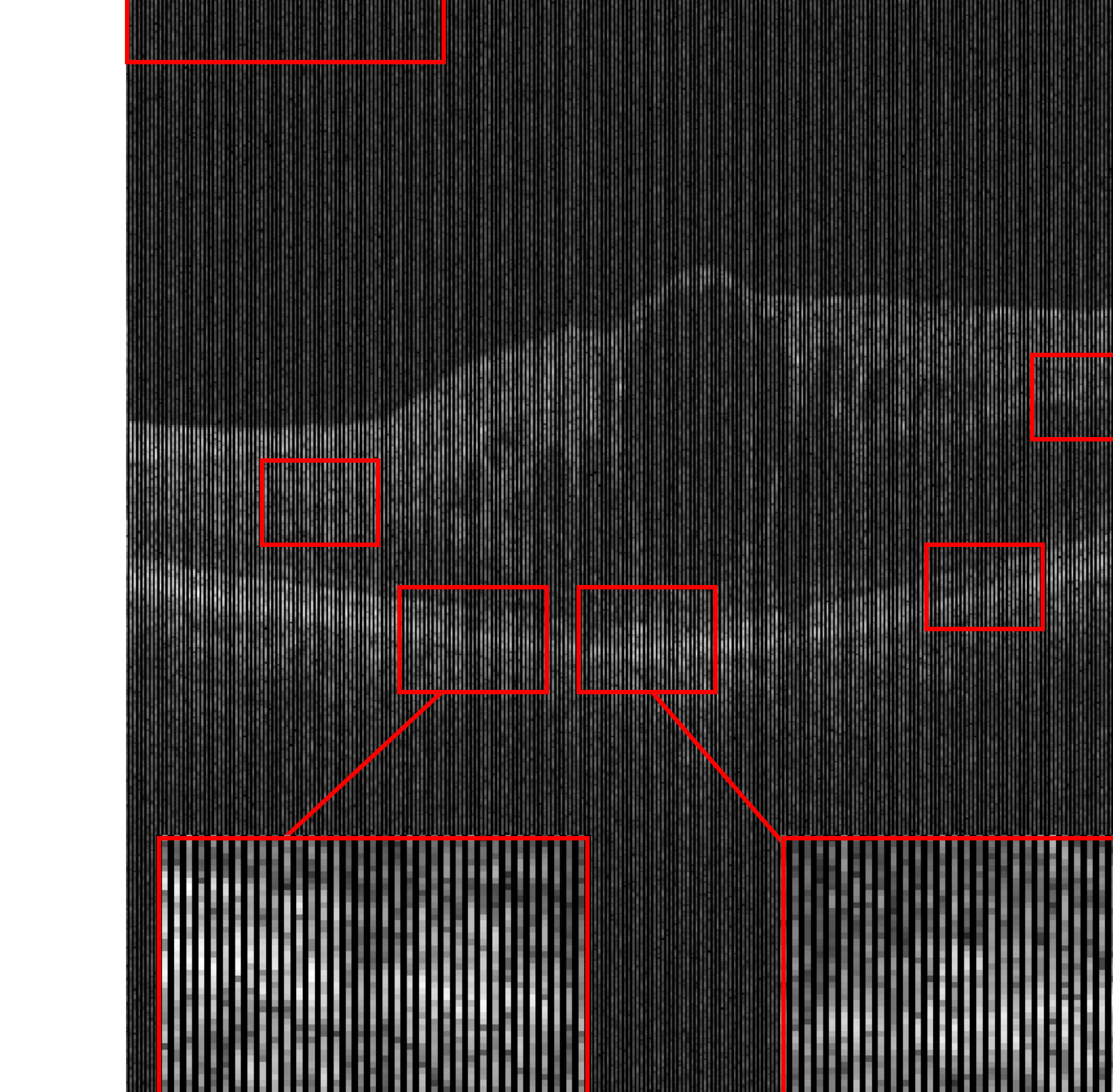}&\includegraphics[scale=0.18, trim={14.5cm 6cm 18.2cm 1.7cm}, clip]{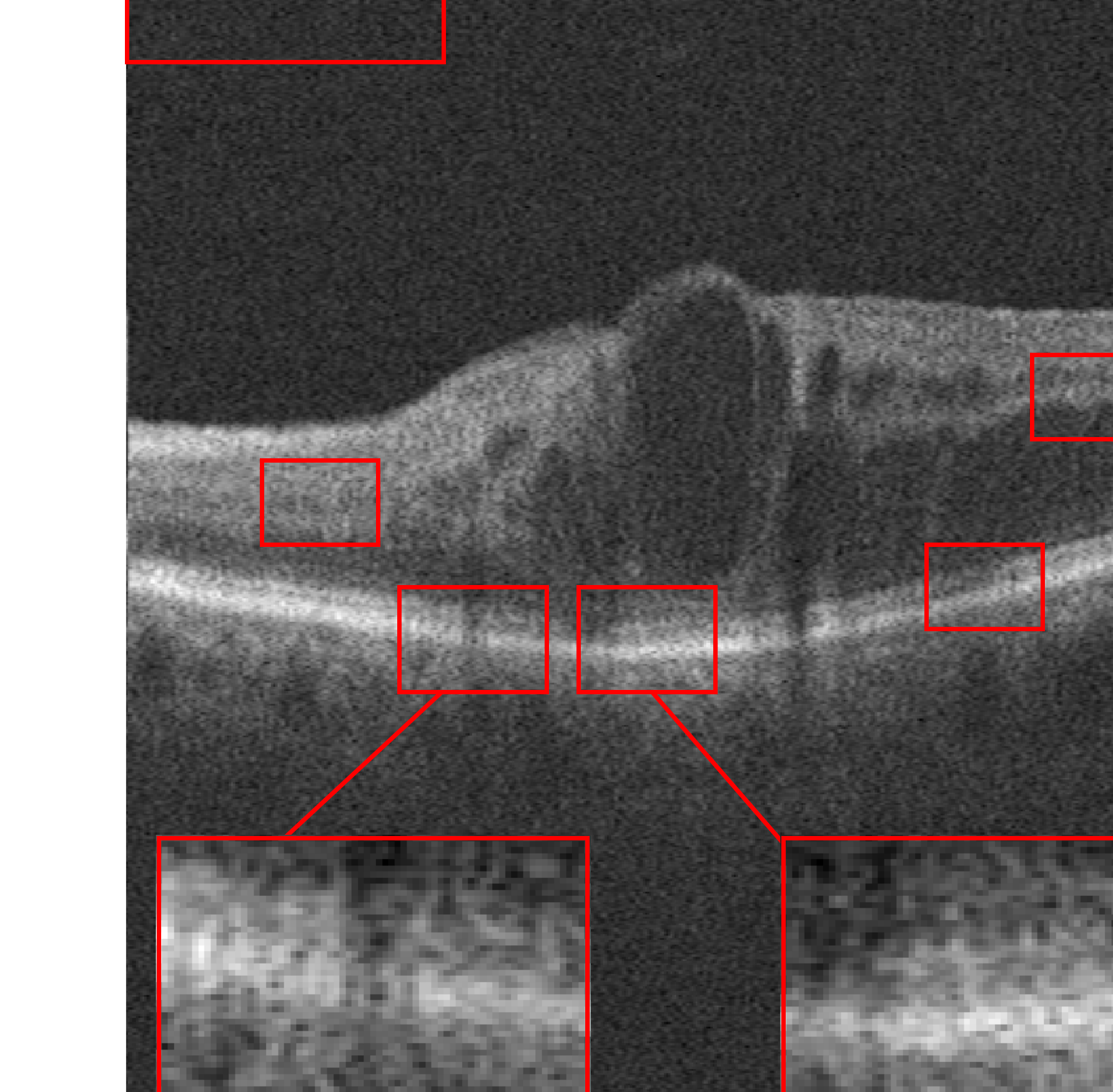}&
			\includegraphics[scale=0.18, trim={14.5cm 6cm 18.2cm 1.7cm}, clip]{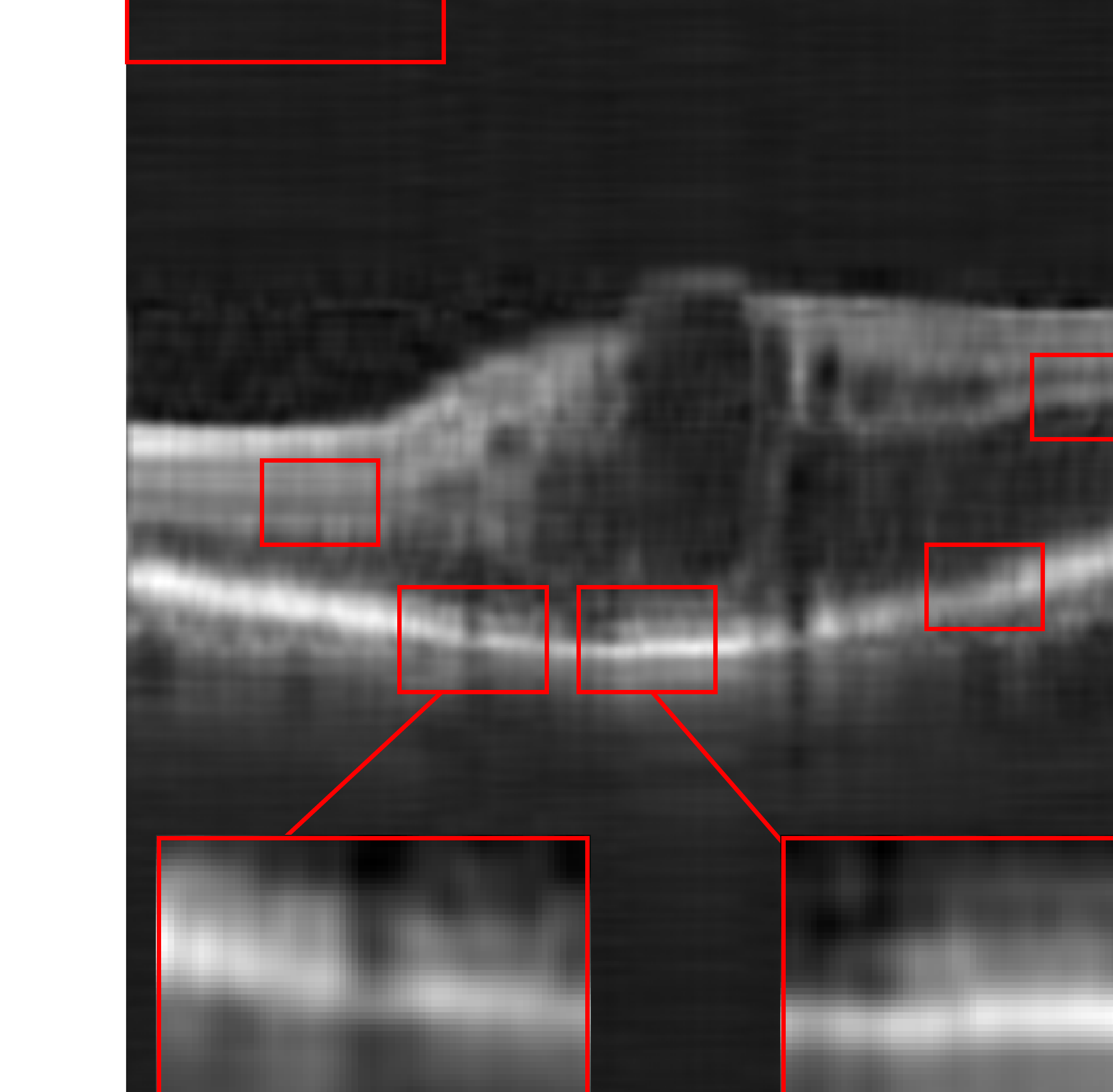}&\includegraphics[scale=0.18, trim={14.5cm 6cm 18.2cm 1.7cm}, clip]{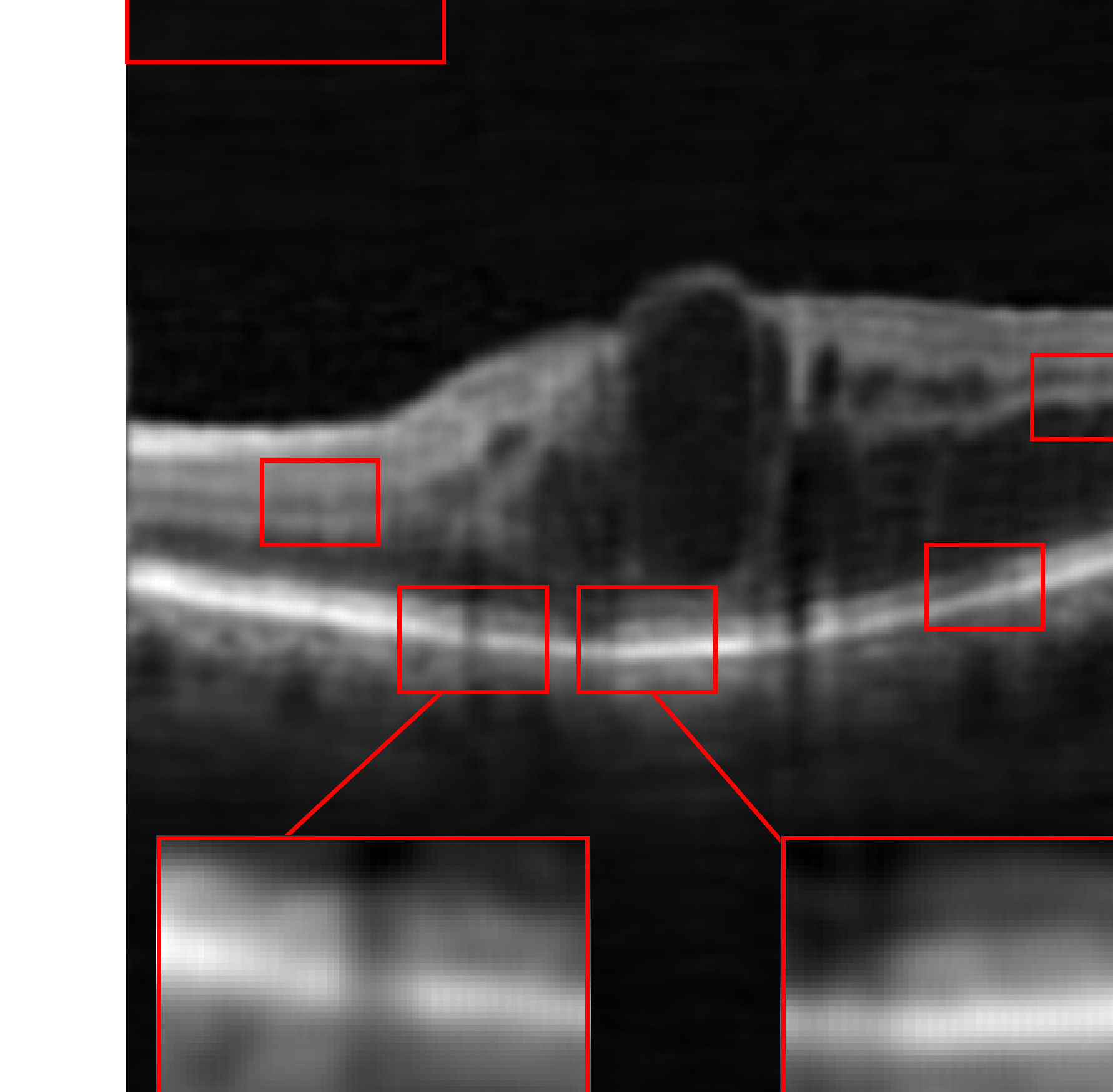}\\
			\quad \quad  CNR=0.4119&CNR=1.7943&CNR=2.8807&CNR=3.7744&\textbf{CNR=3.7885}\\
			\includegraphics[scale=0.18, trim={14.5cm 6cm 18.2cm 1.7cm}, clip]{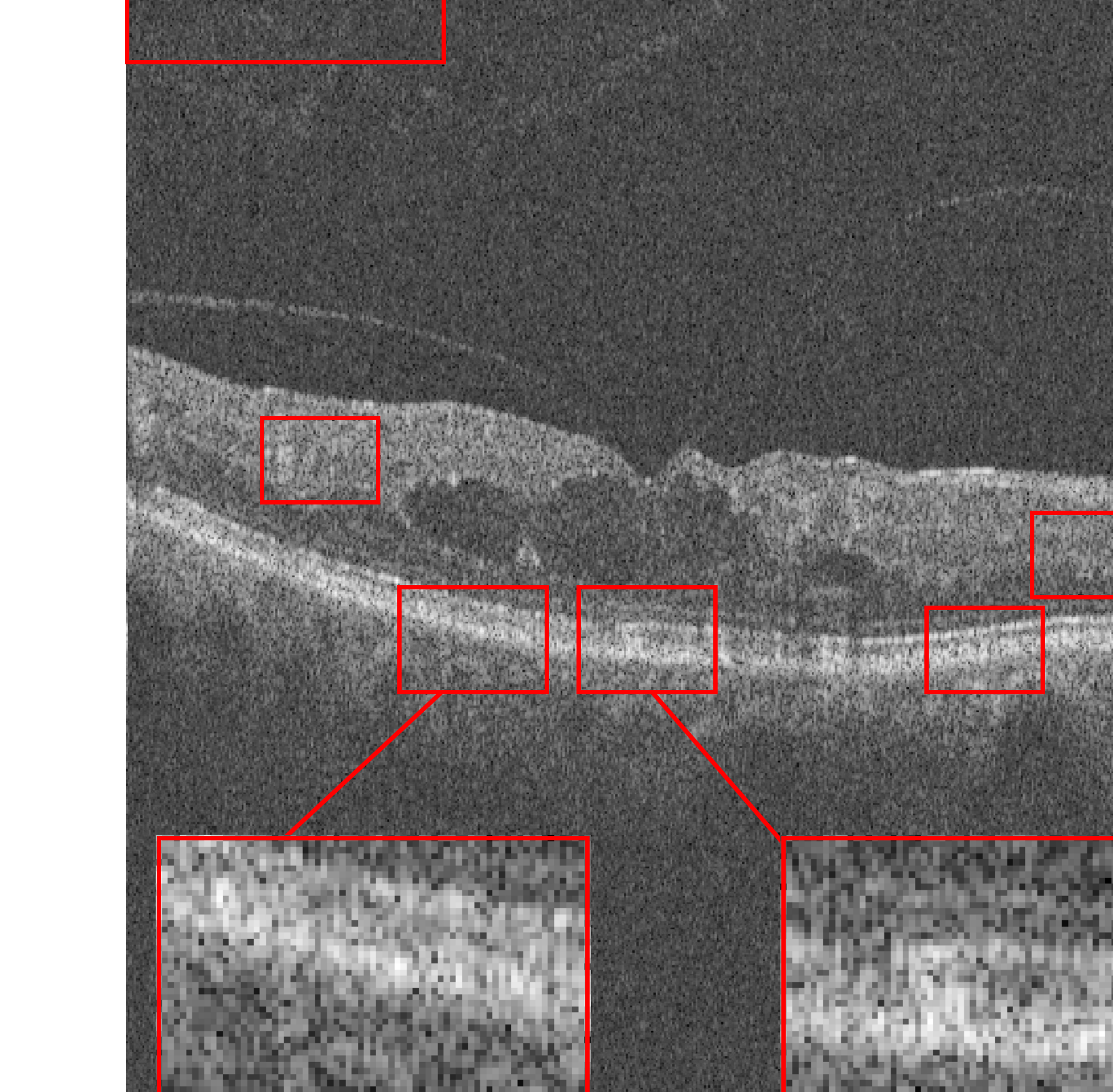}&\includegraphics[scale=0.18, trim={14.5cm 6cm 18.2cm 1.7cm}, clip]{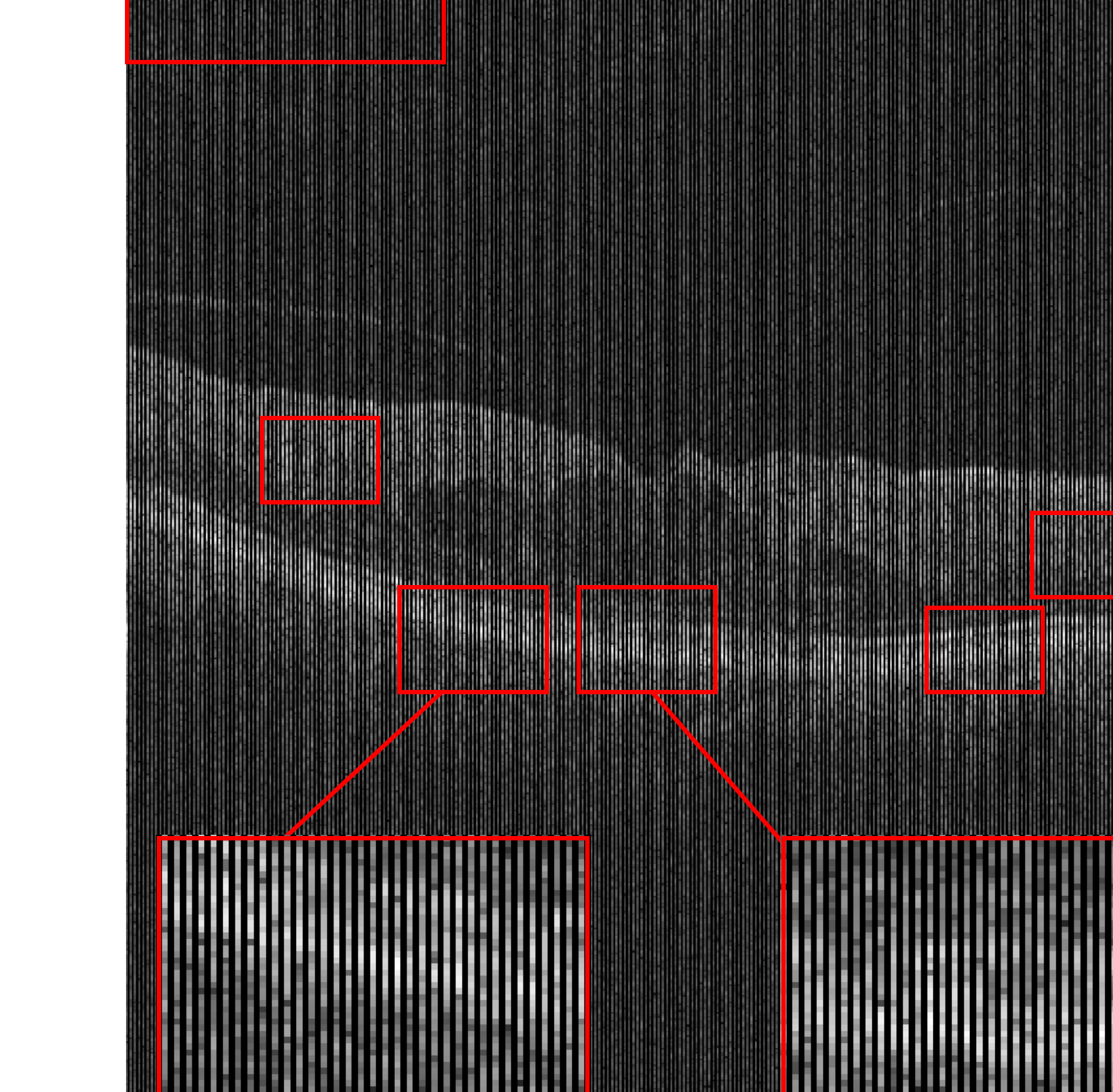}&\includegraphics[scale=0.18, trim={14.5cm 6cm 18.2cm 1.7cm}, clip]{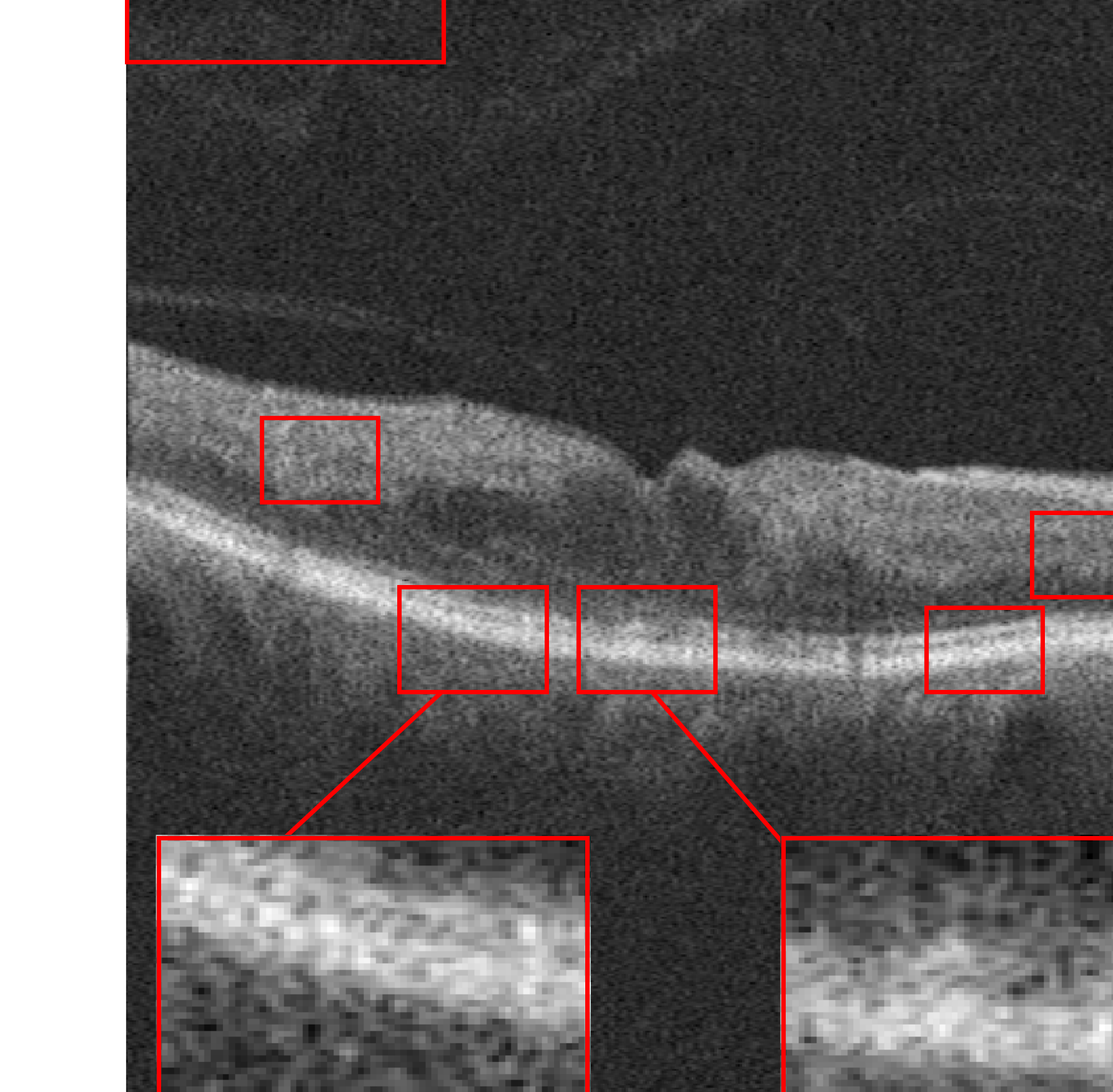}&
			\includegraphics[scale=0.18, trim={14.5cm 6cm 18.2cm 1.7cm}, clip]{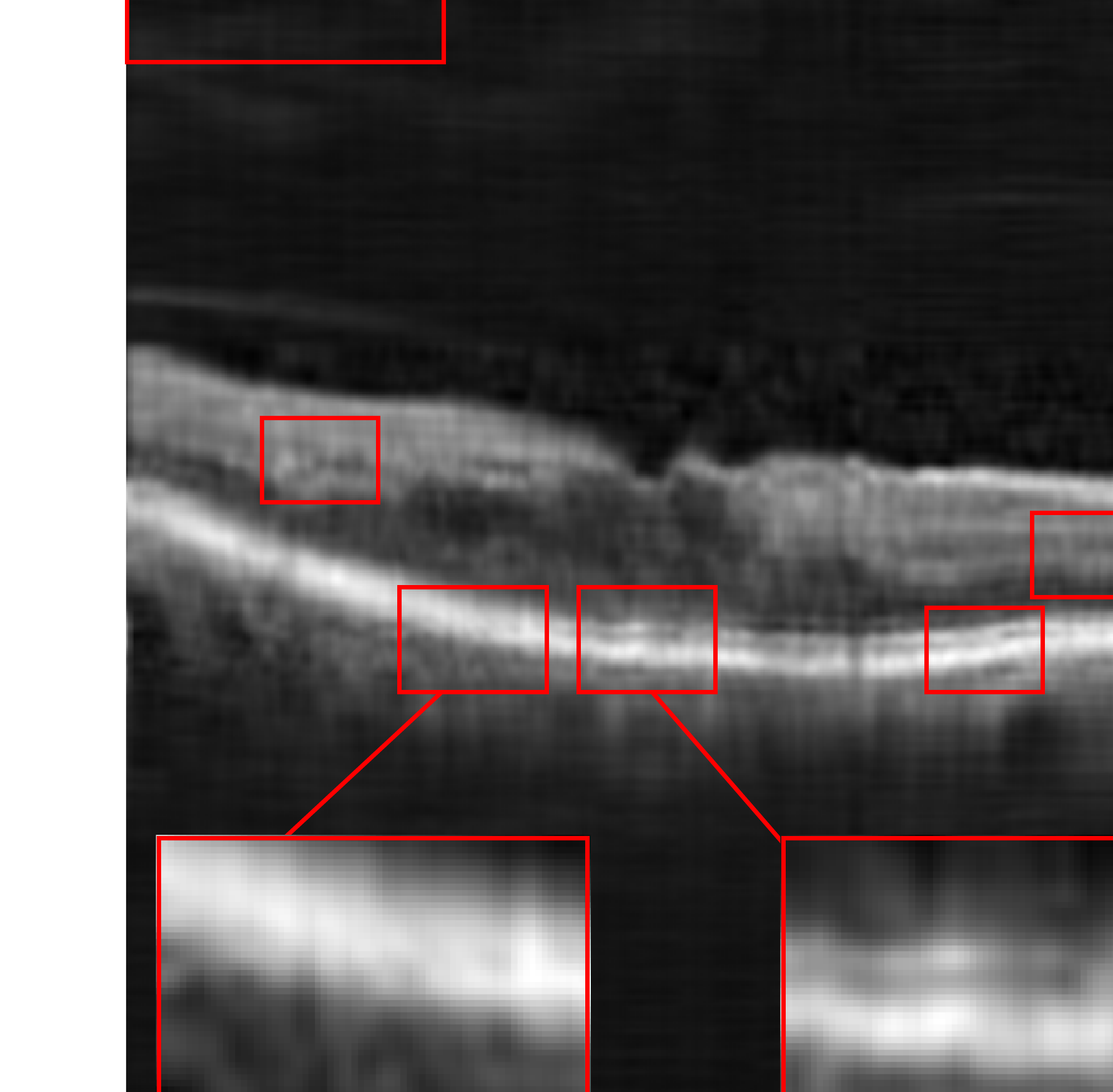}&\includegraphics[scale=0.18, trim={14.5cm 6cm 18.2cm 1.7cm}, clip]{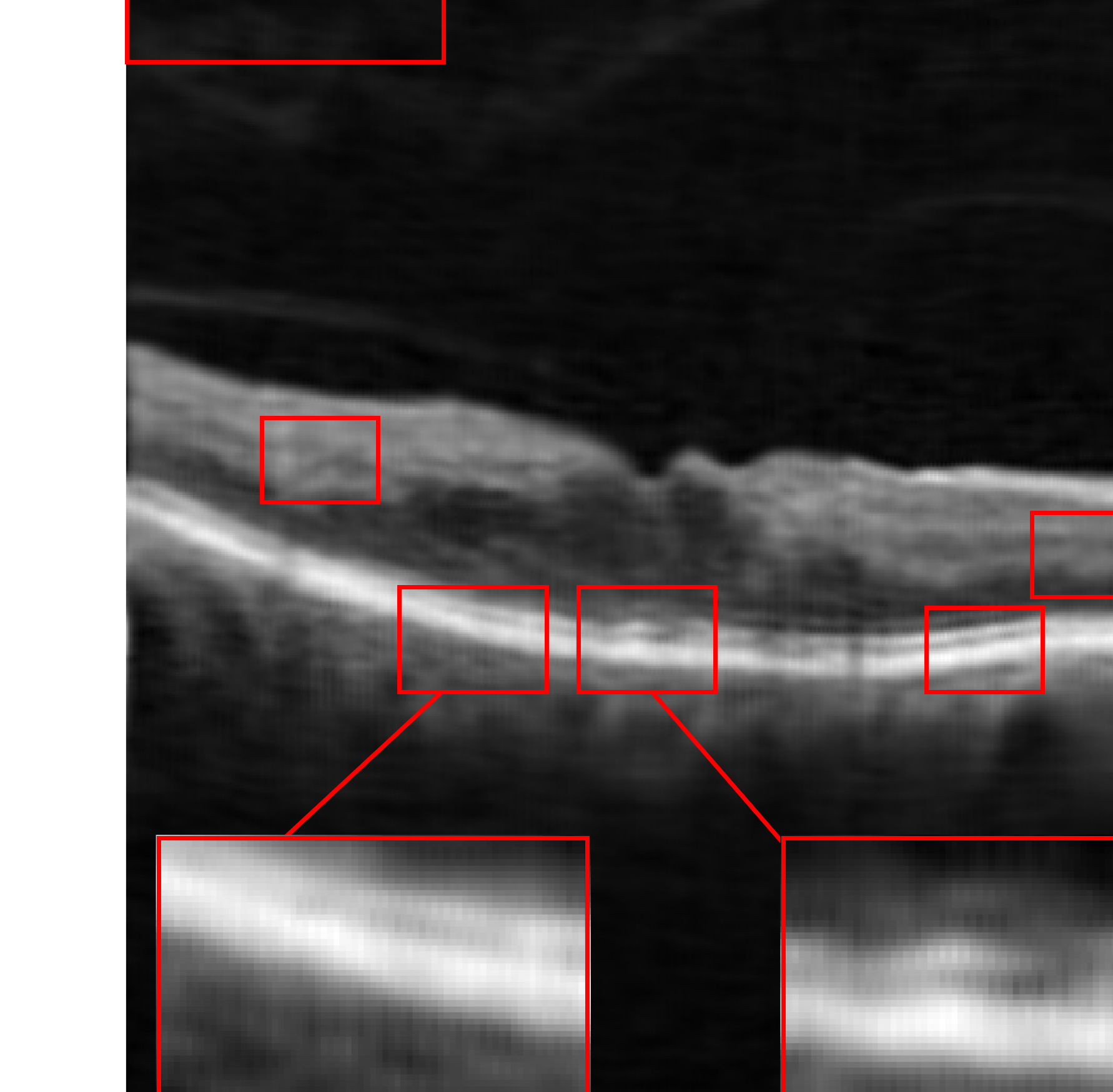}\\
			\quad \quad  CNR=2.1693&CNR=0.4928&CNR=3.2678&CNR=3.9743&\textbf{CNR=4.0468}\\
		\end{tabular}
	}
	\caption{Comparison of the performance of the algorithms for the super-resolution of  OCT images of dataset2 with $50\%$ missing ratio of A-scans.  
	}
	\label{topcon}
\end{figure*}
The proposed algorithm has been also tested for the super-resolution of the two other datasets. For dataset2,  the middle slices of several 3D image have been selected for test and averaged with 2 previous and 2 next slices. The weights $\alpha_i$'s set equal to $0.2$ for each B-scan. For the proposed approach, the patch sizes were set equal to 5 and 7 or (5 and 10) with window size equal to [2,2]. For MDT the window size was set to [2,2]. The results for the super-resolution of artificially missed B-scans of dataset2 with $50\%$ missing ratio have been illustrated in Fig.~\ref{topcon}. Since this dataset does not contain reference images, PSNR and SSIM could not be reported. Hence, only visual comparison and the resulting CNR's, with the following definition, have been reported for each algorithm: 
\begin{equation}
\label{cnreq}	
\textrm{CNR}={\frac{|\mu_f-\mu_b|}{\sqrt{0.5(\sigma_f^2+\sigma_b^2)}}},
\end{equation}
where $\mu_f$  and $\sigma_f$ are the mean and standard deviation of the foreground respectively and $\mu_b$ and  $\sigma_b$ are the  mean and standard deviation of the background region, respectively. Averaged CNR's  for 5 Regions Of Interest (ROI) in each image (relative to an ROI selected in the background region of the image) have been computed. Note that ROI's have been shown with red boxes in each image of Fig.~\ref{topcon}. As the results show, the proposed algorithm results in a better output and higher CNR comparing to the other algorithms.

\begin{figure*}[h!]
	\centering
	\centerline{
		\begin{tabular}{lccccc}
			\quad  \quad Original noisy image& \quad Incomplete image & \quad  Spline interpolation & \quad MDT \cite{yokota2018missing} & \quad Proposed algorithm\\ 
			\includegraphics[scale=0.18, trim={13.5cm 4.5cm 17.2cm 1.7cm}, clip]{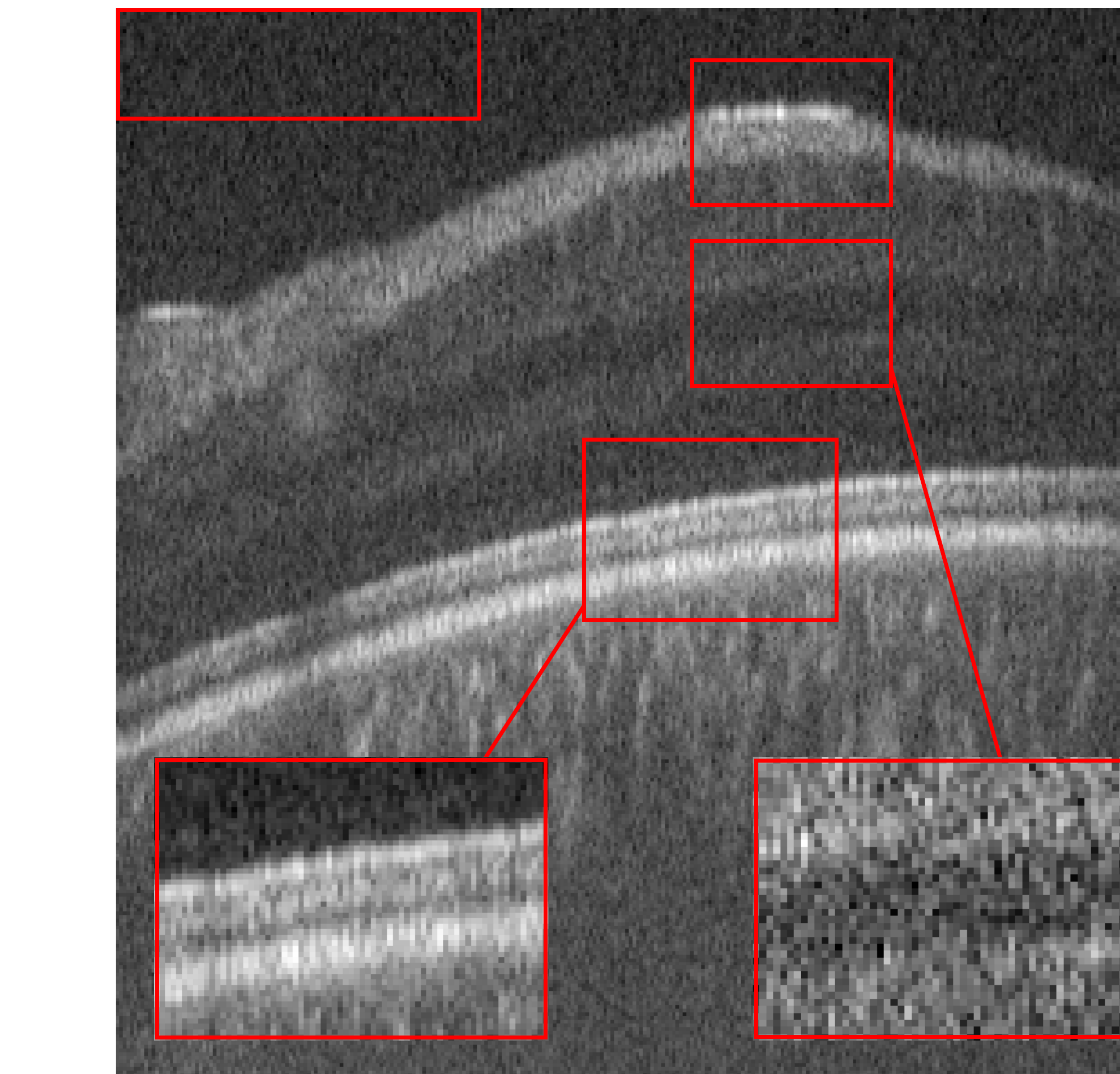}&\includegraphics[scale=0.18, trim={13.5cm 4.5cm 17.2cm 1.7cm}, clip]{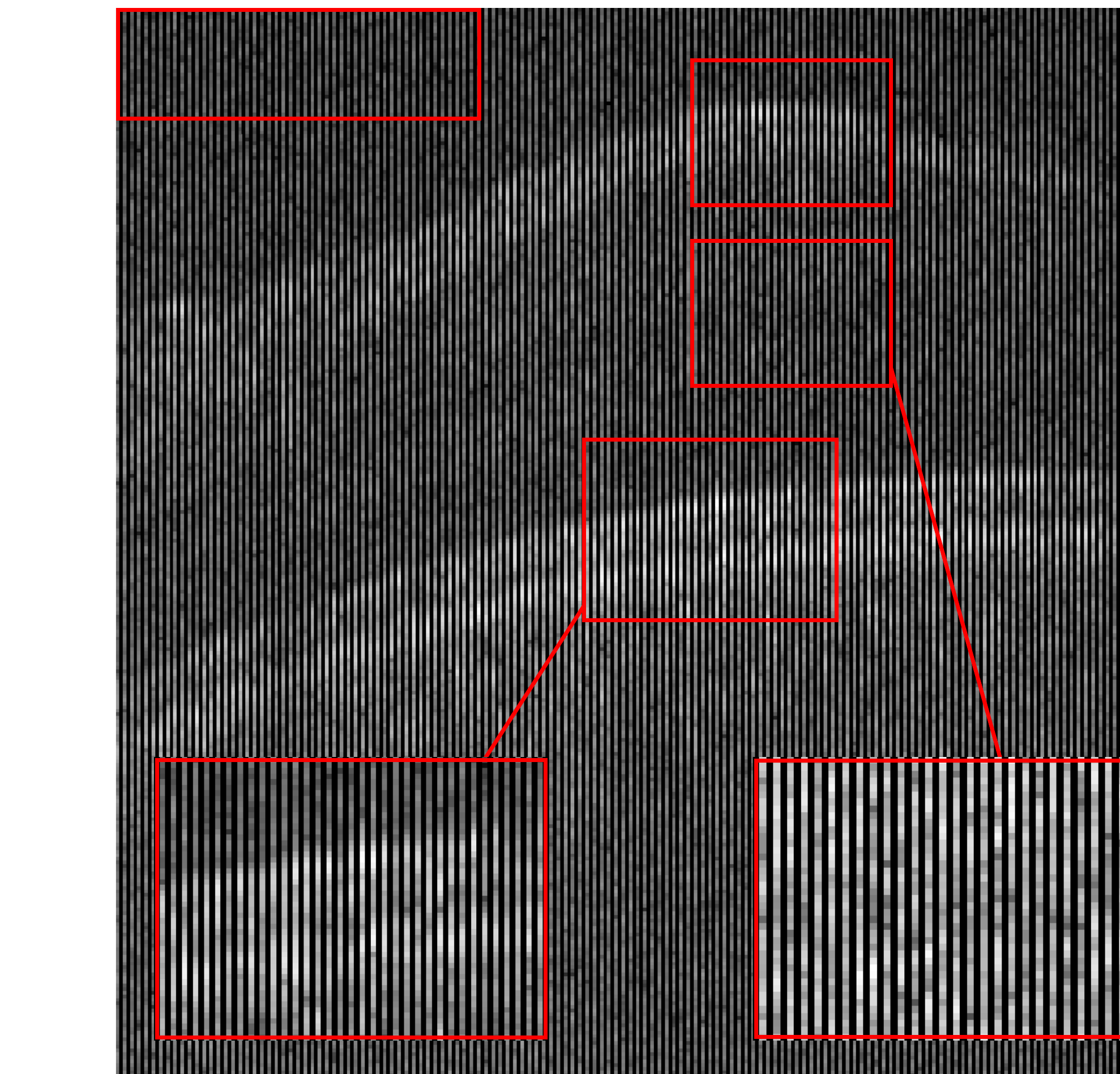}&\includegraphics[scale=0.18, trim={13.5cm 4.5cm 17.2cm 1.7cm}, clip]{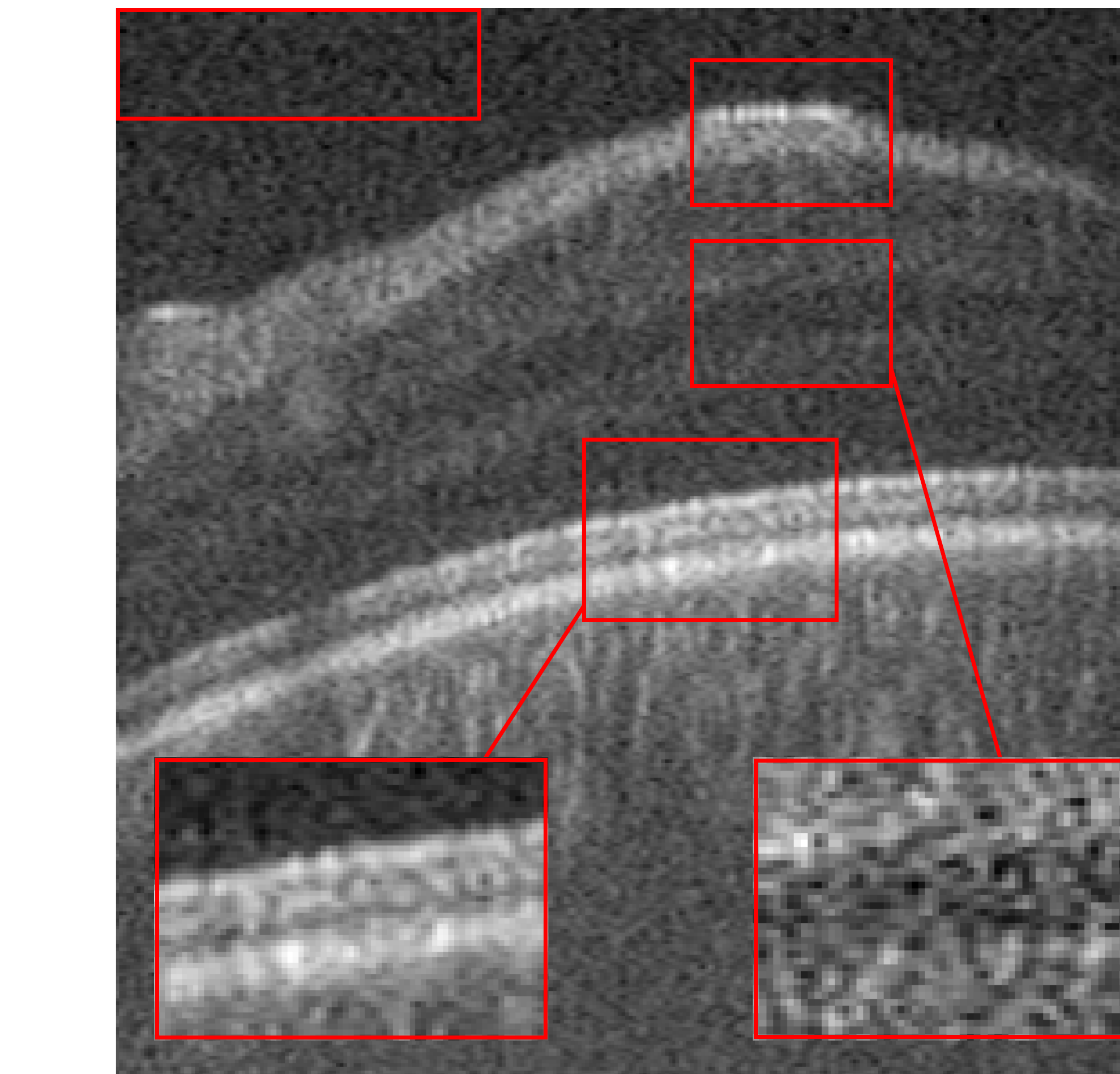}&
			\includegraphics[scale=0.18, trim={13.5cm 4.5cm 17.2cm 1.7cm}, clip]{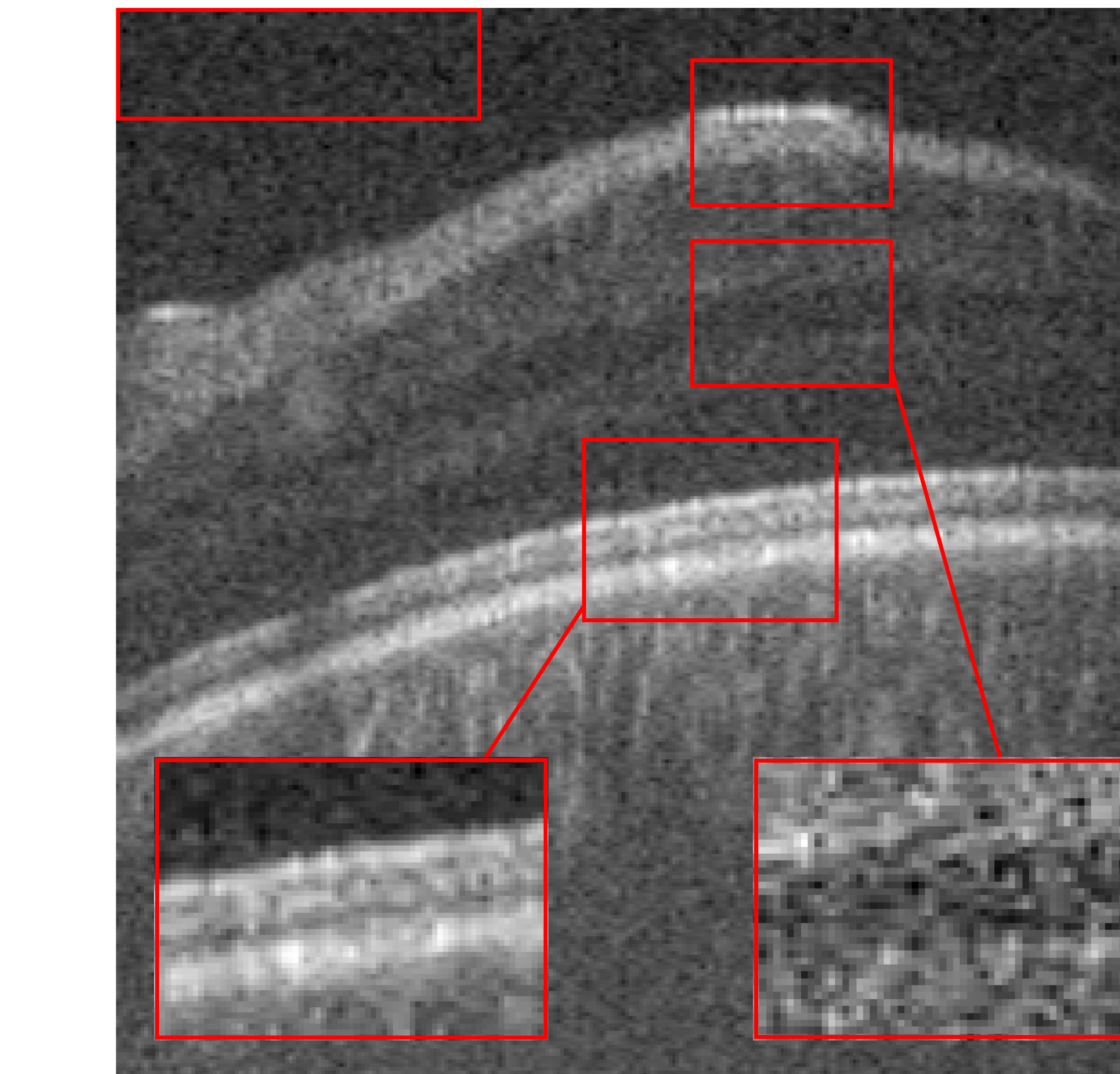}&\includegraphics[scale=0.18, trim={13.5cm 4.5cm 17.2cm 1.7cm}, clip]{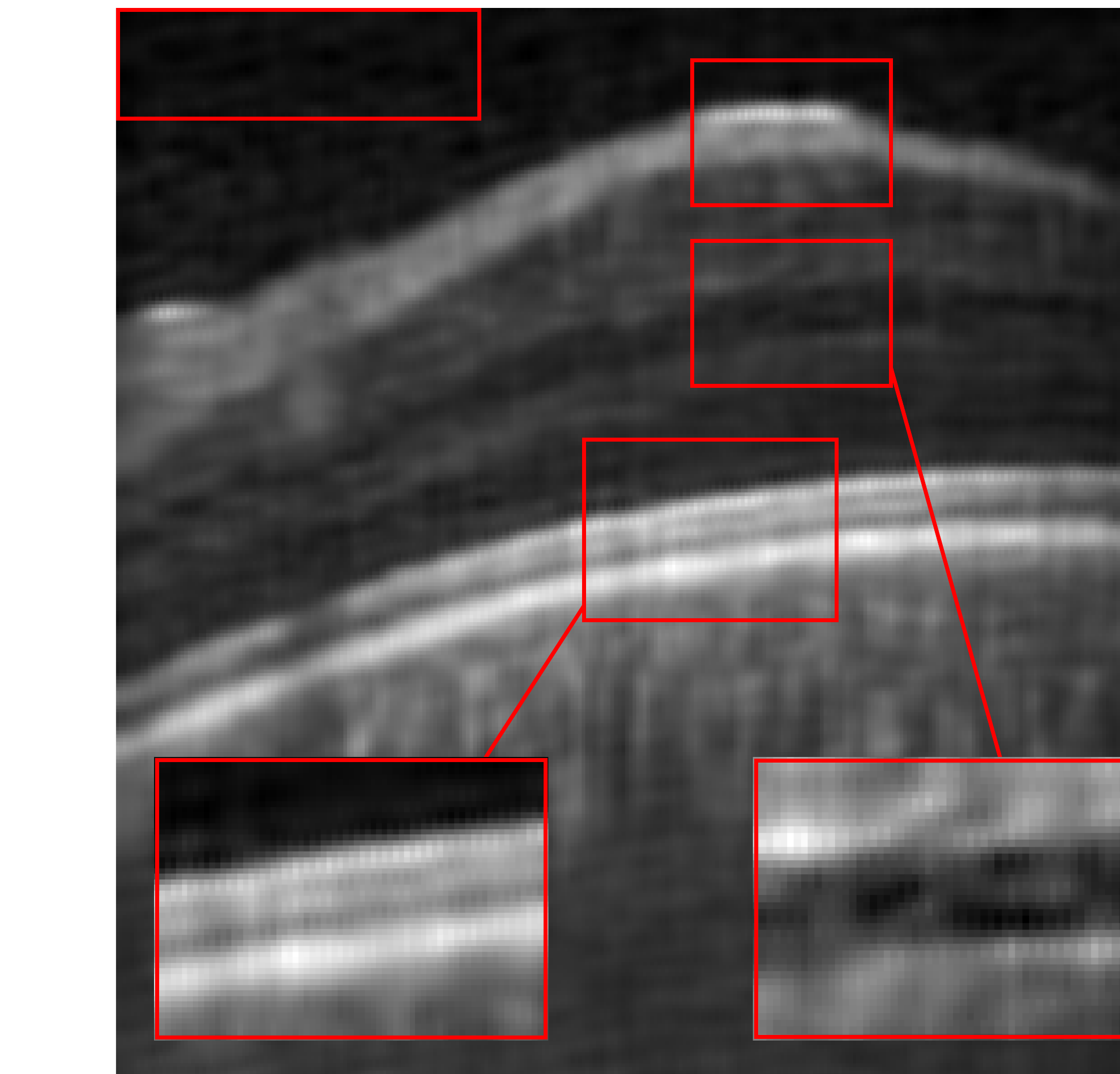}\\
			\quad \quad \quad  CNR = 1.3654&\quad CNR = 0.2807&\quad CNR=1.9968&\quad CNR=2.0549&\quad \textbf{CNR=2.7711}\\
			\includegraphics[scale=0.18, trim={13.5cm 4.5cm 17.2cm 1.7cm}, clip]{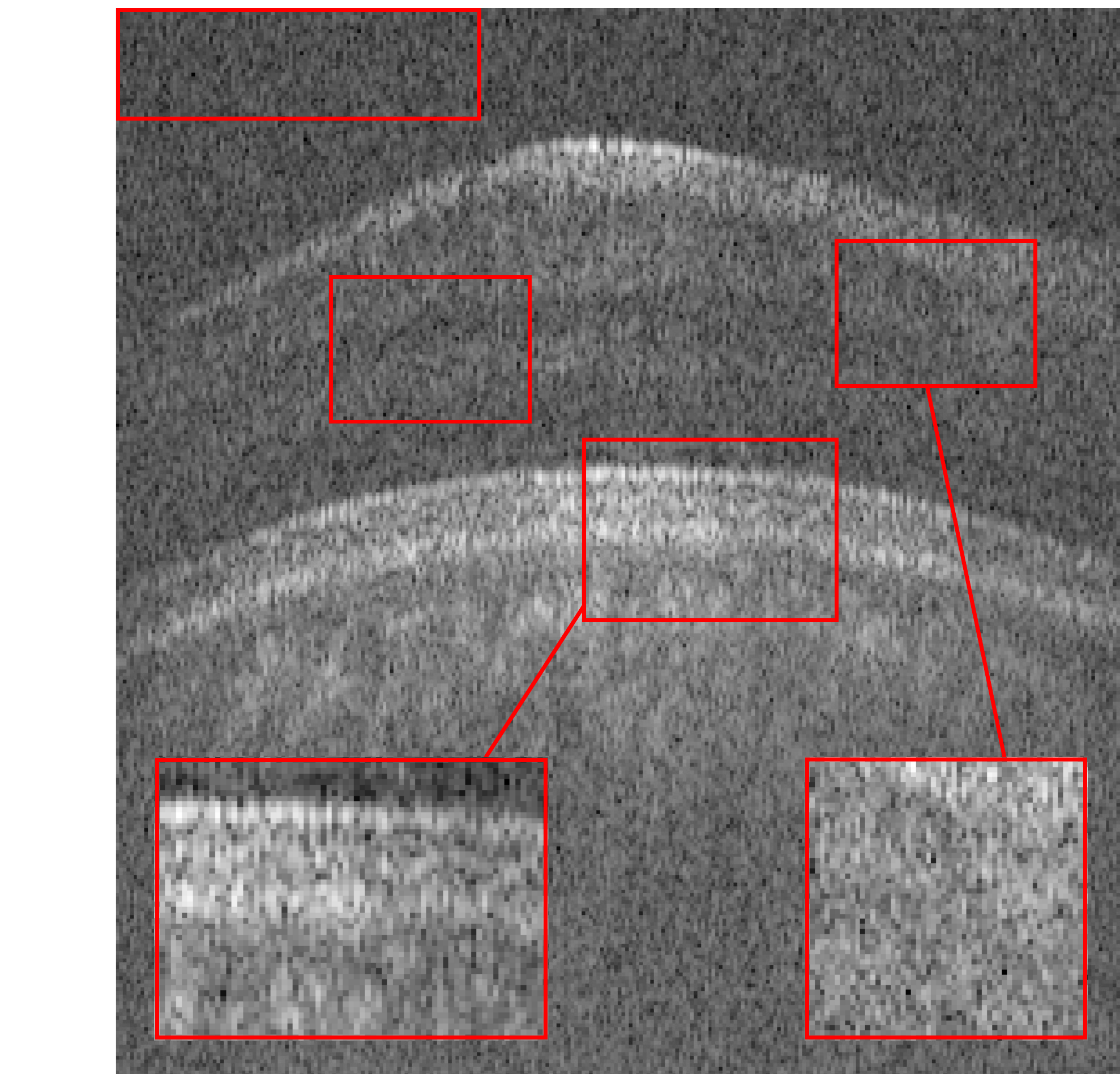}&\includegraphics[scale=0.18, trim={13.5cm 4.5cm 17.2cm 1.7cm}, clip]{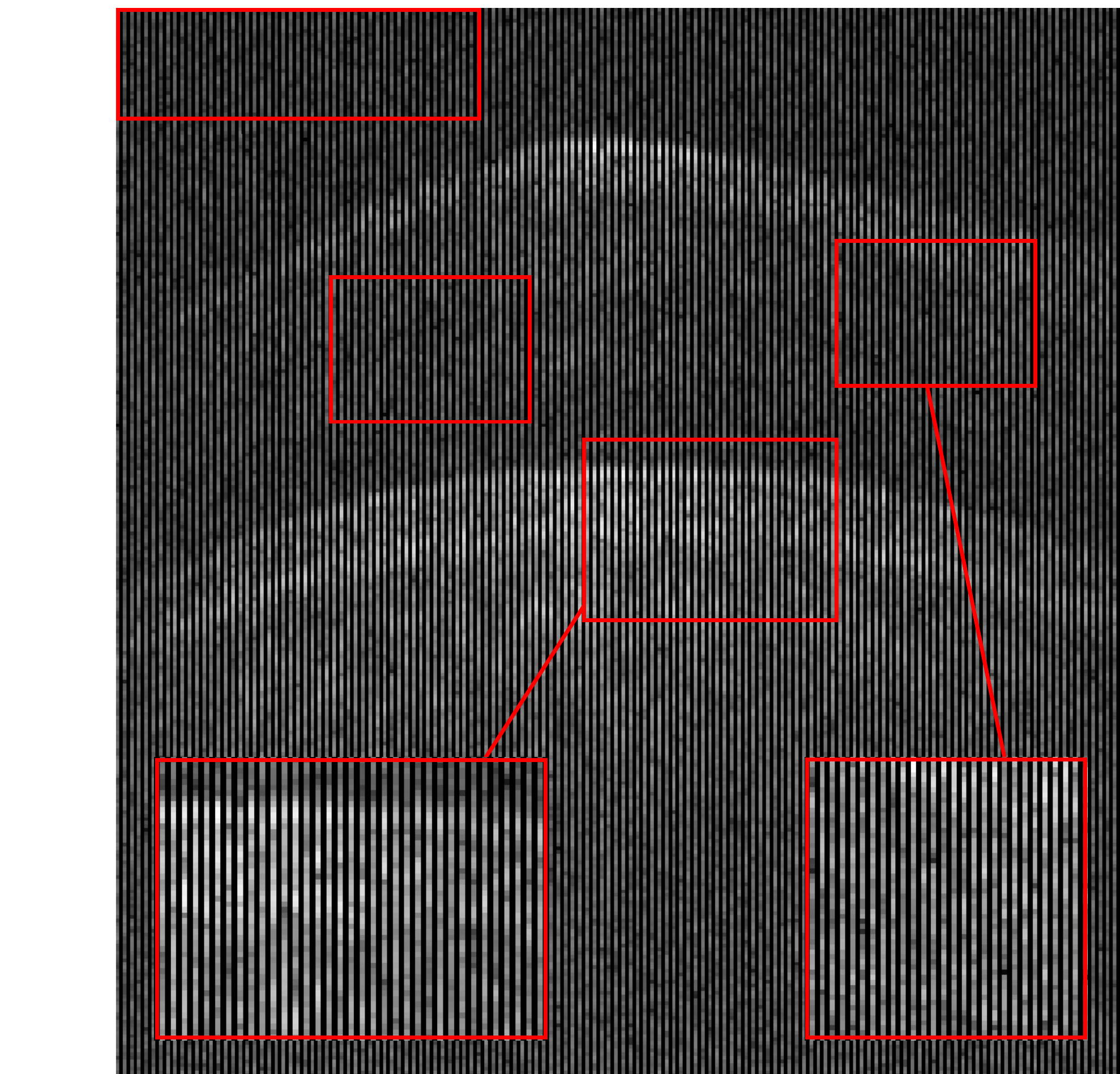}&\includegraphics[scale=0.18, trim={13.5cm 4.5cm 17.2cm 1.7cm}, clip]{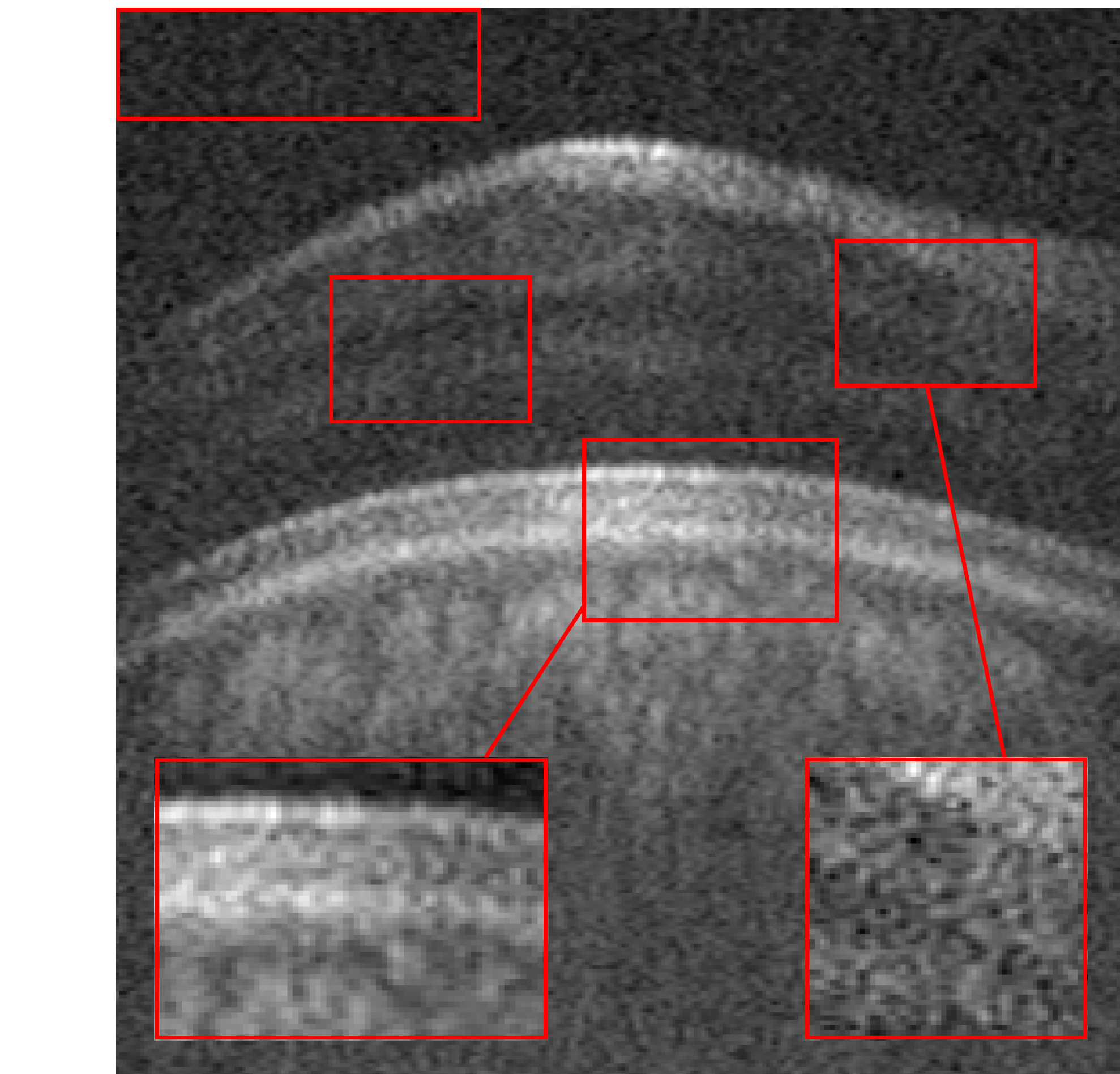}&
			\includegraphics[scale=0.18, trim={13.5cm 4.5cm 17.2cm 1.7cm}, clip]{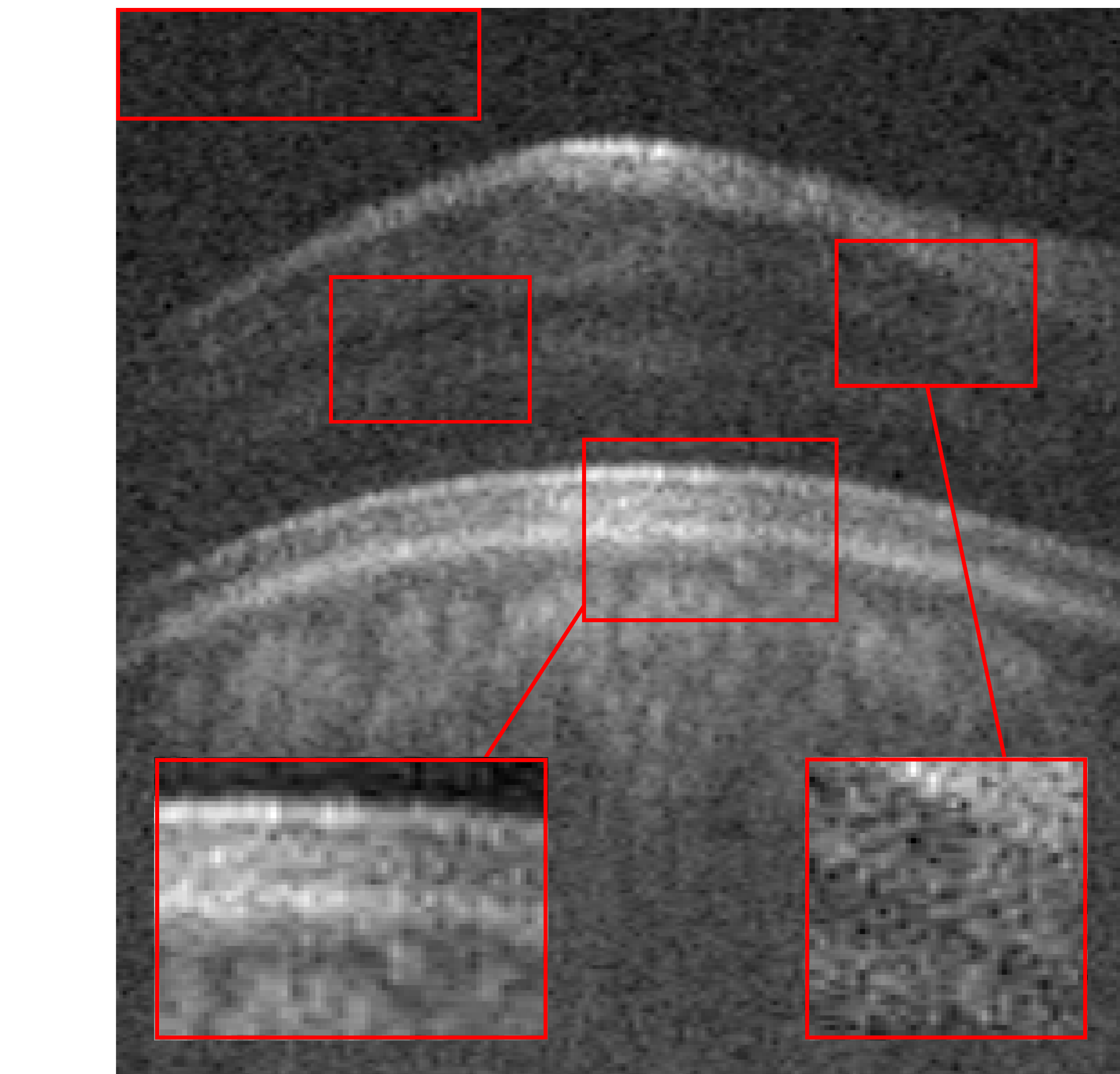}&\includegraphics[scale=0.18, trim={13.5cm 4.5cm 17.2cm 1.7cm}, clip]{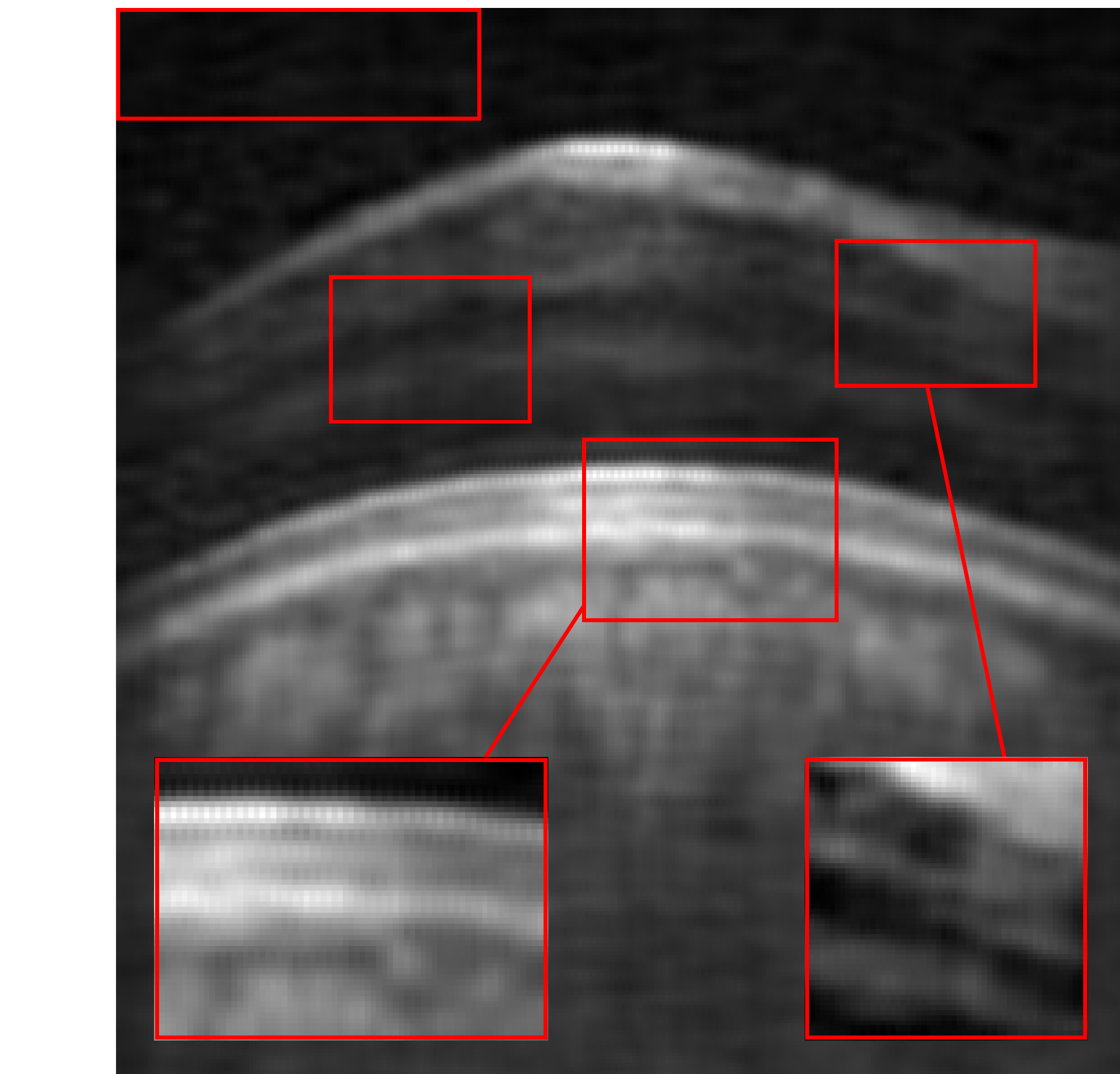}\\
			\quad \quad \quad  CNR=1.493&\quad CNR=0.2835&\quad CNR=2.3996&\quad CNR=2.2534&\quad \textbf{CNR=3.6184}\\
			\includegraphics[scale=0.18, trim={13.5cm 4.5cm 17.2cm 1.7cm}, clip]{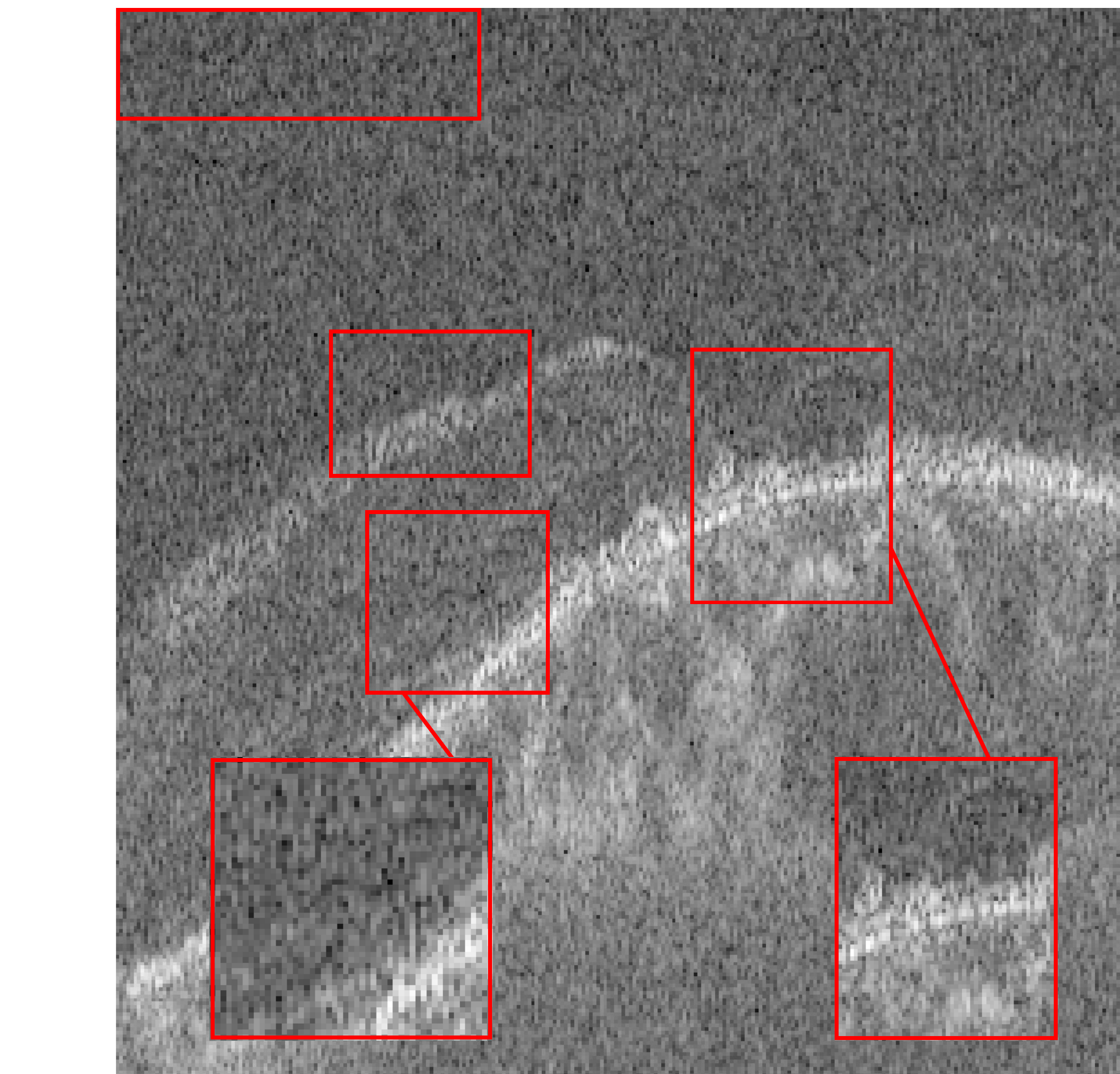}&\includegraphics[scale=0.18, trim={13.5cm 4.5cm 17.2cm 1.7cm}, clip]{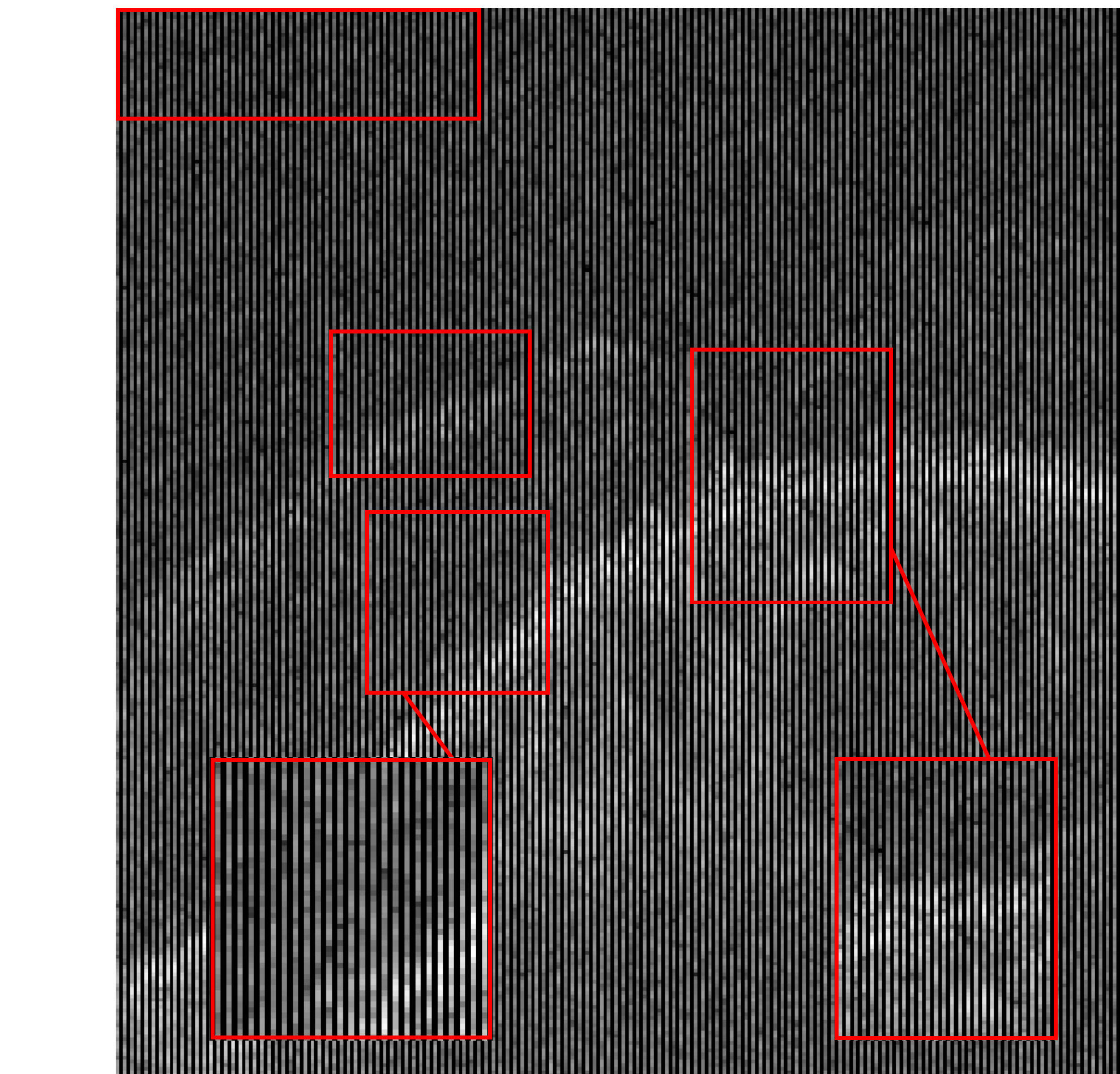}&\includegraphics[scale=0.18, trim={13.5cm 4.5cm 17.2cm 1.7cm}, clip]{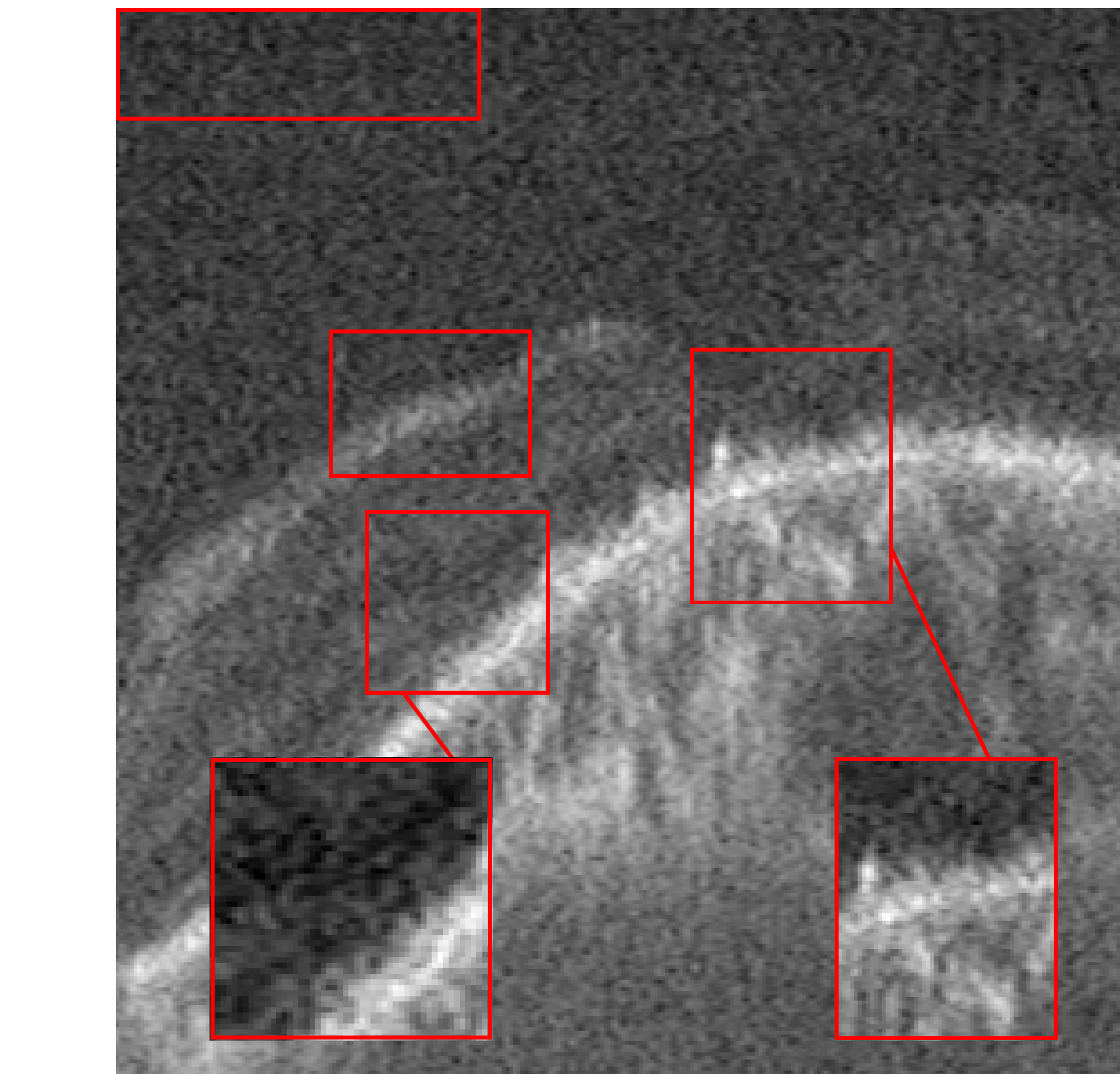}&
			\includegraphics[scale=0.18, trim={13.5cm 4.5cm 17.2cm 1.7cm}, clip]{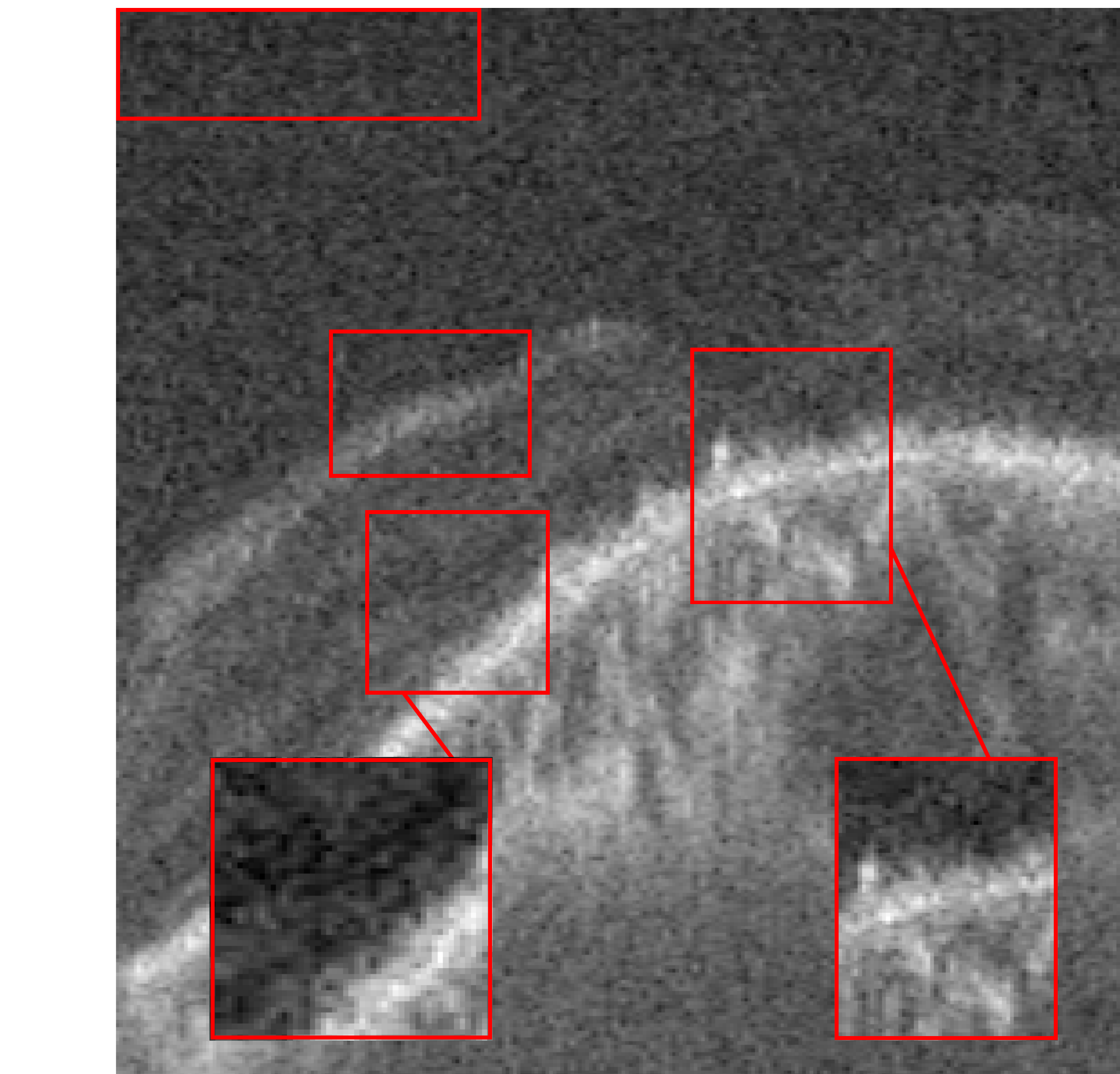}&\includegraphics[scale=0.18, trim={13.5cm 4.5cm 17.2cm 1.7cm}, clip]{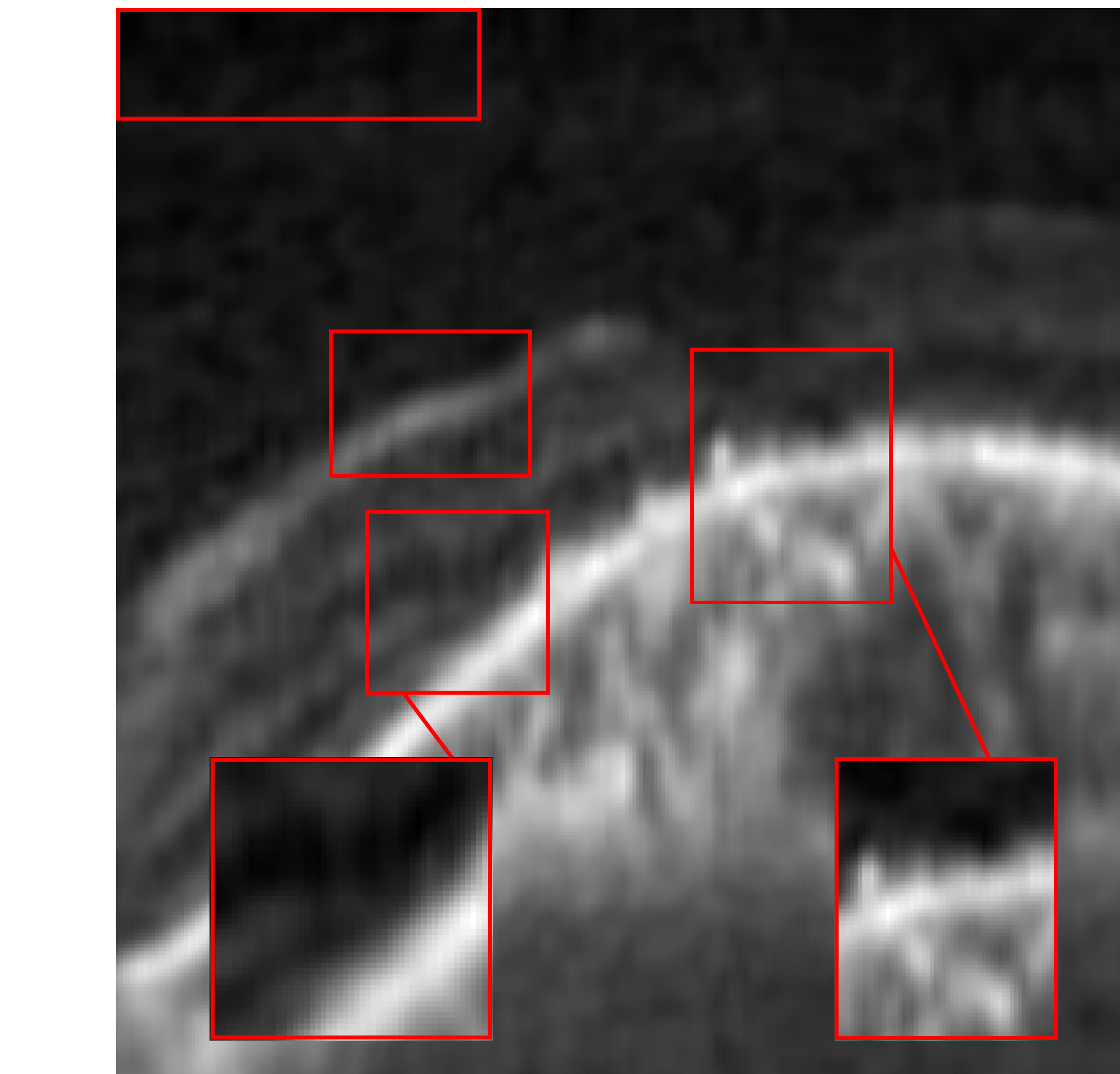}\\
			\quad \quad \quad  CNR=1.1753& \quad CNR=0.2446& \quad CNR=1.7957& \quad CNR=1.8335& \quad \textbf{CNR=2.1343}\\			
		\end{tabular}
	}
	\caption{Comparison of the performance of the algorithms for the super-resolution of OCT images of dataset3 with $50\%$ missing ratio of A-scans. CNR corresponding to each algorithm has been written beneath the resulting image. 
	}
	\label{basel}
\end{figure*}

For testing the algorithm for the super-resolution of images of dataset3, 1 subject from each category has been selected and the algorithms have been applied for the super-resolution of $50\%$ artificially missed B-scans. Patch sizes were set equal to 5 and 7 with window size [2,2] and window size for MDT was set to [4,4] (recall that the window size for MDT has been selected for achieving the best output). The results have been presented in Fig.~\ref{basel}. Similar to the dataset2, in this dataset, no reference image is available and only visual comparison and CNR have been reported.  CNR for each algorithm has been written beneath the resulting image. Visual comparison in addition to the resulting CNR's, show the effectiveness of the proposed algorithm.

\begin{figure*}[h!]
	\centering
	\centerline{
		\small
		\begin{tabular}{lccc}
			\quad \quad \quad  \quad Reference image&Using patches without overlap& Using patches with overlap\\
			\includegraphics[scale=0.2, trim={9.5cm 7.5cm 10.2cm 1.7cm}, clip]{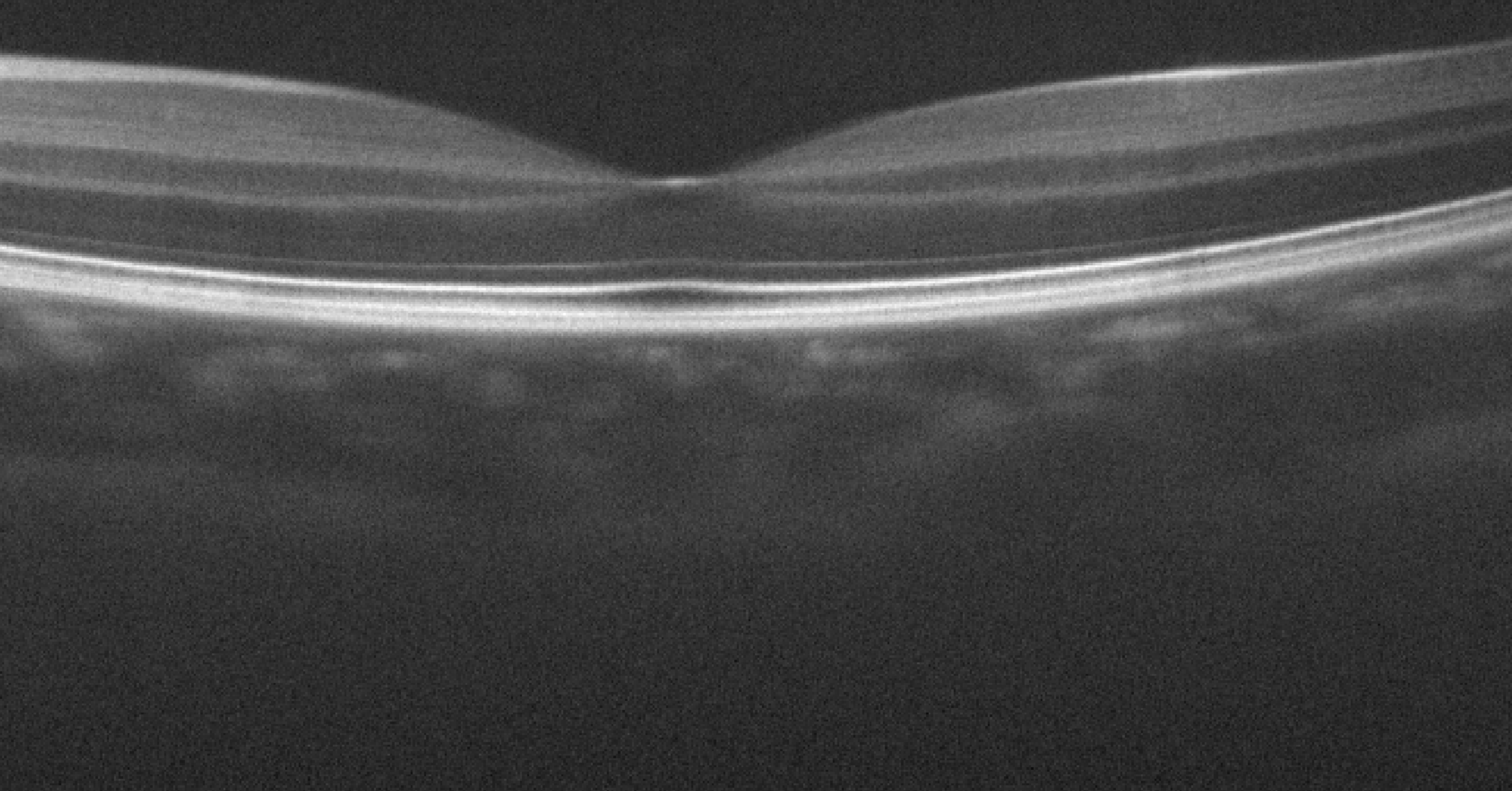}&
			\includegraphics[scale=0.2,trim={9.5cm 7.5cm 10.2cm 1.7cm}, clip]{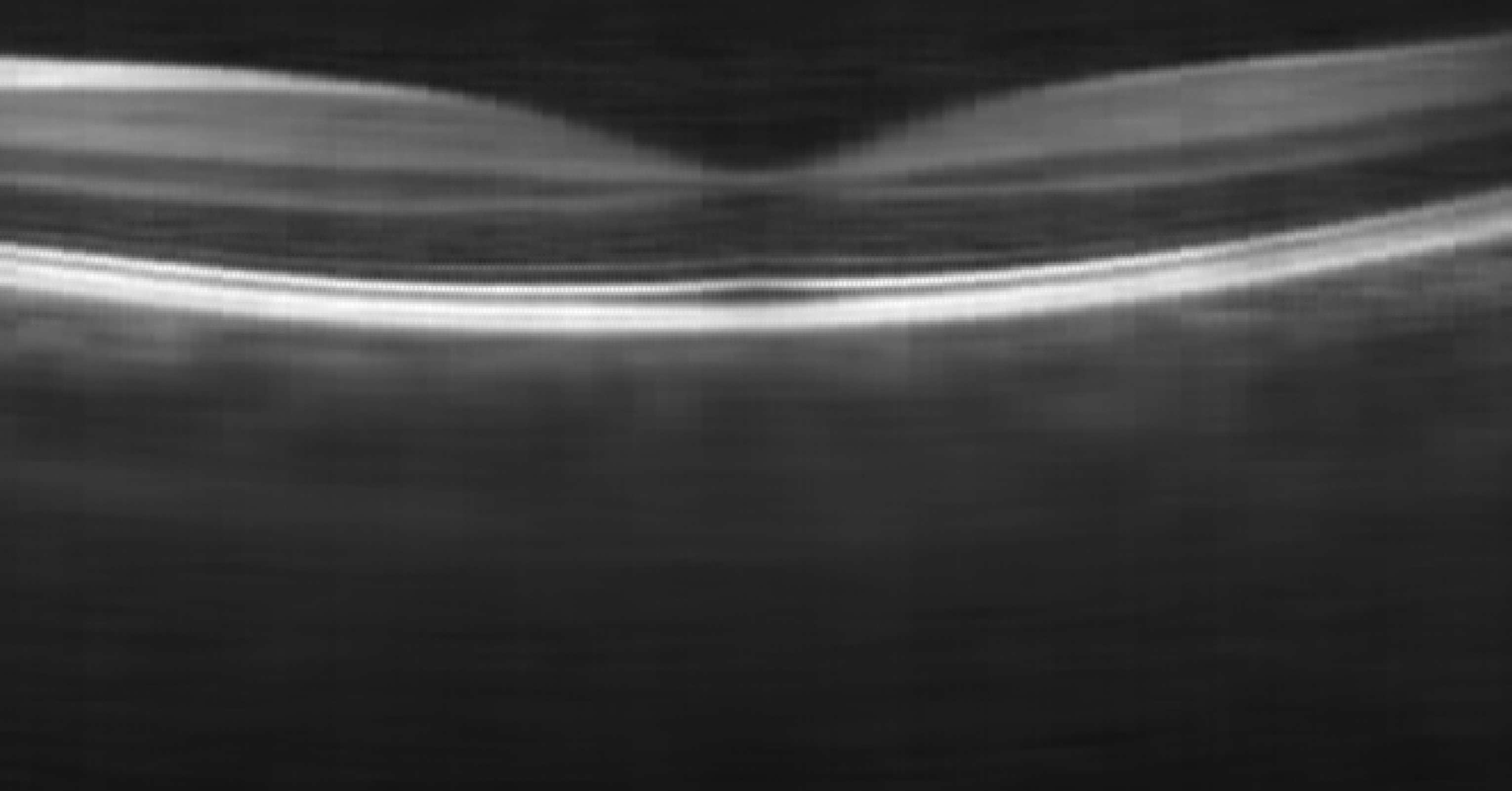}&
			\includegraphics[scale=0.2,trim={9.5cm 7.5cm 10.2cm 1.7cm}, clip]{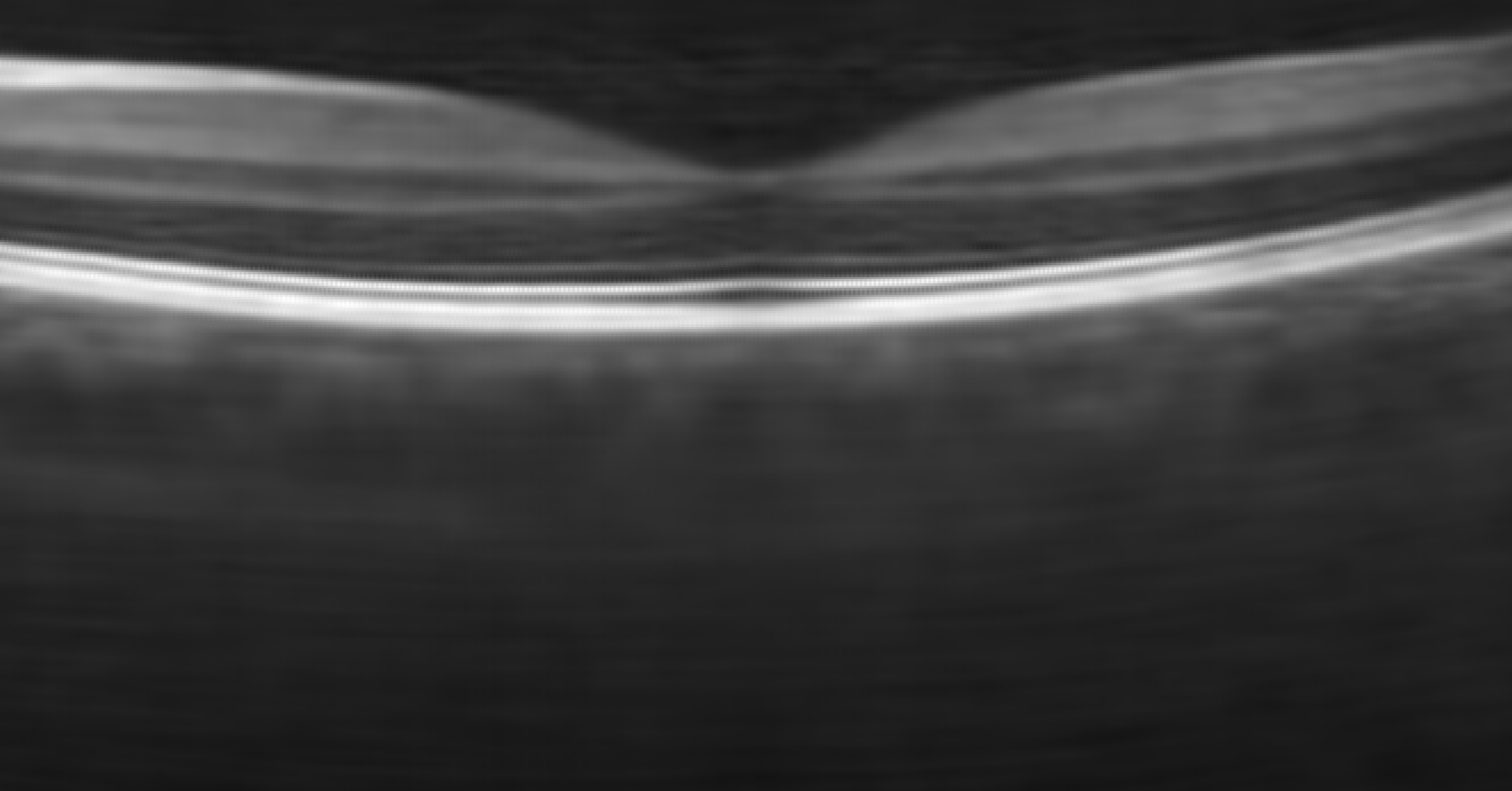}\\
			\quad \quad \quad \quad \quad&$[29.3357,0.7022]$&$\mathbf{[29.3439,0.7041]}$\\
		\end{tabular}
	}
\caption{Investigation of the effect of overlap of the patches on the performance of the super-resolution of an OCT image with $50\%$ missing ratio.
}
	\label{noover}
\end{figure*}

The effect of the overlap of the patches on the final output of the algorithm has been investigated in the next simulation. For this purpose, the outputs of the proposed approach for the situations $O\neq 0$ and $O=0$ have been compared. When $O=0$, the patches have no overlap. 
The resulting higher resolution OCT images for each situation have been illustrated in Fig.~\ref{noover}. Visual comparison in addition to the PSNR and SSIM for each image, show that using patches with overlap, we can increase the performance of the algorithm and consequently improve the quality of the output image. 

\begin{figure*}[t!]
	\centering
	\centerline{
		\small
		\begin{tabular}{lccccc}
		\quad \quad \quad \quad \quad TR Rank=[3,3,3,3,3,3]&TR Rank=[4,4,4,4,4,4]& TR Rank=[5,5,5,5,5,5]\\
		    \includegraphics[scale=0.2, trim={9.5cm 7.5cm 10.2cm 1.7cm}, clip]{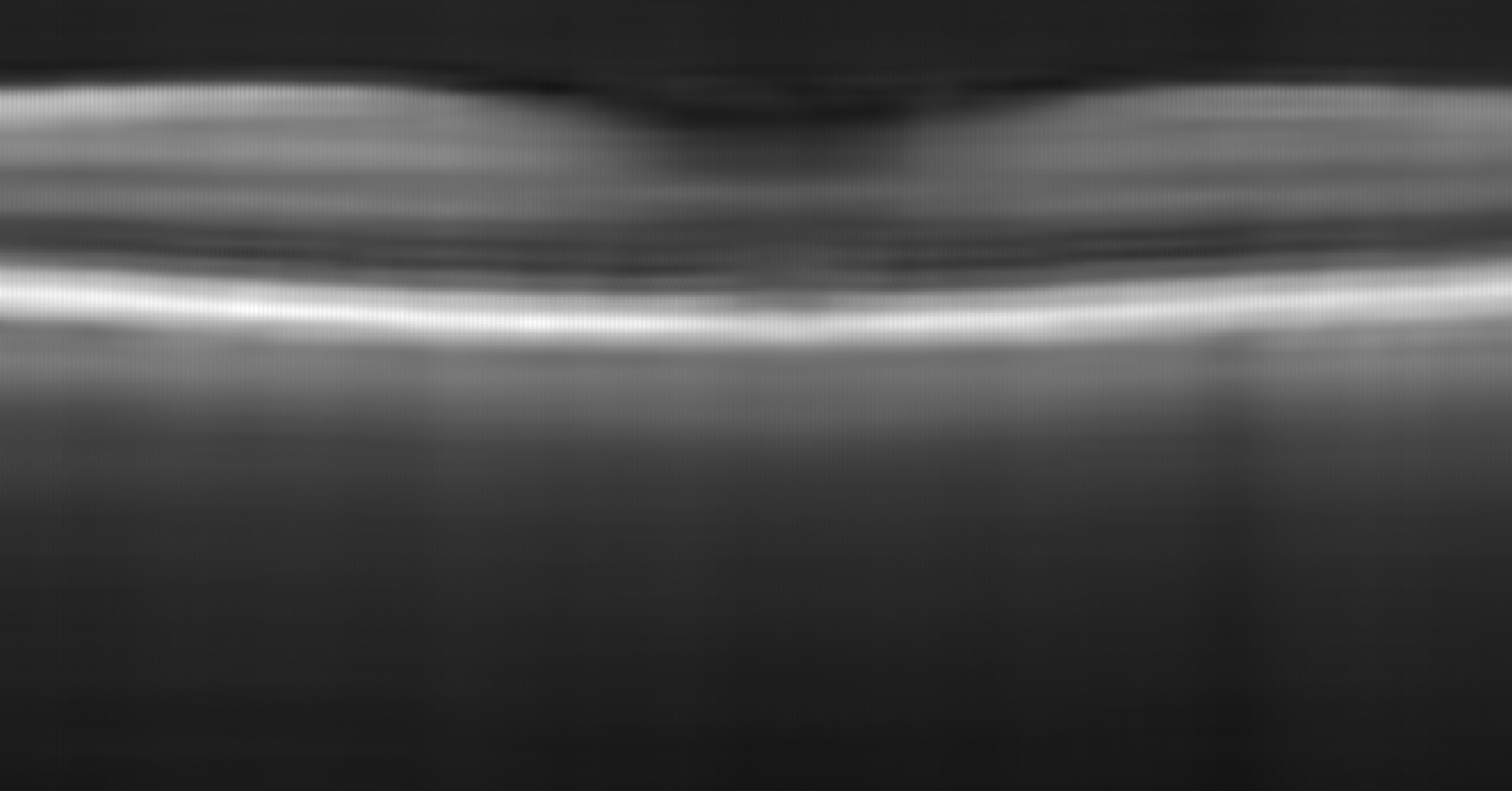}&
			\includegraphics[scale=0.2, trim={9.5cm 7.5cm 10.2cm 1.7cm}, clip]{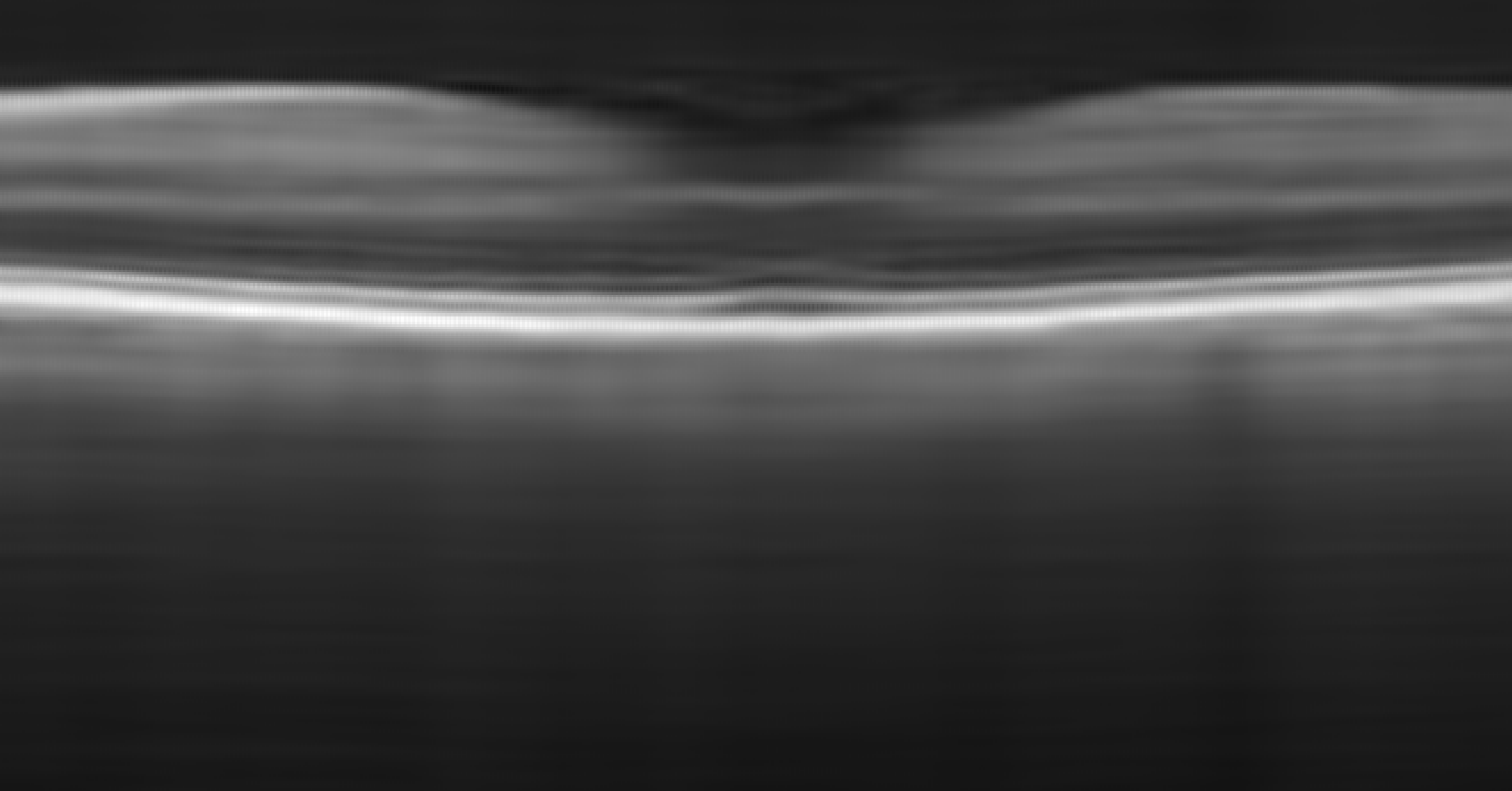}&
			\includegraphics[scale=0.2, trim={9.5cm 7.5cm 10.2cm 1.7cm}, clip]{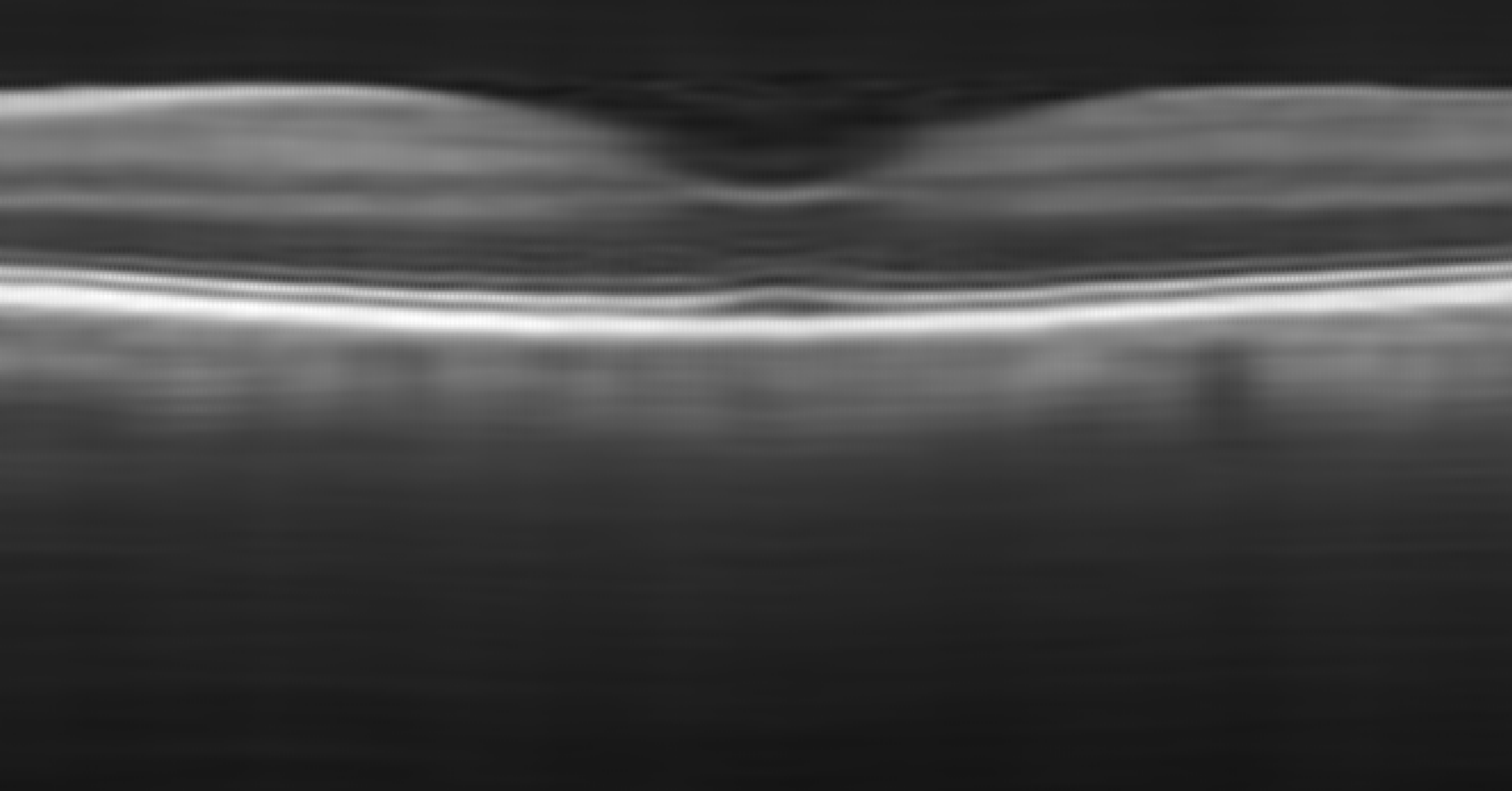}\\
		\quad \quad \quad \quad	\quad TR Rank=[6,6,6,6,6,6]& TR Rank=[7,7,7,7,7,7]& TR Rank=[8,8,8,8,8,8]\\
			\includegraphics[scale=0.2, trim={9.5cm 7.5cm 10.2cm 1.7cm}, clip]{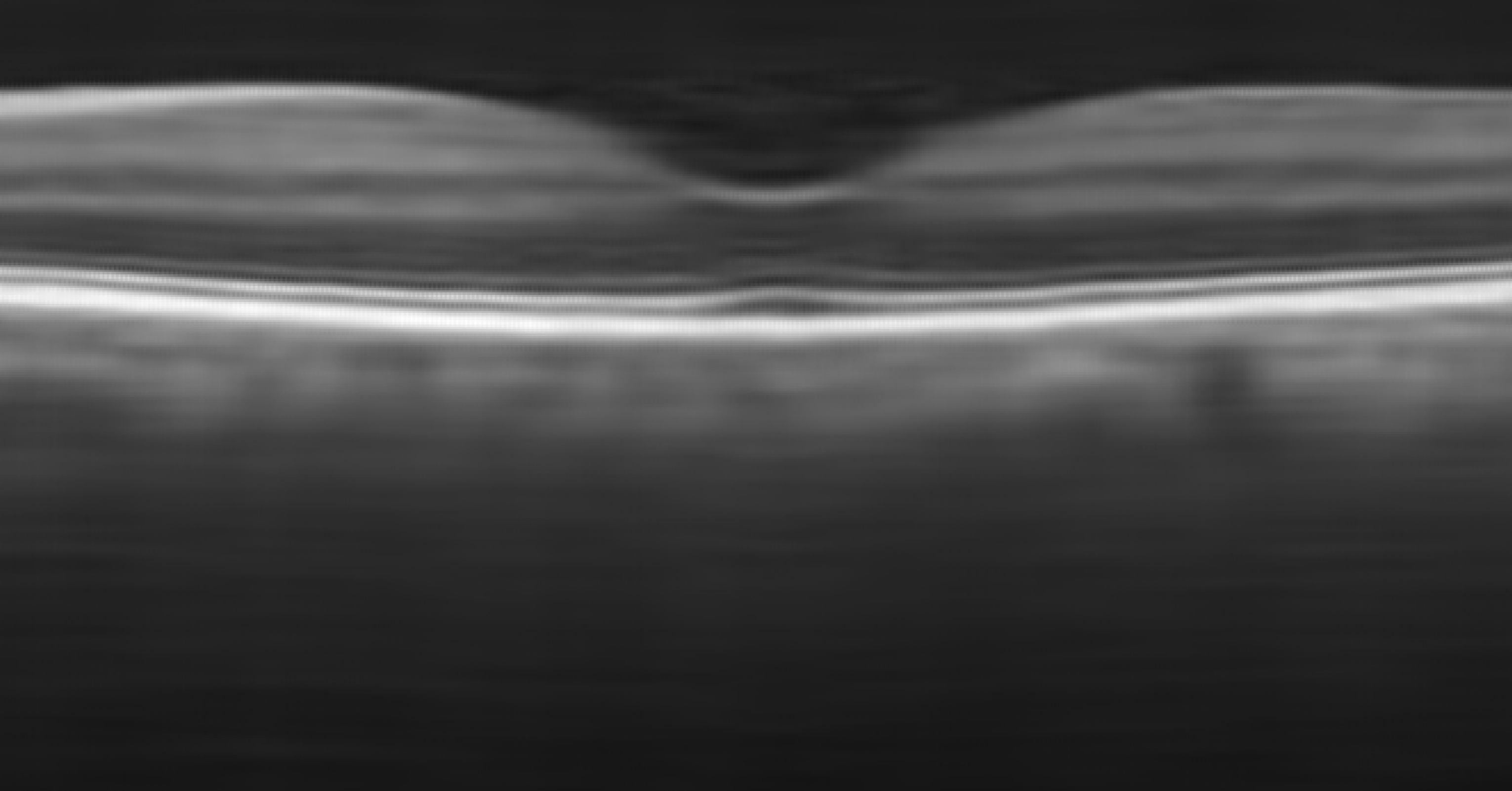}&
			\includegraphics[scale=0.2, trim={9.5cm 7.5cm 10.2cm 1.7cm}, clip]{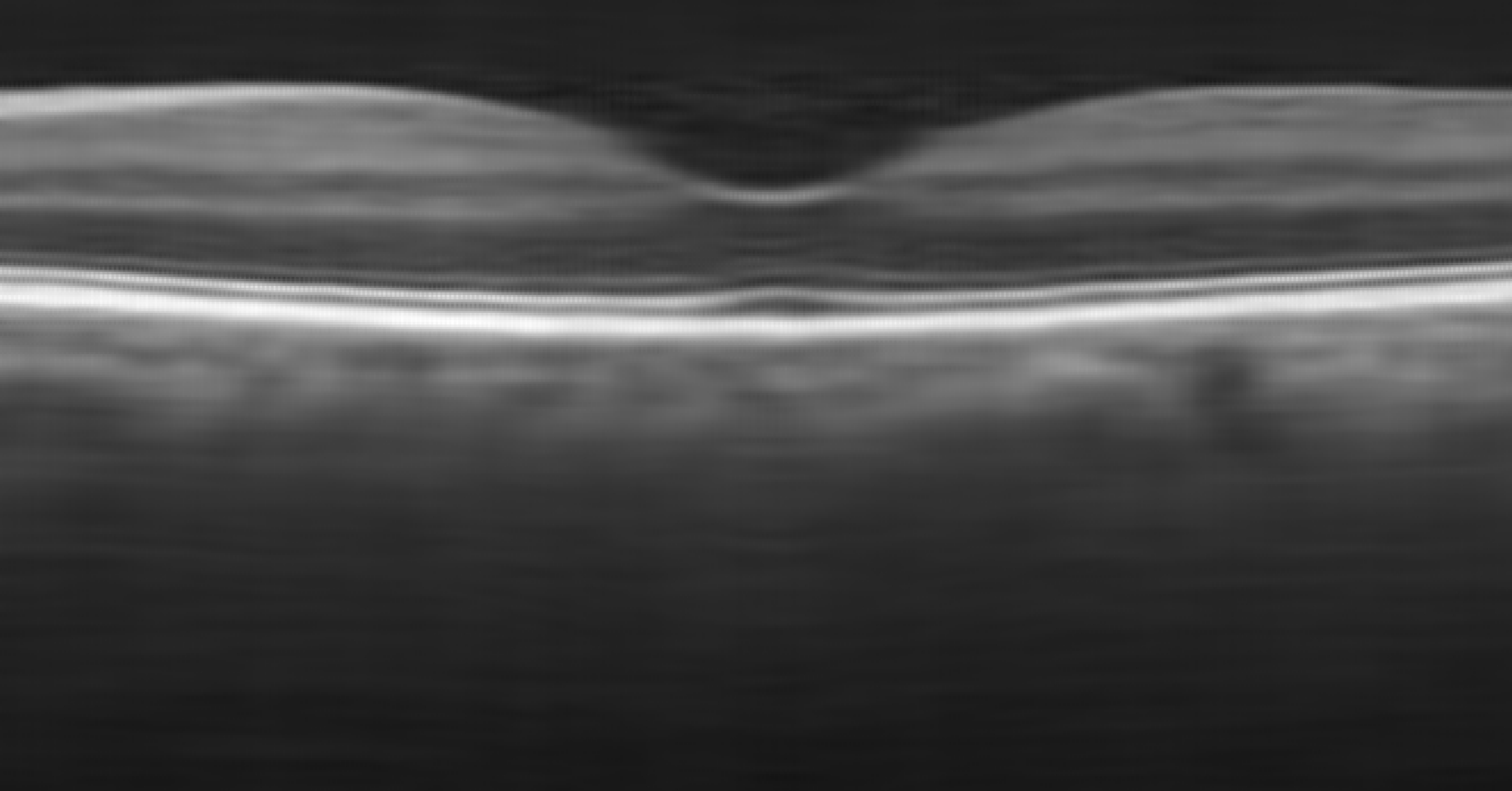}&
			\includegraphics[scale=0.2, trim={9.5cm 7.5cm 10.2cm 1.7cm}, clip]{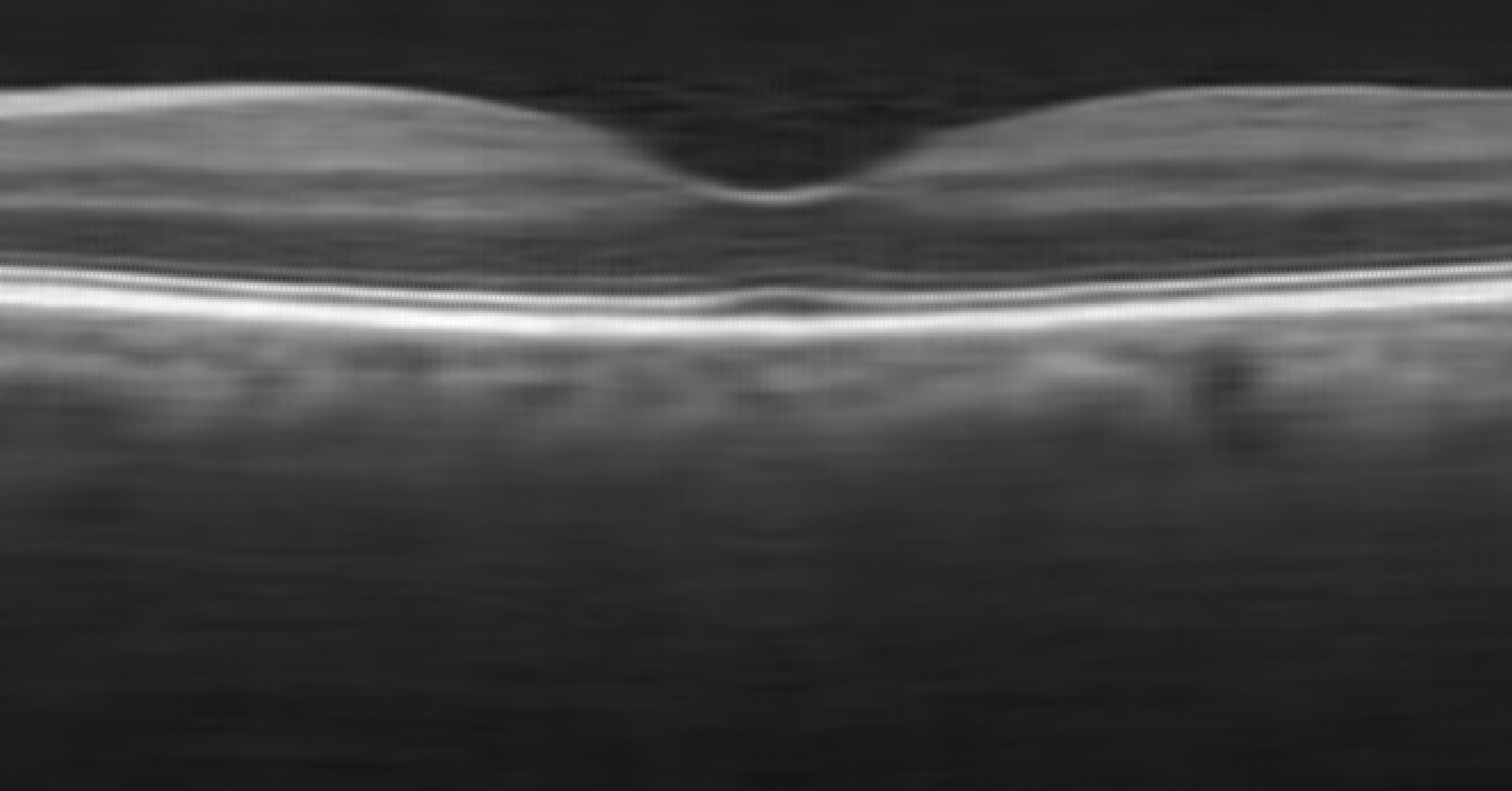}\\
			\end{tabular}
	}
\caption{Illustration of  gradually increasing the quality of the output image in the proposed approach by increasing the TR ranks during iteration process.  It should be noted that   increasing the ranks too much  does not improve performance  and even can result in a noisy output.
}
	\label{evol}
\end{figure*}

\begin{figure}[t!]
	\begin{minipage}[b]{1.0\linewidth}
		\centering
		\centerline{
			\includegraphics[width=10cm]{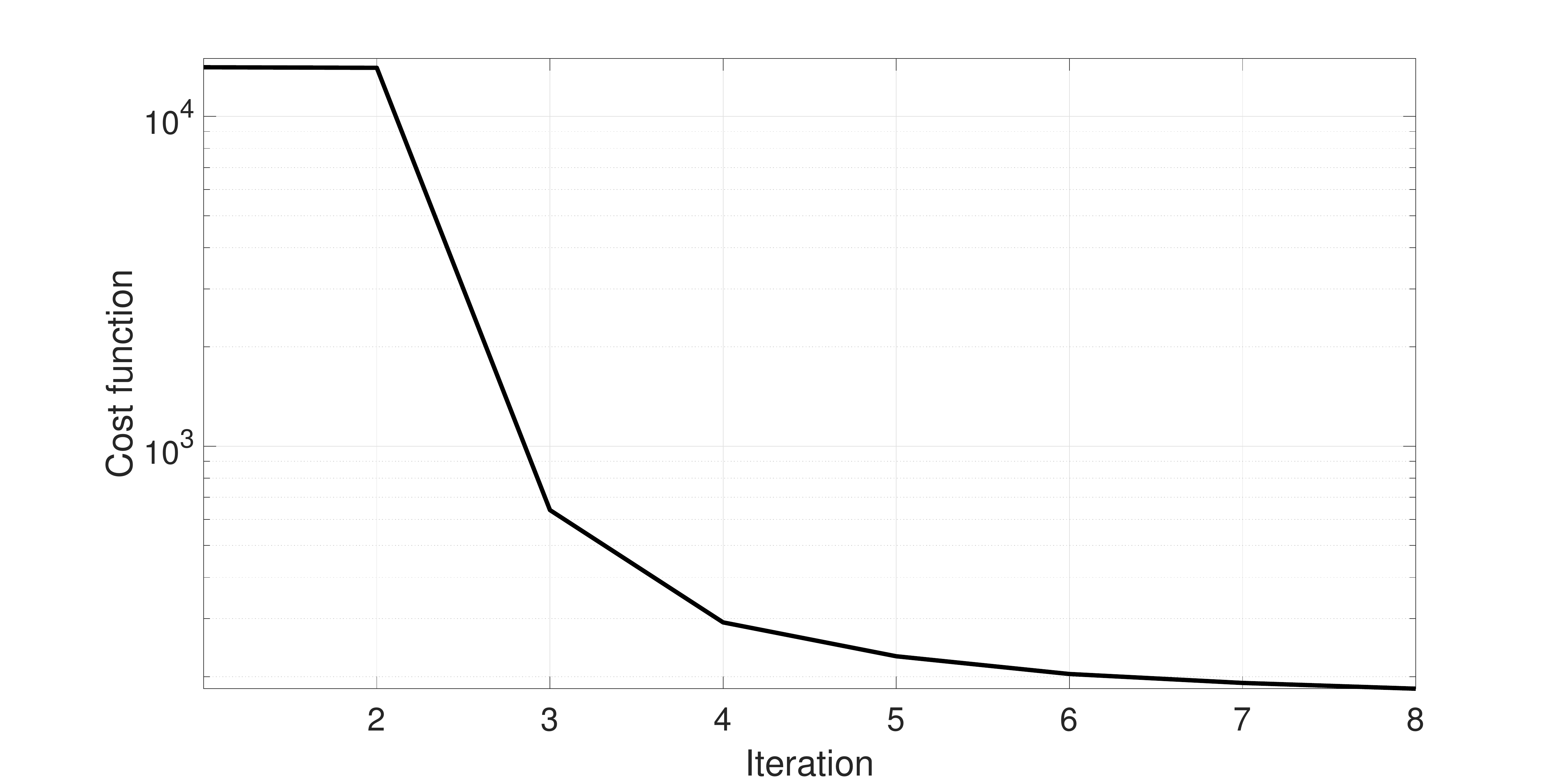}
		}
	\end{minipage}
	\caption{ Decrease in the cost function with gradually increasing ranks of tensor ring during iteration process.}
	\label{costfunction}
\end{figure}

The evolution of the resulting OCT images during rank incremental process has been demonstrated in Fig.~\ref{evol}. It has been shown that by increasing the TR ranks, the clarity and resolution of the output will be increased. However, too much increase of the rank results in increasing the noise in the final image and decreases the resolution of the output image.  The change in the average cost functions (\ref{cost3}) for $P=15$ and $P=10$ for the images of Fig.~\ref{evol} (before smoothing of the output) has been shown in Fig.~\ref{costfunction}. The result confirms the monotonic decreasing of the cost function during iterations, i.e., by gradually increasing TR ranks.


\begin{figure*}[h!]
	\centering
	\centerline{
		\begin{tabular}{lcccccc}
			\quad  Original image& \quad HALRTC \cite{liu2012tensor}& \quad  TRLRF \cite{yuan2019tensor} & \quad TT-WOPT \cite{yuan2019high} & \quad MDT \cite{yokota2018missing} & Proposed algorithm\\ 
			\includegraphics[scale=0.1, trim={10cm 5cm 12cm 1.7cm}, clip]{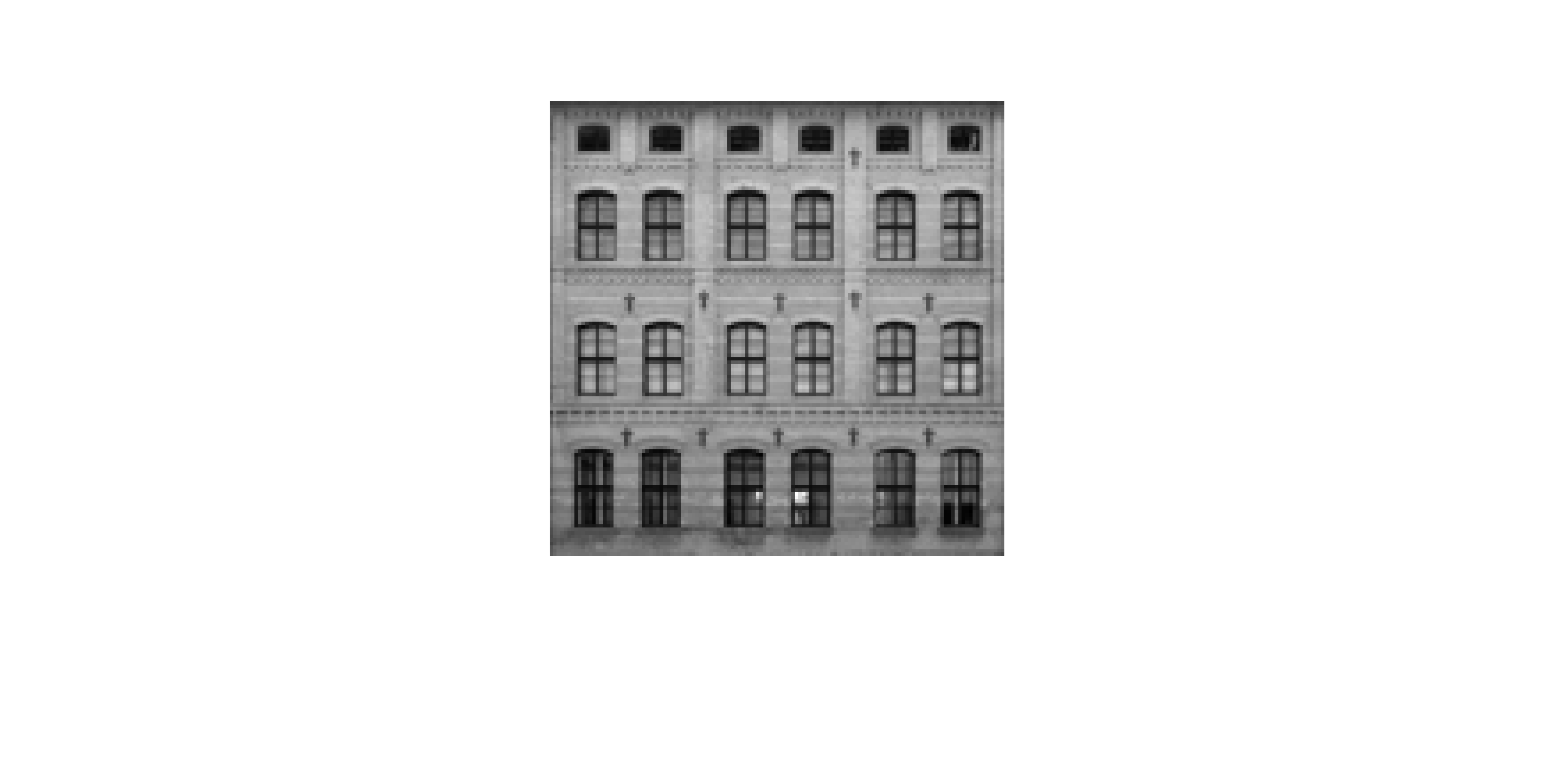}&\includegraphics[scale=0.12, trim={13.5cm 4.5cm 16cm 1.7cm}, clip]{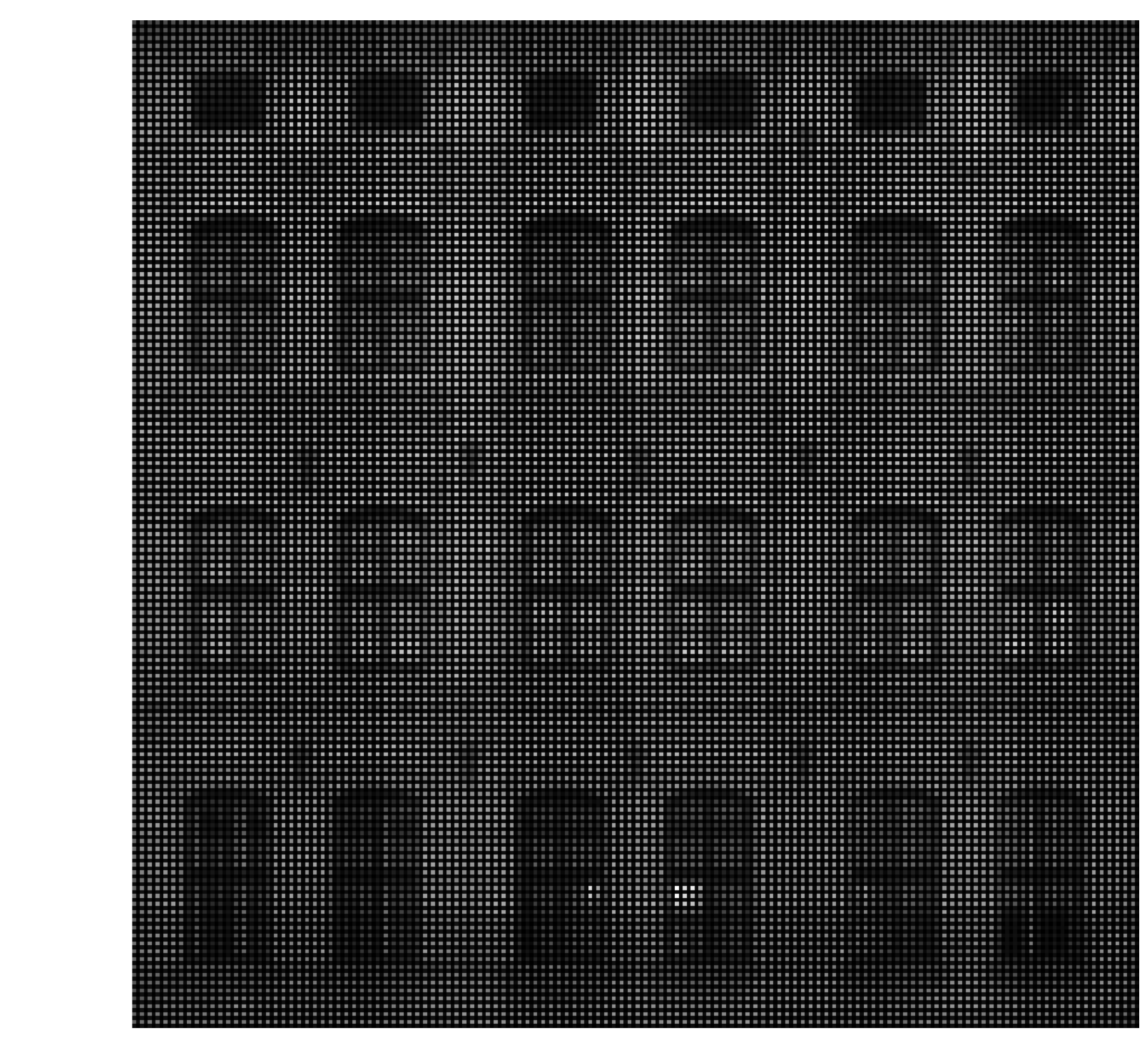}&\includegraphics[scale=0.12, trim={13.5cm 4.5cm 16cm 1.7cm}, clip]{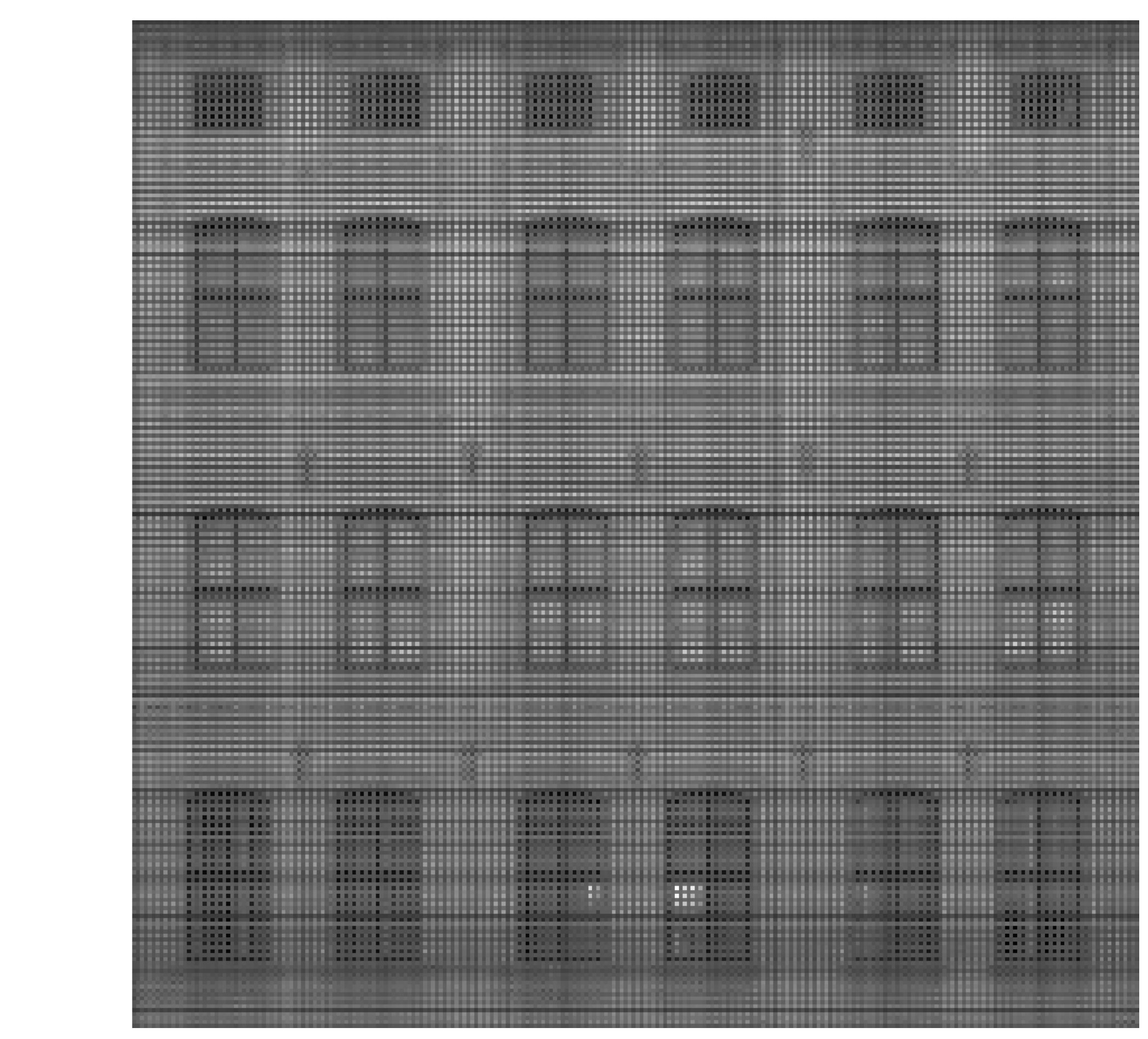}&
			\includegraphics[scale=0.12, trim={13.5cm 4.5cm 16cm 1.7cm}, clip]{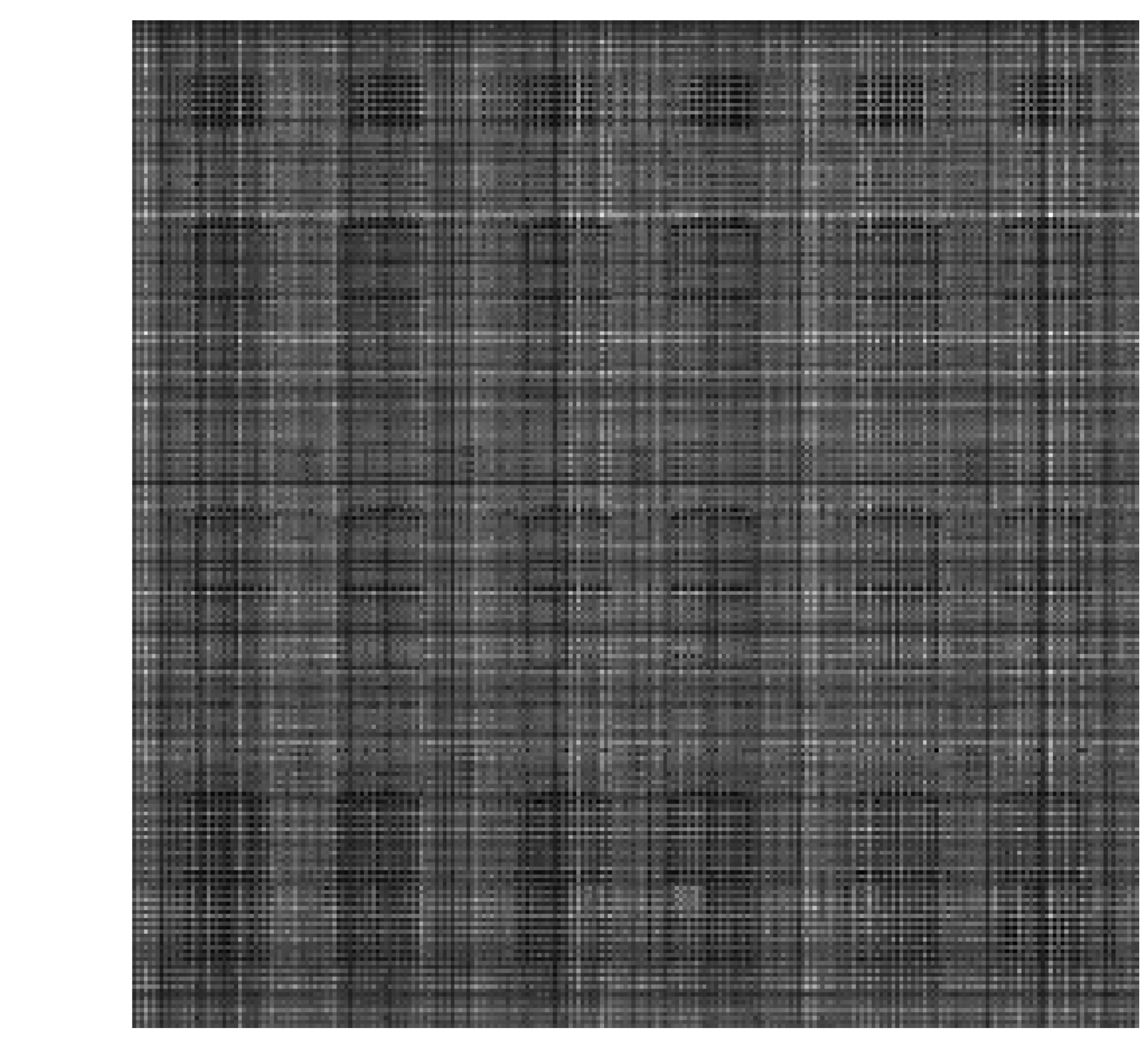}&\includegraphics[scale=0.12, trim={13.5cm 4.5cm 16cm 1.7cm}, clip]{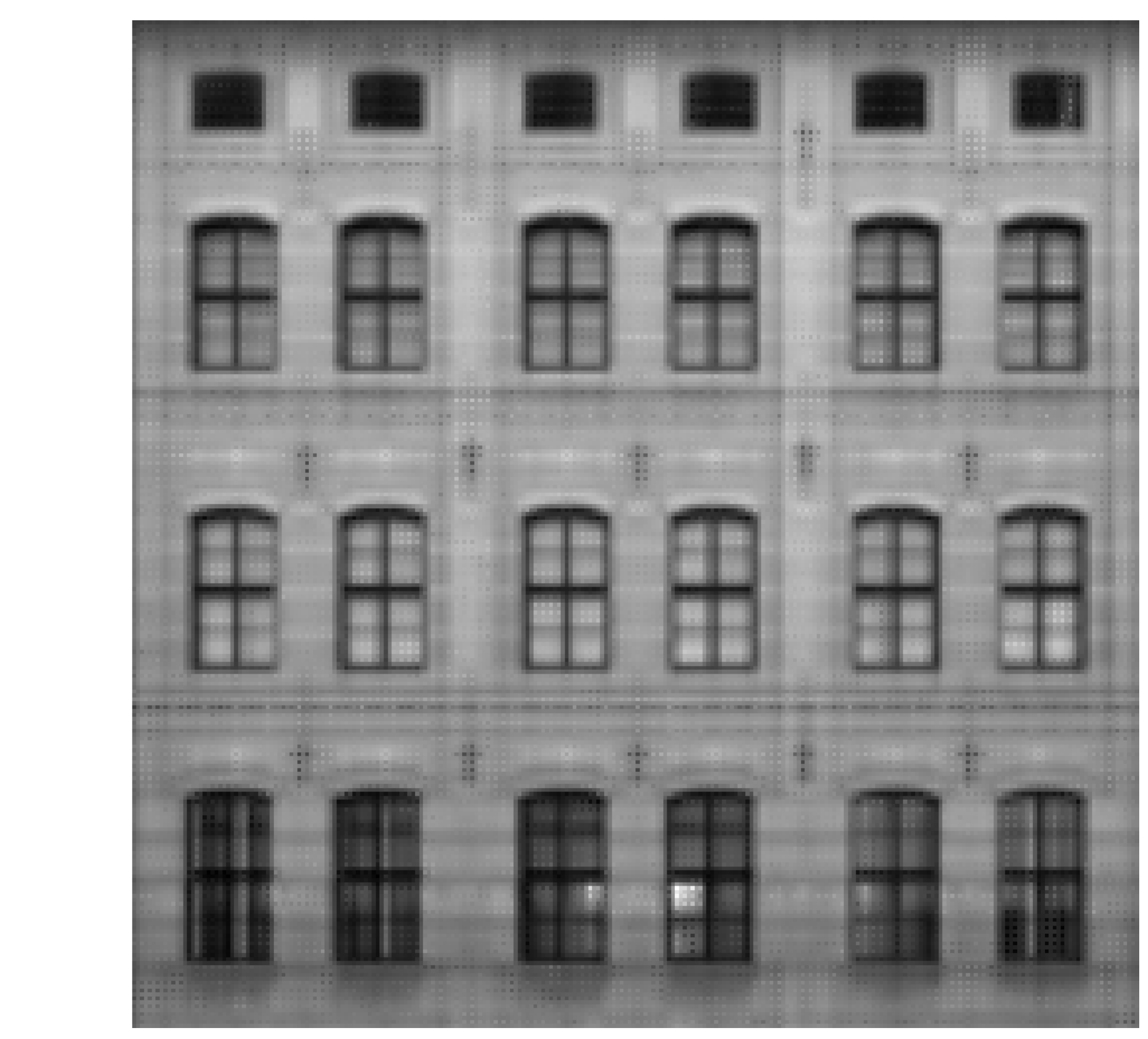}&\includegraphics[scale=0.12, trim={13.5cm 4.5cm 16cm 1.7cm}, clip]{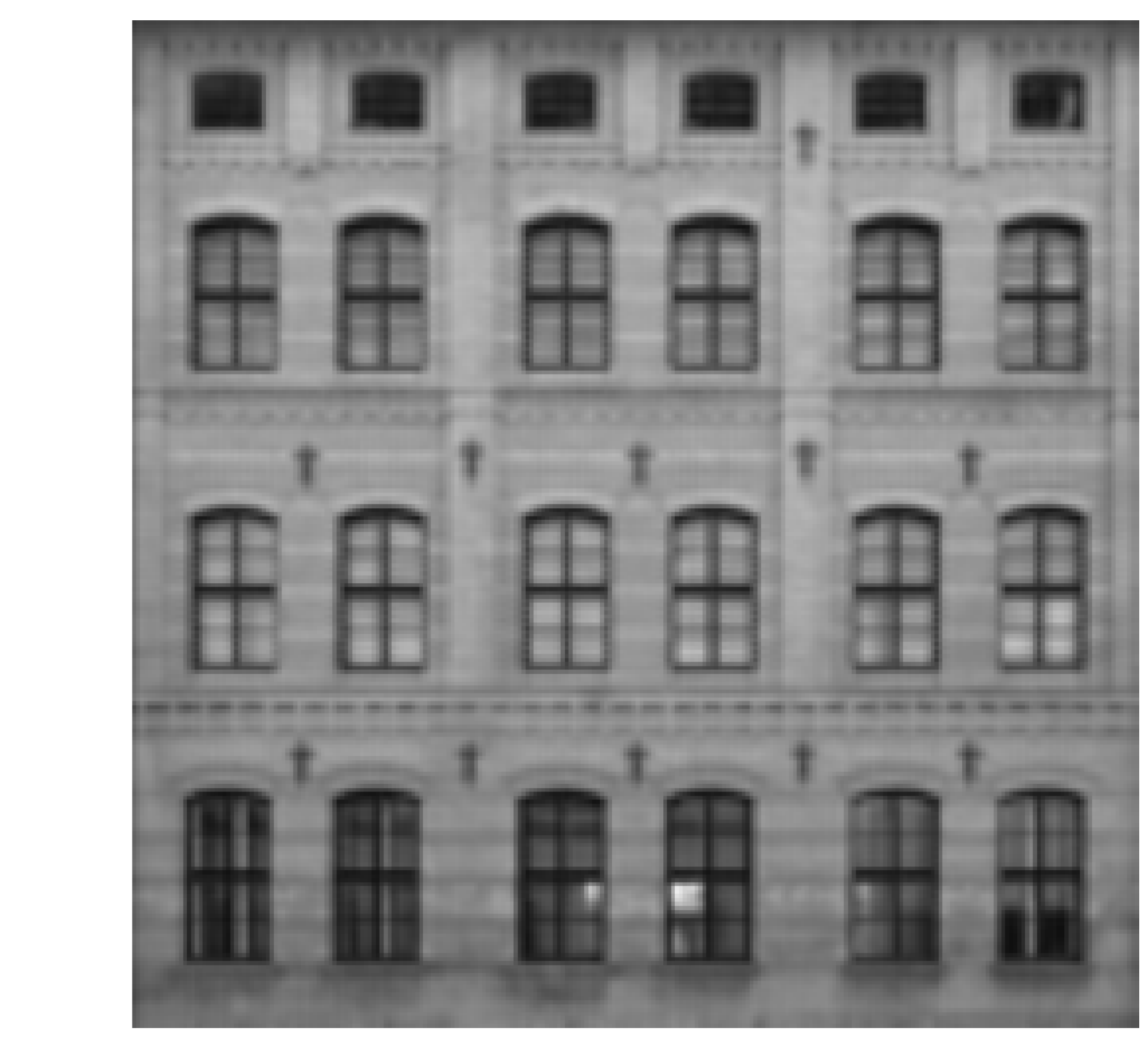}\\
			&[7.6883,0.0503]&[15.9311,0.2476]&[13.7860,0.1499]&[23.4945,0.6572]&[\textbf{24.0870},\textbf{0.7056}]\\
			\includegraphics[scale=0.1, trim={10cm 5cm 12cm 1.7cm}, clip]{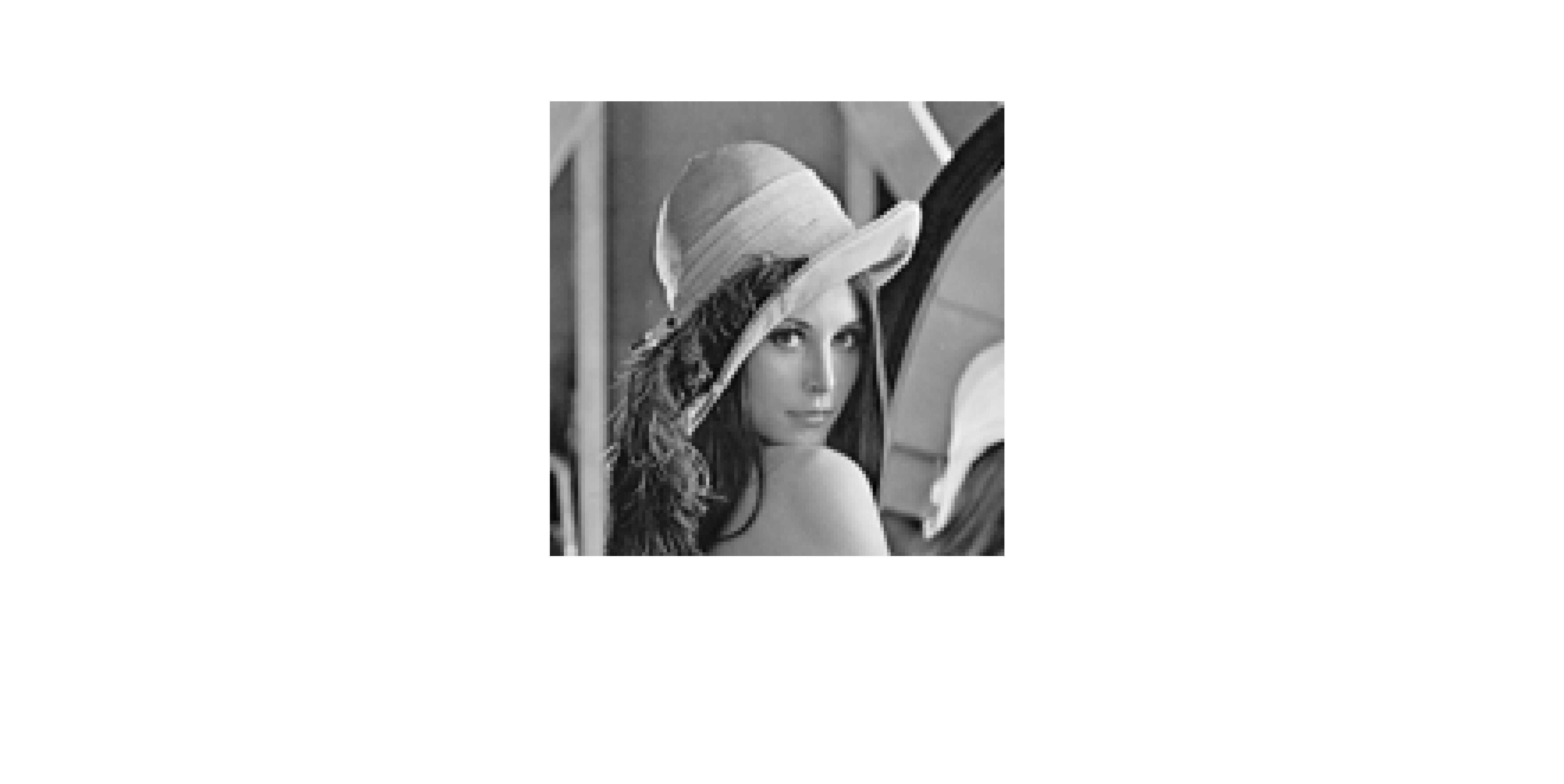}&\includegraphics[scale=0.12, trim={13.5cm 4.5cm 16cm 1.7cm}, clip]{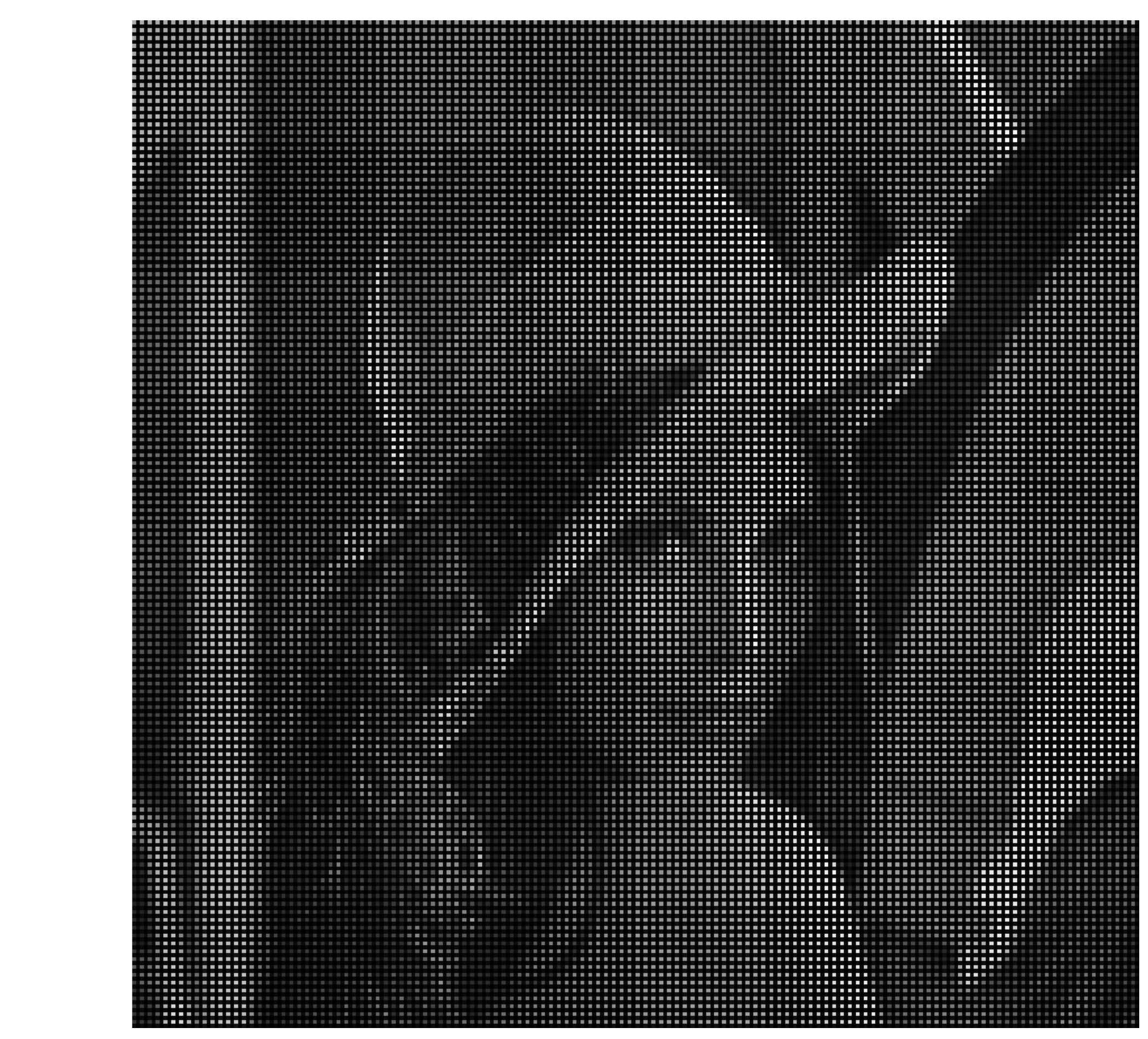}&\includegraphics[scale=0.12, trim={13.5cm 4.5cm 16cm 1.7cm}, clip]{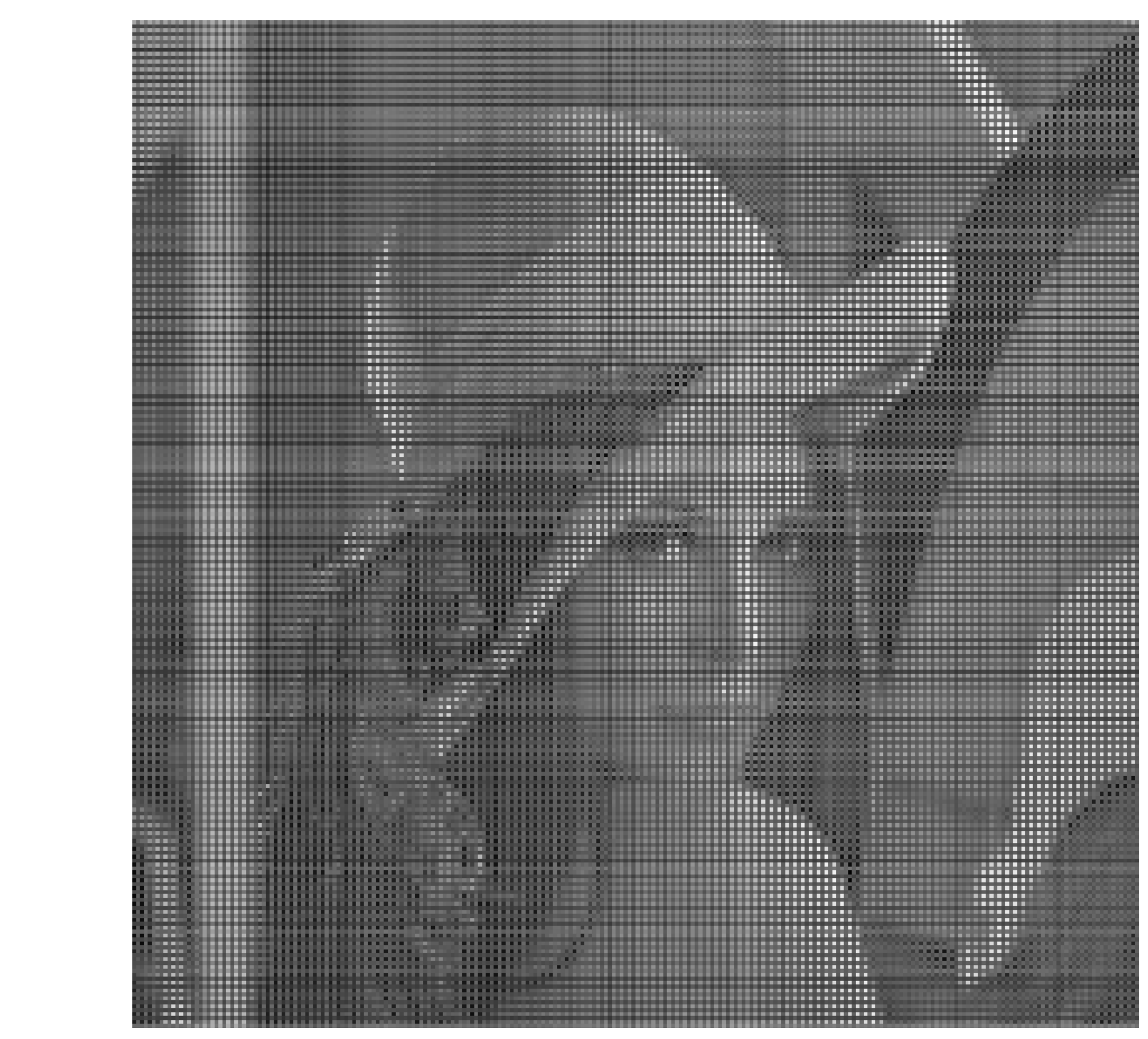}&
			\includegraphics[scale=0.12, trim={13.5cm 4.5cm 16cm 1.7cm}, clip]{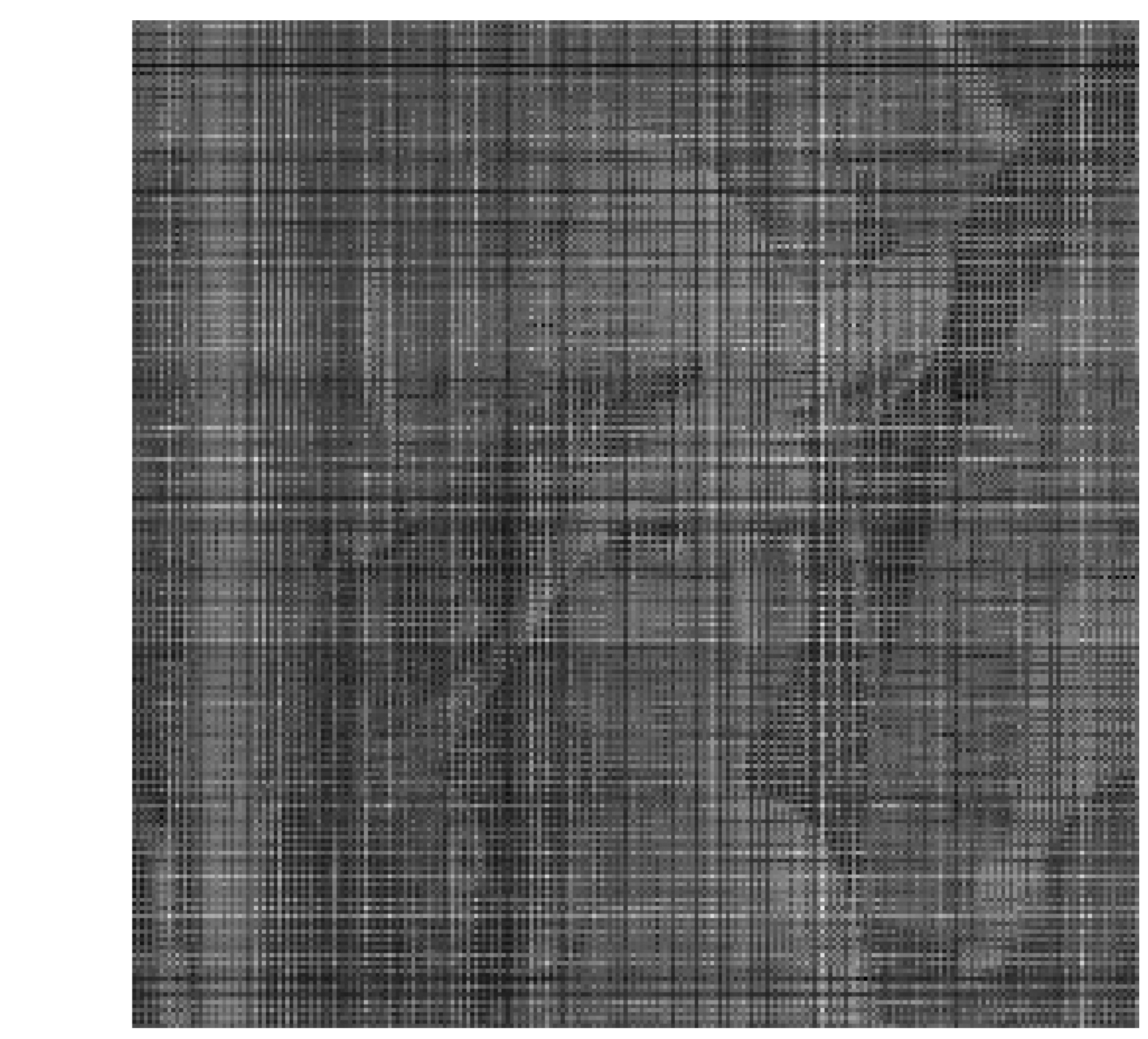}&\includegraphics[scale=0.12, trim={13.5cm 4.5cm 16cm 1.7cm}, clip]{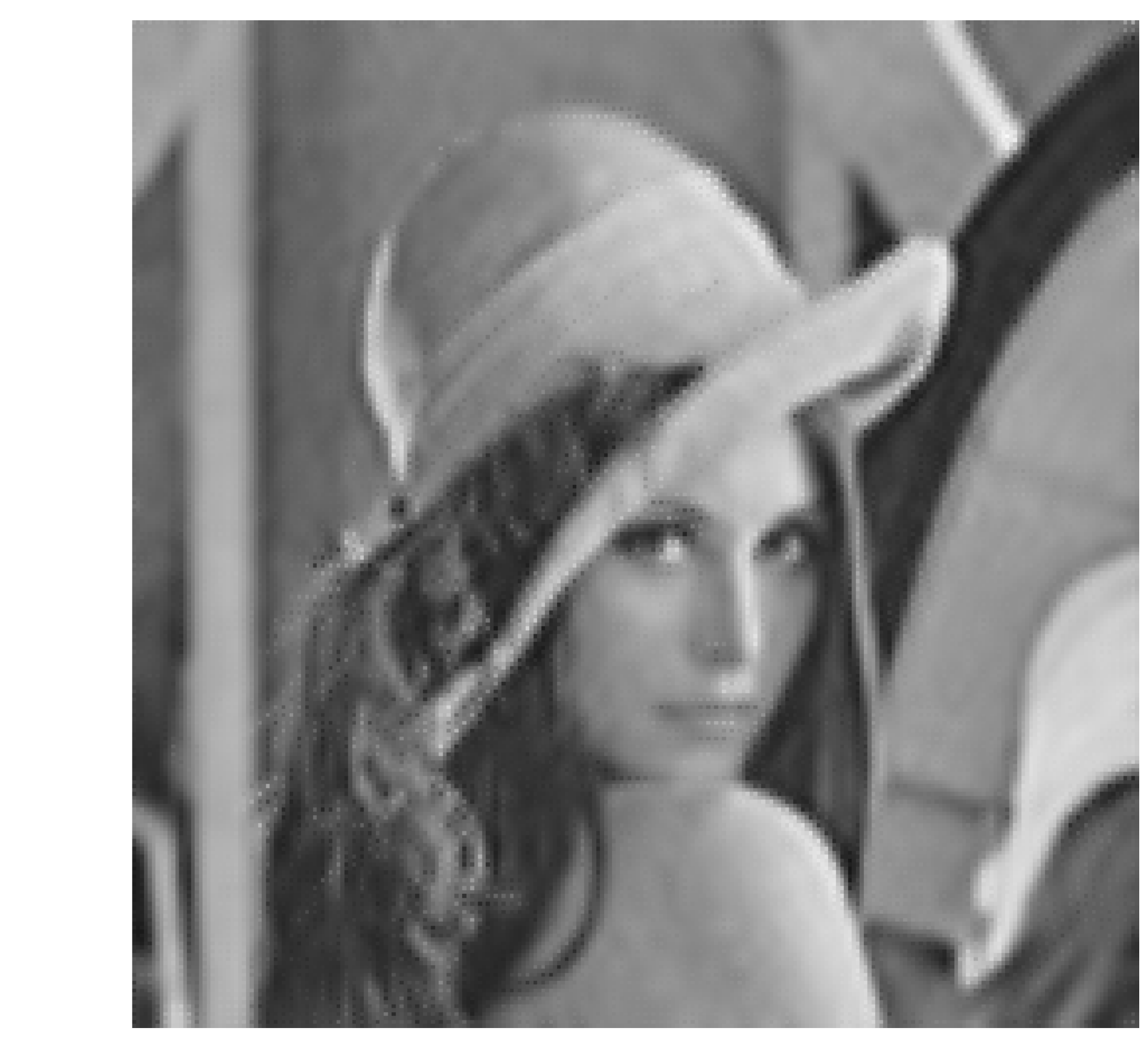}&\includegraphics[scale=0.12, trim={13.5cm 4.5cm 16cm 1.7cm}, clip]{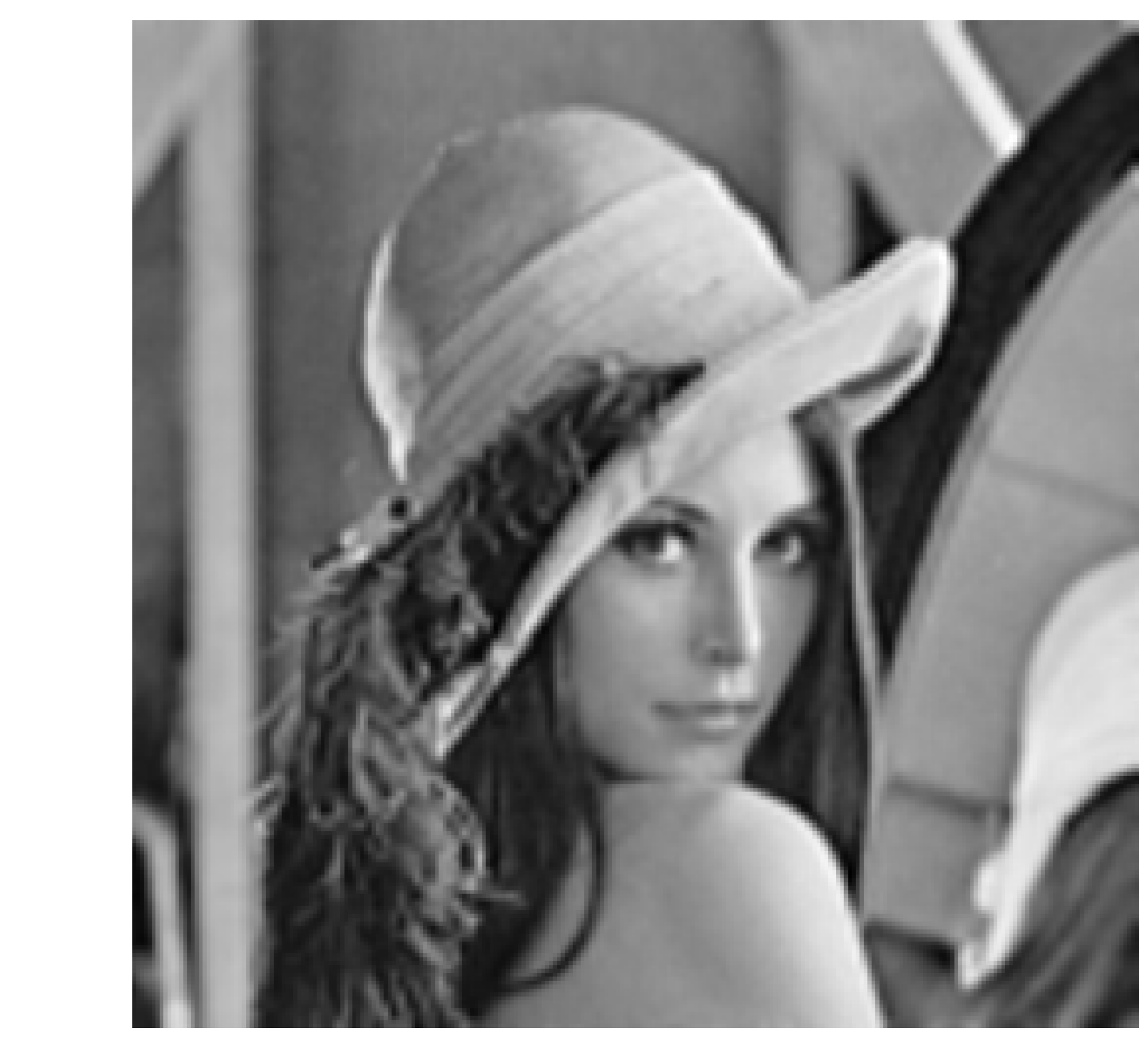}\\
			& [7.5296,0.0431]&[14.759,0.1711]&[13.2664,0.0846]&[25.6123,0.7660]&[\textbf{26.5846},\textbf{0.8238}]\\
			\includegraphics[scale=0.1, trim={10cm 5cm 12cm 1.7cm}, clip]{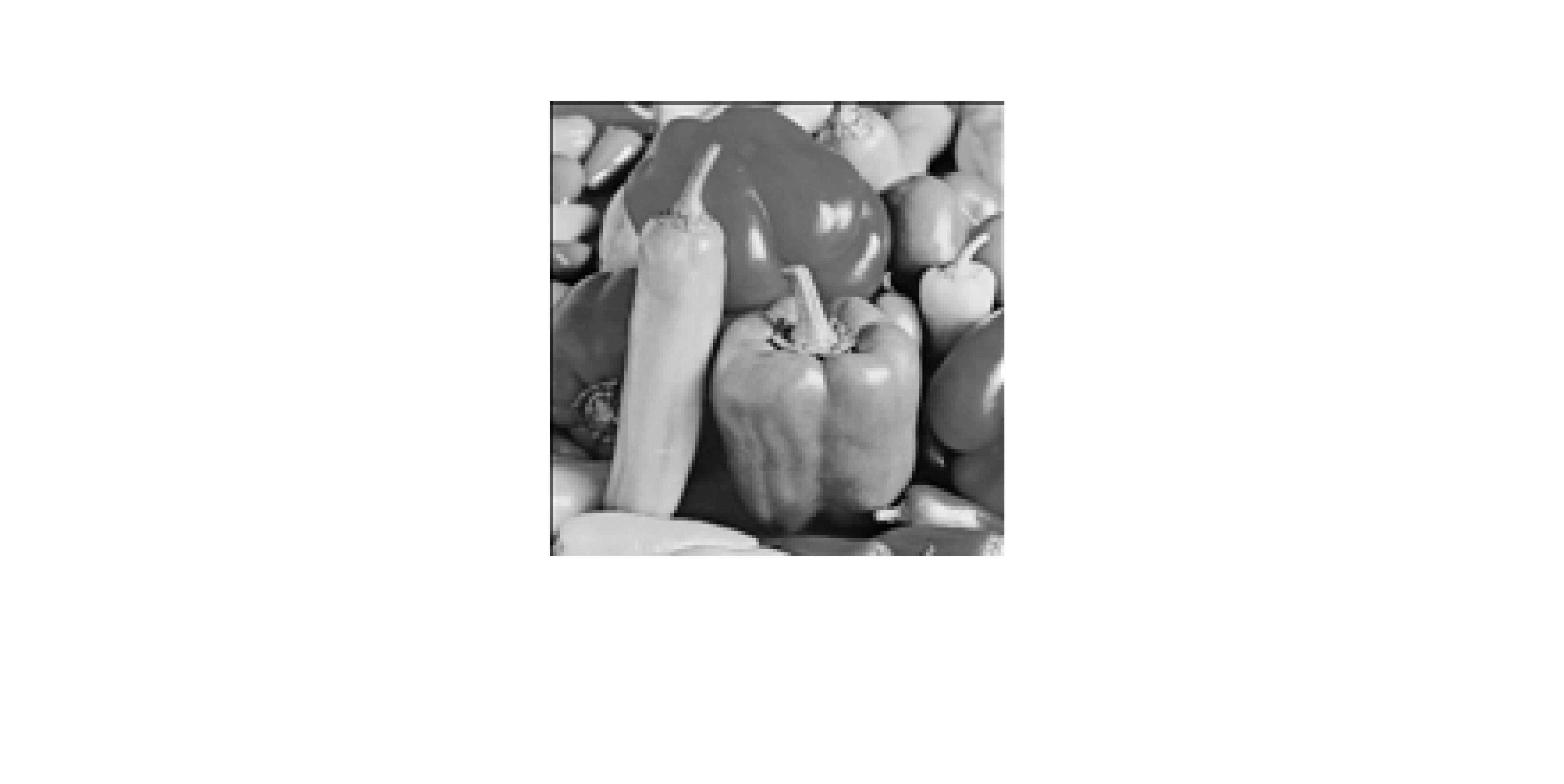}&\includegraphics[scale=0.12, trim={13.5cm 4.5cm 16cm 1.7cm}, clip]{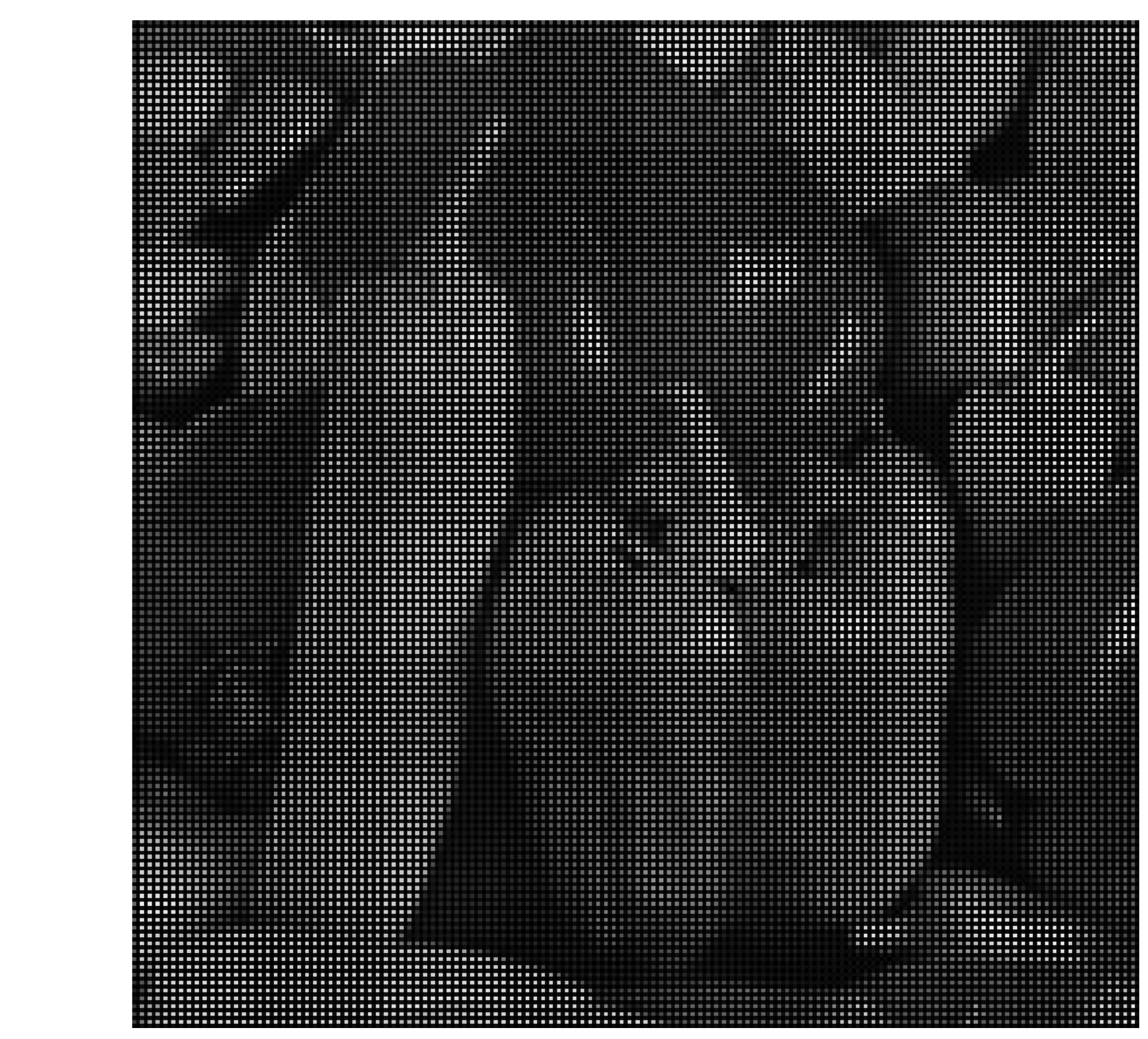}&\includegraphics[scale=0.12, trim={13.5cm 4.5cm 16cm 1.7cm}, clip]{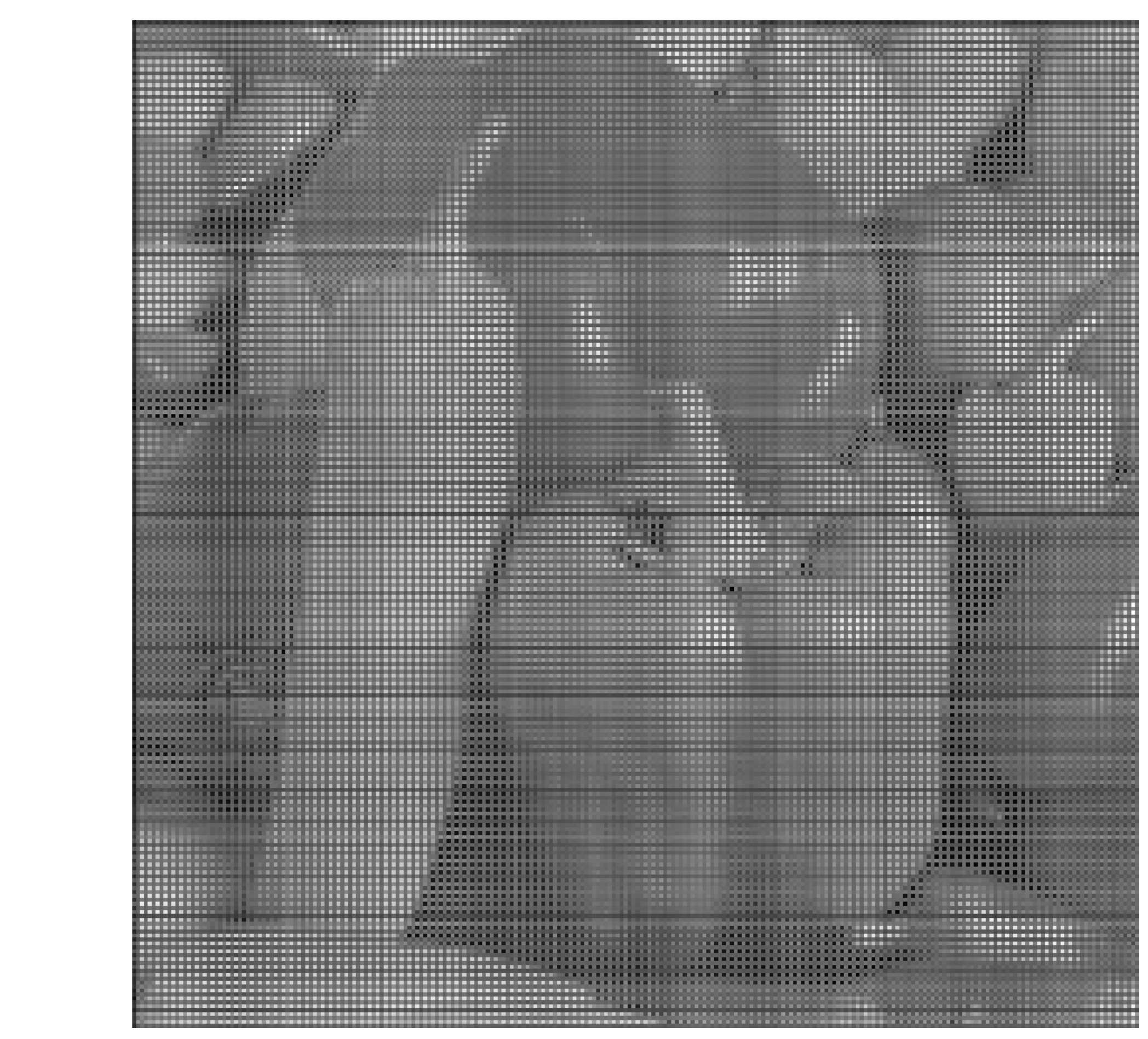}&
			\includegraphics[scale=0.12, trim={13.5cm 4.5cm 16cm 1.7cm}, clip]{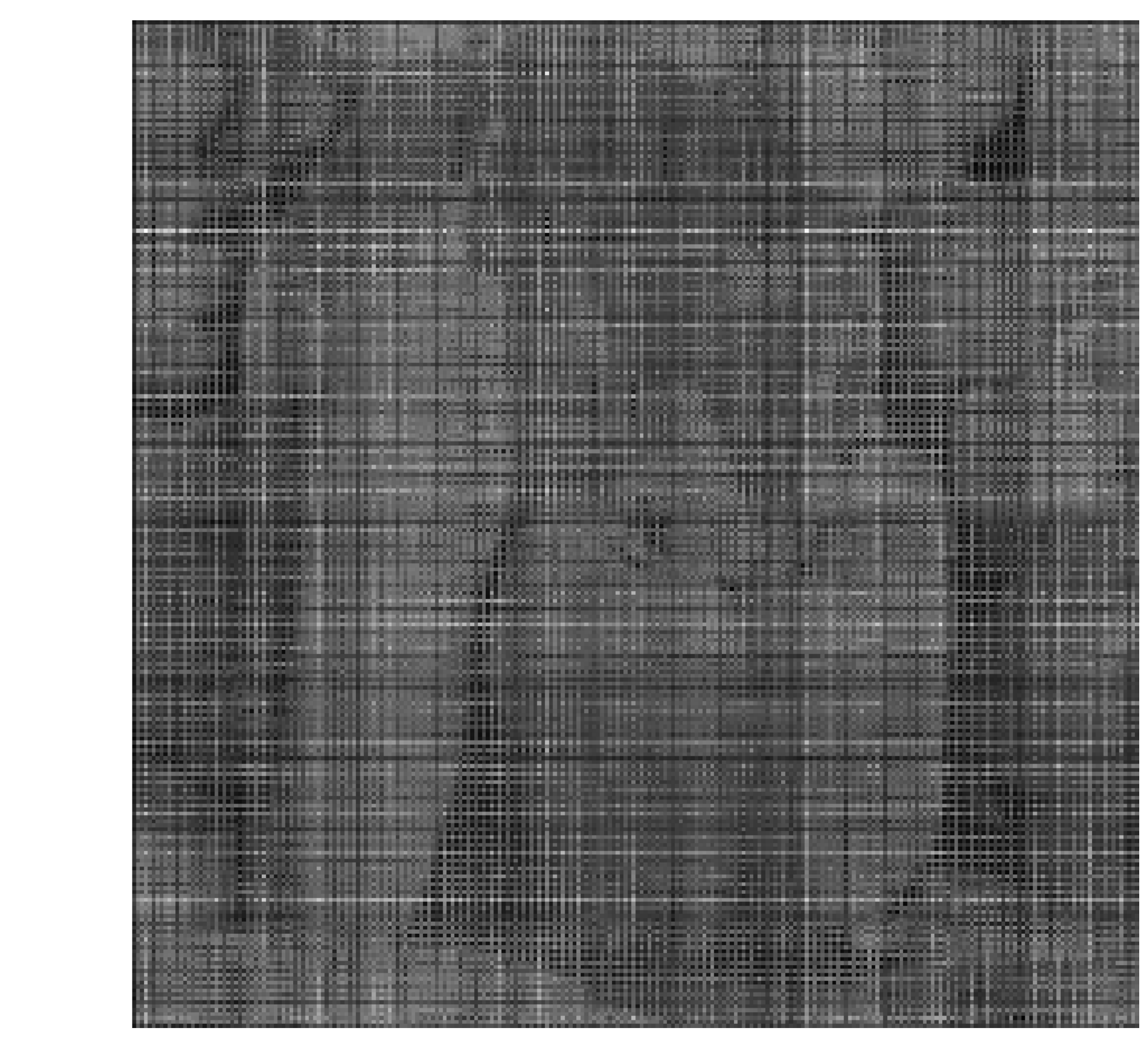}&\includegraphics[scale=0.12, trim={13.5cm 4.5cm 16cm 1.7cm}, clip]{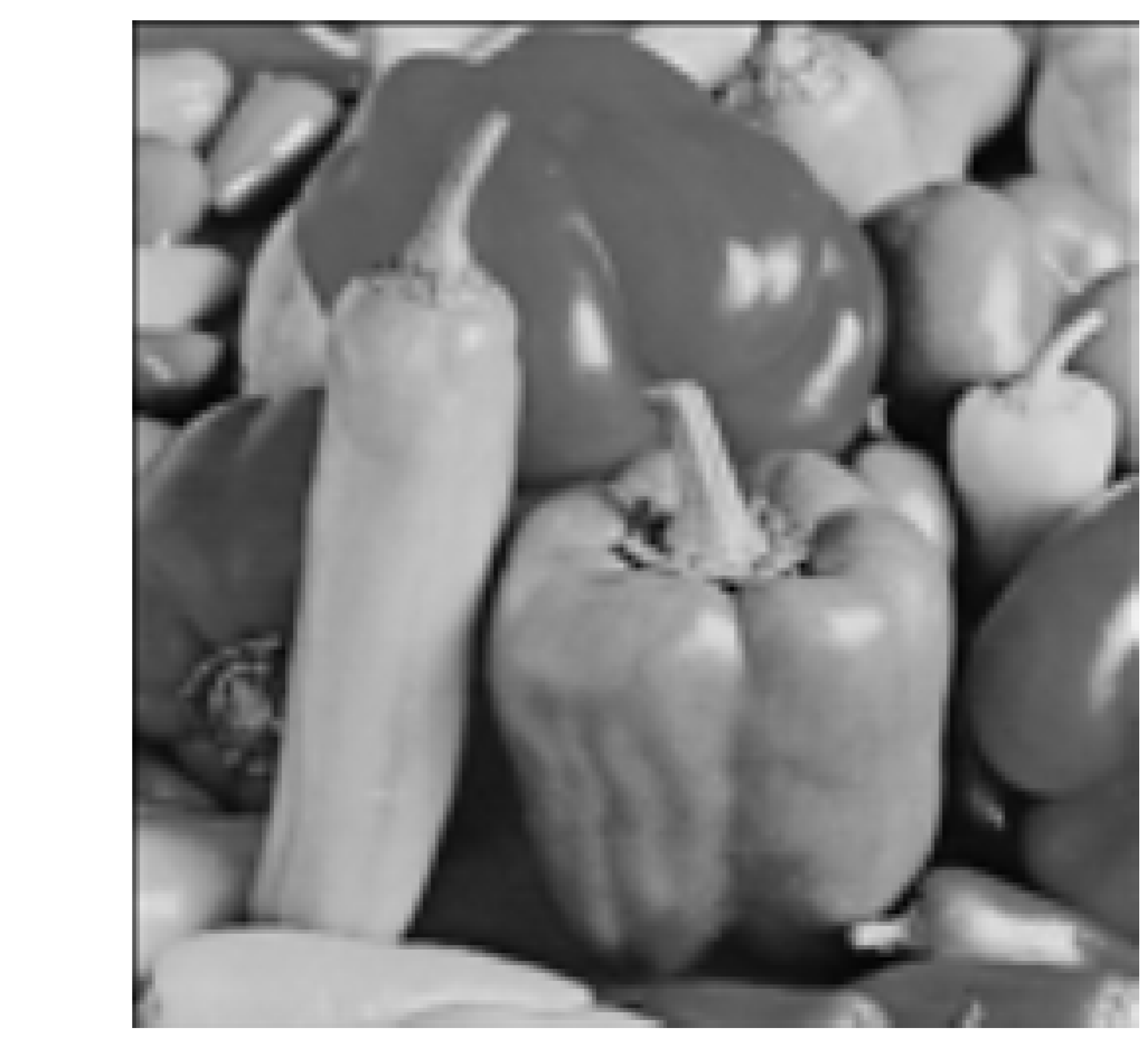}&\includegraphics[scale=0.12, trim={13.5cm 4.5cm 16cm 1.7cm}, clip]{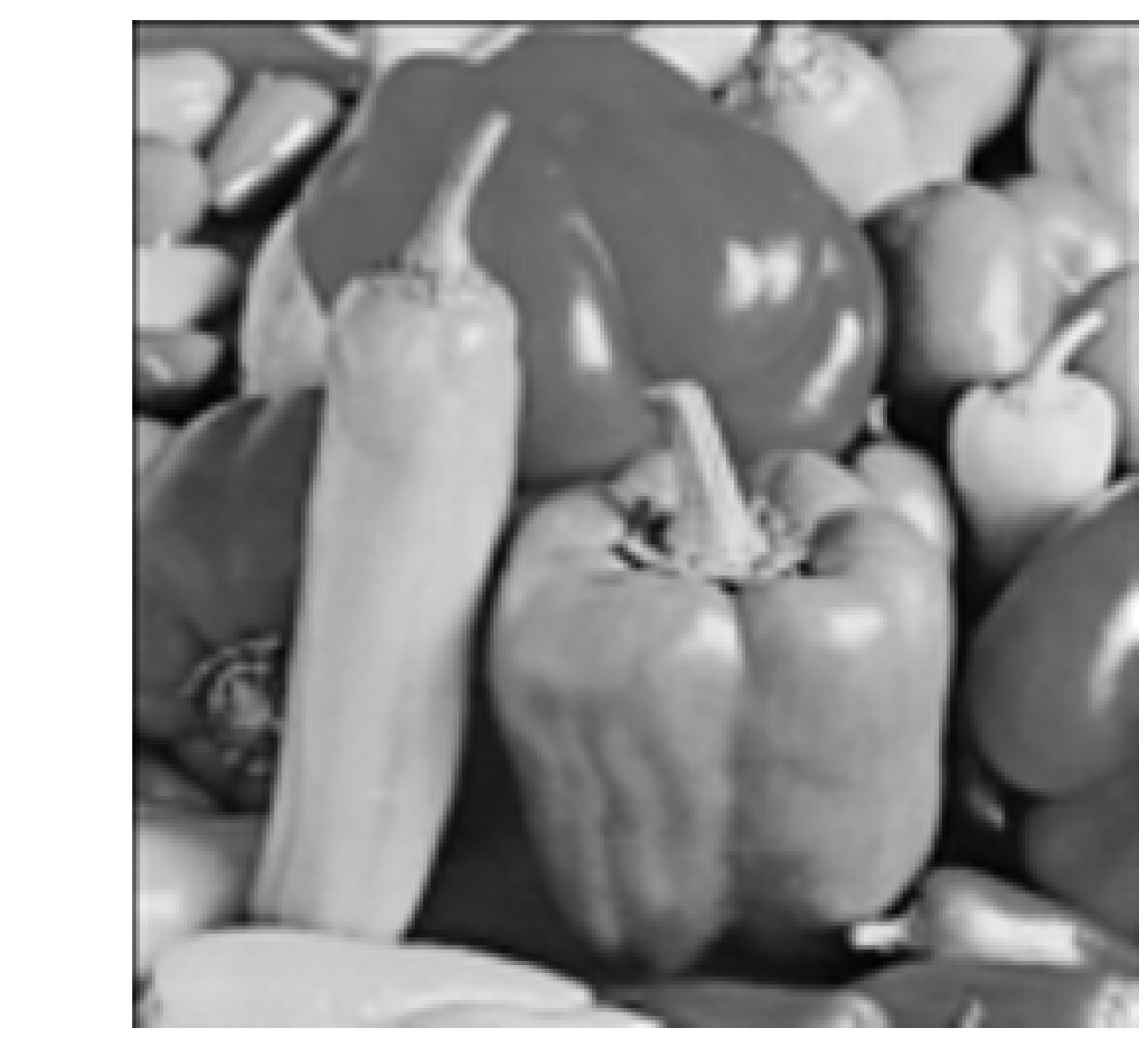}\\
			&[7.3929,0.0543]&[14.0788,0.1893]&[12.6388,0.0958]&[25.9255,0.8680]&[\textbf{26.0394},\textbf{0.8690}]&\\
		\end{tabular}
	}
	\caption{Comparison of the performance of the algorithms for the super-resolution of several gray-scale images. The PSNR and SSIM corresponding to each image have been also reported. The results confirm the quality of the proposed approach for the super-resolution of natural images.  
	}
	\label{natural}
\end{figure*}

Although this algorithm has been derived for OCT super-resolution, as a final simulation, we have demonstrated its quality for the super-resolution of natural images. For this purpose, the results of super-resolution of several $128\times 128$ gray-scale images have been shown in Fig.~\ref{natural}. The proposed approach has been compared with state of the art tensor based completion algorithms as  MDT \cite{yokota2018missing}, HALRTC \cite{liu2012tensor}, TRLRF \cite{yuan2019tensor} and TT-WOPT \cite{yuan2019high}. For the proposed approach, the patch size was set equal to 2 with overlap 1. Note that the final images have not been smoothed. In the MDT approach, the window size was set to [2,2] for the first row and [8,8] fo the second and the third rows. As the results show, the proposed approach has a higher performance in super-resolution of natural images. 

\section{conclusion}
\label{conc}

In this paper, a new approach for OCT super-resolution based on Tensor Ring (TR) decomposition has been developed and extensively tested. To the best of our knowledge, it was the first time of using TR decomposition for OCT super-resolution.   In the proposed approach, the data was first overlapped patch Hankelized and then the resulting higher order tensor has been approximated by TR decomposition. Since determining the TR ranks in advance is difficult, the  TR ranks, or better to say the size of the resulting core tensors, have been increased gradually during each iteration until the desired accuracy achieved or the rank reach its maximum. The simulation results and comparison with the other existing algorithms, confirm the effectiveness and higher performance of the proposed algorithm for the super-resolution of OCT and other natural images. 

\section{Acknowledgment}
This work was supported by Isfahan University of Medical Sciences (Grant number: 2400134).

\bibliographystyle{IEEEtran}
\bibliography{refs}

\end{document}